\documentclass[fleqn,twoside]{article}
\usepackage{epsfig}
\usepackage{xspace}
\usepackage{longtable}
\usepackage{graphicx}
\usepackage[figuresright]{rotating}

\def\Journal#1#2#3#4{{#1} {\bf #2} (#3) #4}

\def\NIM{\rm Nucl. Instrum. Methods}

\newcommand{\GeVc}{\ensuremath{\mbox{GeV}/c}\xspace}
\newcommand{\MeVc}{\ensuremath{\mbox{MeV}/c}\xspace}
\newcommand{\T}{\ensuremath{\mbox{T}}\xspace}

\newcommand{\cm}{\ensuremath{\mbox{cm}}\xspace}
\newcommand{\mm}{\ensuremath{\mbox{mm}}\xspace}

\newcommand{\mrad}{\ensuremath{\mbox{mrad}}\xspace}
\newcommand{\rad}{\ensuremath{\mbox{rad}}\xspace}
\newcommand{\ns}{\ensuremath{\mbox{ns}}\xspace}
\newcommand{\m}{\ensuremath{\mbox{m}}\xspace}

\newcommand{\ms}{\ensuremath{\mbox{ms}}\xspace}
\newcommand{\micros}{\ensuremath{\mu \mbox{s}}\xspace}
\newcommand{\ps}{\ensuremath{\mbox{ps}}\xspace}

\newcommand{\dedx}{\ensuremath{\mbox{d}E/\mbox{d}x}\xspace}
\newcommand{\pip}{\ensuremath{\pi^+}\xspace}
\newcommand{\pim}{\ensuremath{\pi^-}\xspace}
\newcommand{\piz}{\ensuremath{\pi^0}\xspace}
\newcommand{\bfpip}{\ensuremath{\mathbf {\pi^+}}\xspace}
\newcommand{\bfpim}{\ensuremath{\mathbf {\pi^-}}\xspace}

\newcommand{\dzeroprime}{\ensuremath{d'_0}\xspace}
\newcommand{\zzeroprime}{\ensuremath{z'_0}\xspace}
\newcommand{\evtspill}{\ensuremath{N_{\mathrm{evt}}}\xspace}
\newcommand{\pt}{\ensuremath{p_{\mathrm{T}}}\xspace}
\newcommand{\tht}{\ensuremath{\theta}\xspace}

\def\be{\begin{equation}}
\def\ee{\end{equation}}
\def\bea{\begin{eqnarray}}
\def\eea{\end{eqnarray}}

\newcommand{\bfGeVc}{\ensuremath{\mathbf {\mbox{\bf GeV}/c}}\xspace}
\begin{document}
\title{\bf Measurement of the production of charged pions by protons on a tantalum target}

\author{HARP Collaboration}

\maketitle

\begin{abstract}
  A measurement of the double-differential cross-section for the production
  of charged pions in proton--tantalum collisions emitted at large
  angles from the incoming beam direction is presented. 
  The data were taken in 2002 with the HARP detector in the T9 beam
  line of the CERN PS.
  The pions were produced by proton beams in a momentum range from
  3~\GeVc to  12~\GeVc hitting a tantalum target with a thickness of
  5\% of a nuclear interaction length.  
  The angular and momentum range covered by the experiment 
  ($100~\MeVc \le p < 800~\MeVc$ and $0.35~\rad \le \theta <2.15~\rad$)
  is of particular importance for the design of a neutrino factory.
  The produced particles were detected using a small-radius
  cylindrical time projection chamber (TPC) placed in a solenoidal
  magnet. 
  Track recognition, momentum determination and particle
  identification were all performed based on the measurements made with
 the TPC. 
  An elaborate system of detectors in the beam line ensured the
  identification of the incident particles.
  Results are shown for the double-differential cross-sections 
  $
  {{\mathrm{d}^2 \sigma}}/{{\mathrm{d}p\mathrm{d}\theta }}
  $
  at four incident proton beam momenta (3~\GeVc, 5~\GeVc, 8~\GeVc 
  and 12~\GeVc).  
  In addition, the pion yields within the acceptance of typical
  neutrino factory designs are shown as a function of beam momentum.
  The measurement of these yields within a single experiment
  eliminates most systematic errors in the comparison between rates
  at different beam momenta and between positive and negative pion
  production. 
\end{abstract}

\clearpage
\thispagestyle{plain}
\begin{center}
{\large HARP collaboration}\\
\newcommand{\afdoct}{{3}\xspace}
\vspace{0.1cm}
{\small
M.G.~Catanesi, 
E.~Radicioni
\\ 
{\bf Universit\`{a} degli Studi e Sezione INFN, Bari, Italy} 
\\
R.~Edgecock, 
M.~Ellis$^{1}$,          
S.~Robbins$^{2,3}$,      
F.J.P.~Soler$^{4}$
\\
{\bf Rutherford Appleton Laboratory, Chilton, Didcot, UK} 
\\
C.~G\"{o}\ss ling 
\\
{\bf Institut f\"{u}r Physik, Universit\"{a}t Dortmund, Germany} 
\\
S.~Bunyatov, 
A.~Krasnoperov, 
B.~Popov$^5$, 
V.~Serdiouk,        
V.~Tereschenko 
\\
{\bf Joint Institute for Nuclear Research, JINR Dubna, Russia} 
\\
E.~Di~Capua, 
G.~Vidal--Sitjes$^{6}$  
\\
{\bf Universit\`{a} degli Studi e Sezione INFN, Ferrara, Italy}  
\\
A.~Artamonov$^7$,   
P.~Arce$^8$,        
S.~Giani, 
S.~Gilardoni,       
P.~Gorbunov$^{7,9}$,  
A.~Grant,  
A.~Grossheim$^{10}$, 
P.~Gruber$^{11}$,    
V.~Ivanchenko$^{12}$,  
A.~Kayis-Topaksu$^{13}$,
J.~Panman, 
I.~Papadopoulos,  
J.~Pasternak, 
E.~Tcherniaev, 
I.~Tsukerman$^7$,   
R.~Veenhof, 
C.~Wiebusch$^{14}$,    
P.~Zucchelli$^{9,15}$ 
\\
{\bf CERN, Geneva, Switzerland} 
\\
A.~Blondel, 
S.~Borghi$^{16}$,  
M.~Campanelli,       
M.C.~Morone$^{17}$, 
G.~Prior$^{18}$,   
R.~Schroeter
\\
{\bf Section de Physique, Universit\'{e} de Gen\`{e}ve, Switzerland} 
\\
R.~Engel,
C.~Meurer
\\
{\bf Institut f\"{u}r Physik, Forschungszentrum Karlsruhe, Germany}
\\
\newcommand{\afkyot}{{19}\xspace}
I.~Kato$^{10,\afkyot}$ 
\\
{\bf University of Kyoto, Japan} %
\\
U.~Gastaldi
\\
{\bf Laboratori Nazionali di Legnaro dell' INFN, Legnaro, Italy} 
\\
\newcommand{\aflanl}{{20}\xspace}
G.~B.~Mills$^{\aflanl}$  
\\
{\bf Los Alamos National Laboratory, Los Alamos, USA} %
\\
J.S.~Graulich$^{21}$, 
G.~Gr\'{e}goire 
\\
{\bf Institut de Physique Nucl\'{e}aire, UCL, Louvain-la-Neuve,
  Belgium} 
\\
M.~Bonesini, 
A.~De~Min,          
F.~Ferri,           
M.~Paganoni,        
F.~Paleari          
\\
{\bf Universit\`{a} degli Studi e Sezione INFN, Milano, Italy} 
\\
M.~Kirsanov
\\
{\bf Institute for Nuclear Research, Moscow, Russia} 
\\
A. Bagulya, 
V.~Grichine,  
N.~Polukhina
\\
{\bf P. N. Lebedev Institute of Physics (FIAN), Russian Academy of
Sciences, Moscow, Russia} 
\\
V.~Palladino
\\
{\bf Universit\`{a} ``Federico II'' e Sezione INFN, Napoli, Italy} 
\\
\newcommand{\afclmb}{{20}\xspace}
L.~Coney$^{\afclmb}$, 
D.~Schmitz$^{\afclmb}$
\\
{\bf Columbia University, New York, USA} %
\\
G.~Barr, 
A.~De~Santo$^{22}$, 
C.~Pattison, 
K.~Zuber$^{23}$  
\\
{\bf Nuclear and Astrophysics Laboratory, University of Oxford, UK} 
\\
F.~Bobisut, 
D.~Gibin,
A.~Guglielmi, 
M.~Mezzetto
\\
{\bf Universit\`{a} degli Studi e Sezione INFN, Padova, Italy} 
\\
J.~Dumarchez, 
F.~Vannucci 
\\
{\bf LPNHE, Universit\'{e}s de Paris VI et VII, Paris, France} 
\\
U.~Dore
\\
{\bf Universit\`{a} ``La Sapienza'' e Sezione INFN Roma I, Roma,
  Italy} 
\\
D.~Orestano, 
F.~Pastore, 
A.~Tonazzo, 
L.~Tortora
\\
{\bf Universit\`{a} degli Studi e Sezione INFN Roma III, Roma, Italy}
\\
C.~Booth, 
C.~Buttar$^{4}$,  
P.~Hodgson, 
L.~Howlett
\\
{\bf Dept. of Physics, University of Sheffield, UK} 
\\
M.~Bogomilov, 
M.~Chizhov, 
D.~Kolev, 
R.~Tsenov
\\
{\bf Faculty of Physics, St. Kliment Ohridski University, Sofia,
  Bulgaria} 
\\
S.~Piperov, 
P.~Temnikov
\\
{\bf Institute for Nuclear Research and Nuclear Energy, 
Academy of Sciences, Sofia, Bulgaria} 
\\
M.~Apollonio, 
P.~Chimenti,  
G.~Giannini, 
G.~Santin$^{24}$  
\\
{\bf Universit\`{a} degli Studi e Sezione INFN, Trieste, Italy} 
\\
J.~Burguet--Castell, 
A.~Cervera--Villanueva, 
J.J.~G\'{o}mez--Cadenas, 
J. Mart\'{i}n--Albo,
P.~Novella,
M.~Sorel,
A.~Tornero
\\
{\bf  Instituto de F\'{i}sica Corpuscular, IFIC, CSIC and Universidad de Valencia,
Spain} 
}
\end{center}
\thispagestyle{plain}
\vfill
\rule{0.3\textwidth}{0.4mm}
\newline
\newpage
$^{~1}${Now at FNAL, Batavia, Illinois, USA.}
\newline
$^{~2}$Jointly appointed by Nuclear and Astrophysics Laboratory,
            University of Oxford, UK.
\newline
$^{~3}${Now at Codian Ltd., Langley, Slough, UK.}
\newline
$^{~4}${Now at University of Glasgow, UK.}
\newline
$^{~5}${Also supported by LPNHE, Paris, France.}
\newline
%
$^{~6}${Now at Imperial College, University of London, UK.}
\newline
$^{~7}${ITEP, Moscow, Russian Federation.}
\newline
$^{~8}${Permanently at Instituto de F\'{\i}sica de Cantabria,
            Univ. de Cantabria, Santander, Spain.} 
\newline
$^{~9}${Now at SpinX Technologies, Geneva, Switzerland.}
\newline
$^{10}${Now at TRIUMF, Vancouver, Canada.}
\newline
$^{11}${Now at University of St. Gallen, Switzerland.}
\newline
$^{12}${On leave of absence from Ecoanalitica, Moscow State University,
Moscow, Russia.}
\newline
$^{13}${Now at \c{C}ukurova University, Adana, Turkey.}
\newline
$^{14}${Now at III Phys. Inst. B, RWTH Aachen, Aachen, Germany.}
\newline
$^{15}$On leave of absence from INFN, Sezione di Ferrara, Italy.
\newline
$^{16}${Now at CERN, Geneva, Switzerland.}
\newline
$^{17}${Now at Univerity of Rome Tor Vergata, Italy.}
\newline
$^{18}${Now at Lawrence Berkeley National Laboratory, Berkeley, California, USA.}
\newline
$^{19}${K2K Collaboration.}
\newline
$^{20}${MiniBooNE Collaboration.}
\newline
$^{21}${Now at Section de Physique, Universit\'{e} de Gen\`{e}ve, Switzerland, Switzerland.}
\newline
$^{22}${Now at Royal Holloway, University of London, UK.}
\newline
$^{23}${Now at University of Sussex, Brighton, UK.}
\newline
$^{24}${Now at ESA/ESTEC, Noordwijk, The Netherlands.}

\clearpage

\section{Introduction}

The HARP experiment aims at a systematic study of hadron
production for beam momenta from 1.5~\GeVc to 15~\GeVc for a large
range of target nuclei~\cite{harp-prop}. 
The main motivations are the measurement of pion yields for a quantitative
design of the proton driver of a future neutrino factory, a
substantial improvement of the calculation of the atmospheric neutrino
flux~\cite{Battistoni,Stanev,Gaisser,Engel,Honda} 
and the measurement of particle yields as input for the flux
calculation of accelerator neutrino experiments, 
such as K2K~\cite{ref:k2k,ref:k2kfinal},
MiniBooNE~\cite{ref:miniboone} and SciBooNE~\cite{ref:sciboone}. 


In this paper we address one of the main motivations of the HARP
experiment: the measurement of the yields of positive and negative
pions for a quantitative design of a proton driver and a target 
station of a future neutrino factory. 
In order to achieve the highest number of potentially collected pions
of both charge signs per unit of energy a pion production measurement
should give the information necessary to optimize both proton beam
energy and target material. 
At the moment the CERN scenario makes provision for a 3~\GeVc \ -- \ 
5~\GeVc proton linac with a target using a high-$Z$
material~\cite{ref:nufact}. 
Other scenarios are contemplated and may call for higher energy
incident beams.
In most cases targets are foreseen with high-$Z$ materials.
For this reason it was decided to analyse first a series of settings
taken with a range of different beam momenta  incident on a tantalum
target. 
The different settings have been taken within a short period so that
in their comparison detector variations are minimized.
Also similar data sets on lead, tin, copper, aluminium, carbon and
beryllium have been collected.
These will be presented in future papers.

Here, the measurement of the double-differential cross-section, 
$
{{\mathrm{d}^2 \sigma^{\pi}}}/{{\mathrm{d}p\mathrm{d}\theta }}
$
for $\pi^{\pm}$ production by
protons of 3~\GeVc, 5~\GeVc, 8~\GeVc and 12~\GeVc momentum impinging
on a thin Ta target of 5\% nuclear interaction length
($\lambda_{\mathrm{I}}$) is presented. 

%
%

 The HARP experiment~\cite{harp-prop,ref:harpTech}
 makes use of a large-acceptance spectrometer consisting of a
 forward and large-angle detection system. 
 The forward spectrometer covers polar angles up to 250~\mrad which
 is well matched to the angular range of interest for the 
 measurement of hadron production to calculate the properties of
 conventional neutrino beams.
 The HARP publications devoted to the measurements of the $\pi^+$
 production cross-sections in proton interactions with 
 aluminium~\cite{ref:alPaper,ref:pidPaper} and
 beryllium~\cite{ref:bePaper}
 targets are relevant for the K2K and MiniBooNE neutrino
 oscillation experiments.
 The large-angle spectrometer has a large acceptance in the momentum
 and angular  range for the pions relevant to the production of the
 muons in a  neutrino factory. 
 It covers the large majority of the pions accepted in the focusing
 system of a typical design.
 The neutrino beam of a neutrino factory originates from
 the decay of muons which are in turn the decay products of pions
 produced by a proton beam hitting a production target.
For this programme of measurements data were taken with high-$Z$ nuclear
 targets such as tantalum and lead.

 The results reported here are based on data taken in 2002 in
 the T9 beam of the CERN PS.
 About one million incoming protons were selected which gave an
 interaction trigger in the Large Angle spectrometer collected at four
 distinct beam momenta.
 After cuts, 150,000 secondary pion tracks reconstructed in the large-angle
 spectrometer were used in the analysis.

%
%

The analysis proceeds by selecting tracks in the Time Projection
Chamber (TPC) in events with incident beam protons.  
Momentum and polar angle measurements and particle identification are
based on the measurements of track position and energy deposition in
the TPC.
An unfolding method is used to correct for experimental resolution,
efficiency and acceptance and to obtain the double-differential pion
production cross-sections.

 The experimental apparatus is outlined in Section~\ref{sec:apparatus}.
 Section~\ref{sec:tracking} describes track reconstruction and
 measurement of \dedx with the large-angle spectrometer. 
 The event and track selection for the analysis is described in
 Section~\ref{sec:selection}. 
 The performance of the detector and the methods employed to
 characterise the performance are shown in
 Section~\ref{sec:performance}. 
 Section~\ref{sec:xsec} describes details of the cross-section calculation.  
 Results are discussed in Section~\ref{sec:results}.
 A comparison with previous data is presented in Section~\ref{sec:compare}.
 An approximate calculation of the yield of pions within the
 acceptance of typical focusing systems of some neutrino factory
 designs is given in Section~\ref{sec:factory}.
 The conclusions are presented in Section~\ref{sec:summary}.
 Tables with all cross-section data and a comparison with an alternative
 analysis of the data are given in appendices.

\section{Experimental apparatus}
\label{sec:apparatus}
\begin{figure}[tbp]
  \begin{center}
    \hspace{0mm} \epsfig{file=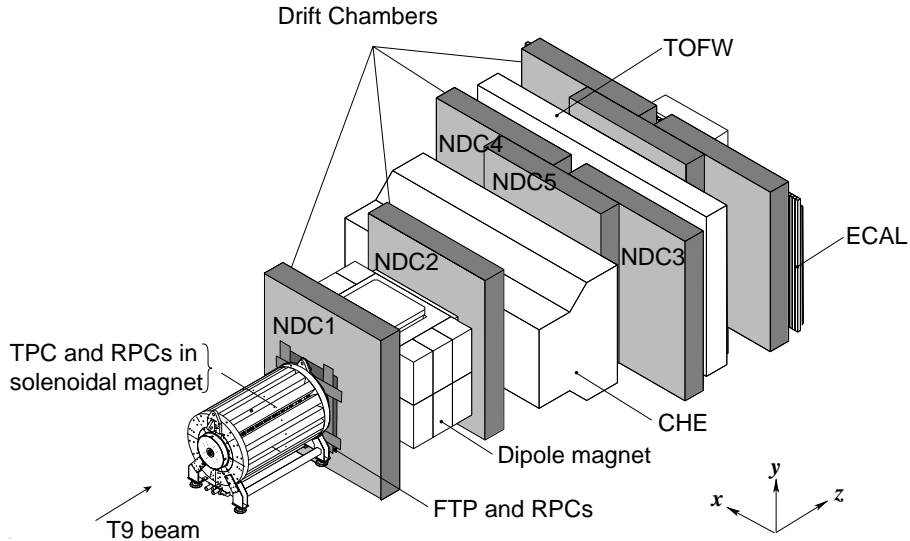,width=12cm}
  \end{center}
\caption{Schematic layout of the HARP detector. 
The convention for the coordinate system is shown in the lower-right
corner. 
The three most downstream (unlabelled) drift chamber modules are only partly
equipped with electronics and are not used for tracking.
The detector covers a total length of 13.5~m along the beam direction
 and has a maximum width of 6.5~m perpendicular to the beam.
}
\label{fig:harp}
\end{figure}

 The HARP detector is shown in Fig.~\ref{fig:harp}.
 The forward spectrometer is built around a dipole magnet 
 for momentum analysis, with large planar drift chambers
 (NDC)~\cite{NOMAD_NIM_DC} for particle tracking which had been used
 originally in the NOMAD experiment~\cite{NOMAD_NIM}, and  a
 time-of-flight wall 
 (TOFW)~\cite{ref:tofPaper}, a threshold Cherenkov detector (CHE), 
 and an electromagnetic calorimeter (ECAL)~\cite{ref:chorus-cal} 
 used for particle identification.
 In the large-angle region a cylindrical TPC with a radius of 408~\mm
 is  positioned in a solenoidal magnet with a field of 0.7~\T. 
 The TPC is used
 for tracking, momentum determination and the measurement of the
 energy deposition \dedx for particle identification~\cite{ref:tpc:ieee}.
 A set of resistive plate chambers (RPC) form a barrel inside the solenoid 
 around the TPC to measure the time of arrival of the secondary
 particles~\cite{ref:rpc}. 
 Beam instrumentation provides identification of the incoming
 particle, the determination of the time when it hits the target, 
 and the impact point and direction of the beam particle
 on the target.  
 Several trigger detectors are installed to select events with an
 interaction and to define the normalization.
 
Data were taken with a number of beam momentum settings and with
different target materials and thicknesses.
In addition to the data taken with the thin tantalum target of
5\%~$\lambda_{\mathrm{I}}$,
runs were also taken with an empty target holder, a 
thin 2\%~$\lambda_{\mathrm{I}}$ target and a
thick 100\%~$\lambda_{\mathrm{I}}$ target.
Data taken with a liquid hydrogen target at 3~\GeVc, 5~\GeVc and
8~\GeVc incident beam momentum together with cosmic-ray data were used 
to provide an absolute calibration of the efficiency, momentum scale and
resolution of the detector. 
In addition, the tracks produced in runs with Pb, Sn and Cu targets in
the same period and with the same beam settings were used for
the calibration of the detector, event reconstruction and analysis
procedures. 
The momentum definition of the T9 beam is known with a precision of
the order of 1\%~\cite{ref:t9}. 
The absolute normalization of the number of incident protons was
performed using 250,000 `incident-proton' triggers. 
These are triggers where the same selection on the beam particle was
applied but no selection on the interaction was performed.
The rate of this trigger was down-scaled by a factor 64.
A cross-check of the absolute normalization was provided by counting
tracks in the forward spectrometer.

A detailed description of the HARP apparatus is
given in Ref.~\cite{ref:harpTech}. 
In this analysis primarily the detector components
of the large-angle spectrometer and the beam instrumentation are employed.
Below, the detector elements which are important for this analysis will be
briefly described.

\subsection{Beam, target and trigger detectors}
\label{sec:beamtrigger}

\begin{figure}[tbp]
\begin{center}
  \epsfig{file=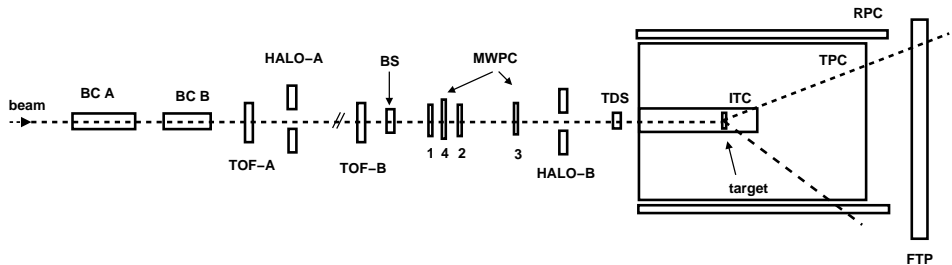,width=0.8\textwidth}
\caption{ Schematic view of the trigger and beam
equipment. The description is given  in the text. The beam
enters from the left. The MWPCs are numbered: 1, 4, 2, 3 from left to
right. On the right, the position of the target inside the inner field
cage of the TPC is shown.}
\label{fig:beamline}
\end{center}
\end{figure}

A sketch of the equipment in the beam line is shown in
Fig.~\ref{fig:beamline}. 
A set of four multi-wire
proportional chambers (MWPCs) measures the position and direction of
the incoming beam particles with an accuracy of $\approx$1~\mm in
position and $\approx$0.2~\mrad in angle per projection.
At low momenta the precision of the prediction at the target is
limited by multiple scattering and increases to $\approx$1~\mrad at
3~\GeVc. 
A beam time-of-flight system (BTOF)
measures the time difference of particles over a $21.4$~m path-length. 
It is made of two
identical scintillation hodoscopes, TOFA and TOFB (originally built
for the NA52 experiment~\cite{ref:NA52}),
which, together with a small target-defining trigger counter (TDS,
also used for the trigger and described below), provide particle
identification at low energies. This provides separation of pions, kaons
and protons up to 5~\GeVc and determines the initial time at the
interaction vertex ($t_0$). 
The timing resolution of the combined BTOF system is about 70~\ps.
A system of two N$_2$-filled Cherenkov detectors (BCA and BCB) is
used to tag electrons at low energies and pions at higher energies. 
The electron and pion tagging efficiency is found to be close to
100\%.
The fraction of protons compared to all hadrons in the beam is
approximately 35\%, 43\%, 66\% and 92\% in the 3~\GeVc, 5~\GeVc, 8~\GeVc
and 12~\GeVc beam, respectively.

The length of the accelerator spill is 400~ms with a typical intensity
of 15~000 beam particles per spill.
The average number of events recorded by the data acquisition ranges
from 300 to 350 per spill for the four different beam momenta.

The target is placed inside the inner field cage (IFC) of the TPC such that,
in addition to particles produced in the forward direction, 
backward-going tracks can be measured.
It has a cylindrical shape with a nominal diameter of 30~\mm.
The tantalum (99.95\% pure) target used for the 
measurement described here has a
nominal thickness of 5\%~$\lambda_{\mathrm{I}}$.
Precise measurements of the thickness and diameter have been performed
at different locations on its surface.
These show a maximum variation of the thickness between 
5.55~\mm and 5.66~\mm and of the diameter between 30.135~\mm and
30.15~\mm. 
A set of trigger detectors completes the beam instrumentation: a
thin scintillator slab covering the full aperture of the last
quadrupole magnet in the beam line to start the trigger logic
decision (BS); a small scintillator disk, TDS mentioned above, positioned
upstream of the target to ensure that only 
particles hitting the target cause a trigger; and `halo' counters
(scintillators with a hole to let the beam particles pass) to veto
particles too far away from the beam axis. 
The TDS is designed to have a very high efficiency (measured to be 99.9\%).
It is located as near as possible to the entrance of
the TPC and has a 20~mm diameter, smaller than the target.
Its time resolution ($\sim 130 $~\ps) is sufficiently good to be used
as an additional detector for the BTOF system.
A cylindrical detector (inner trigger cylinder, ITC) made of six
layers of 1~\mm thick scintillating fibres is positioned inside the
inner field cage of the TPC and surrounds the target.
It provides full coverage of the acceptance of the TPC.
The efficiency of the ITC was measured using events which had been
taken simultaneously using incident-proton triggers which did not
require the ITC and amounts to $>$99.5\%. 
For the incident-proton triggers, also the interaction trigger bits were
stored by the DAQ, although they were not required to record the event.
The recorded TPC data in events taken with the incident proton beam were
passed through the track finding algorithm and for each event with at least
one TPC track the ITC trigger decision was checked.
The efficiency per single ITC layer was found to be typically 80\%,
giving a large redundancy for the OR-signal to reach the quoted overall
efficiency.

\subsection{Large-angle spectrometer}
The large-angle spectrometer consists of a TPC and a set of RPC 
detectors inside the solenoidal magnet. 
The TPC detector was designed to measure and identify tracks in the
angular region from 0.25~\rad to 2.5~\rad from the beam axis.
Charged particle identification (PID) can be achieved by measuring the 
ionization per unit length in the gas (\dedx) as a function of the total
momentum of the particle. 
Additional PID can be performed through a time-of-flight 
measurement with the RPCs.

\begin{figure}[tbp]
\vspace{9pt}
\begin{center}
  \epsfig{figure=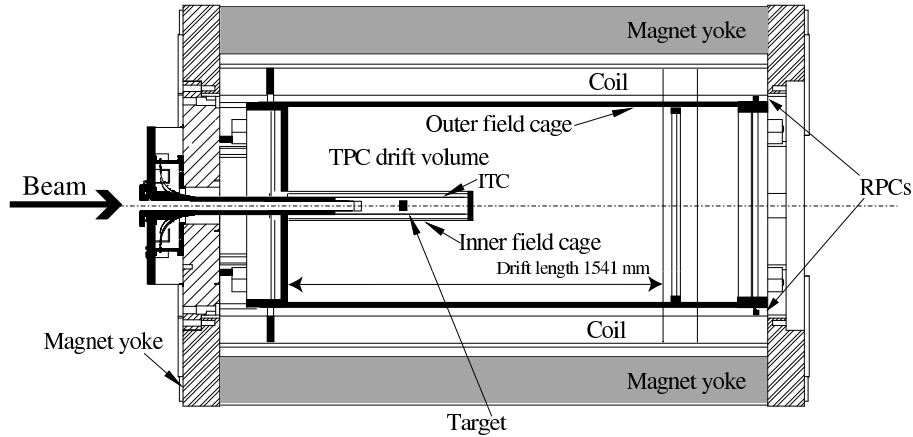,width=0.75\textwidth,angle=0}
\end{center}
\caption{ Schematic layout of the  TPC. The beam enters from the left.
Starting from the outside, first the return yoke of the magnet is
seen, closed with an end-cap at the upstream end, and open at the
downstream end.  The cylindrical coils are drawn inside the yoke. 
The field cage is positioned inside this magnetic volume.
The inner field cage is visible as a short cylinder entering from the left.
The ITC trigger counter and the target holder are inserted in the inner
 field cage.  
The RPCs (not drawn) are positioned between the outer field cage and the
 coil.  
}
\label{fig:tpc}
\end{figure}

Figure~\ref{fig:tpc} shows the schematic layout of the HARP TPC.
The TPC is positioned inside the solenoid magnet. 
The solenoid provides  a magnetic volume with
a diameter of 0.9~\m, a length of 2.25~\m  and a field of 0.7~\T. 
Secondary particles enter the forward spectrometer through the
downstream end of the return yoke which is left open.
At the upstream end there is a small cylindrical hole in the end-cap
for the passage of the incident beam and to insert the ITC and target
holder inside the IFC.
The magnet was previously used for R\&D for the TPC of the ALEPH
experiment and modified for this experiment. 
The induced charge from the gas amplification at the anode wires is
measured using a plane with twenty concentric rows of pads, each
connected to a pre-amplifier.
The pad plane is subdivided into six sectors.  
The anode wires are strung onto six spokes subdividing the six sectors.
The pad plane is subdivided radially into 20 rows of pads.
The pad dimensions are $6.5 \ \mm \ \times 15 \ \mm$ and there are
from 11 (at the inner radius) to 55 (at the outer radius) such pads per
row per sector. 
The drift volume is 1541~\mm long with a field gradient of 111~V/\cm,
resulting in a maximum drift time of approximately 30~\micros.
The pad-charges are sampled by an FADC (one per pad) each 100~\ns.
The total DAQ readout time is 500~\micros to 1000~\micros per event
depending on the event size.
 
Thirty RPC chambers are arranged in the shape of a barrel around the
TPC providing full coverage in azimuth and covering polar angles from
0.3~\rad to 2.5~\rad
with respect to the beam axis.  
The individual chambers are 10~\mm thick, 150~\mm wide and 2~\m long. 
Together with the timing measurement of the beam detectors the RPC
system provides a measurement of time-of-flight of particles produced
at large angles from the beam axis.

In the present analysis, the TPC provides the measurement for the
pattern recognition to find the particle tracks, and to measure their
momentum through the curvature of their trajectory. 
It also provides PID using the measurement of energy deposition.
The RPC system is used in this analysis to provide a calibration of
the PID capabilities of the TPC.

Besides the usual need for calibration of the detector, a number of
hardware shortfalls, discovered mainly after the end of data-taking,
had to be overcome to use the TPC data reliably in the analysis.
The TPC contains a relatively large number of dead or noisy pads.
Noisy pads are disregarded in the analysis and therefore equivalent to
dead pads.  
The problem of dead channels present during operation ($\approx 15\%$) 
necessitates a day--by--day determination of the dead channel map.
The same map is used in the simulation, providing a description of the
performance of the TPC adjusted to the conditions of each short period
of data taking.
A method based on the tracks measured during normal data
taking was developed to measure the variations of
the overall gain of each pad, including the gas gain,  by
accumulating for each pad all the data taken during a period in
time over which the performance of the chamber can be considered
constant (typically a few hours)~\cite{ref:harpTech}.
In addition, this method allows dead and noisy channels to be
identified.
It is used to reduce the fluctuation in the response between pads down
to a 3\% level. 

The well-known position of the target and of
the end-flange of the IFC are used to determine the drift velocity by
reconstructing tracks emerging from these materials.
Since the drift velocity varies as a function of operational parameters
such as pressure, temperature and gas-mixture, it is determined for
each relatively short data taking period. 
Variations of up to 4\% were observed~\cite{ref:harpTech}.
The precision of the calibration for individual periods is better than
0.5\%.

Static distortions of the reconstructed trajectories are observed in
the TPC. 
The most important ones are caused by the inhomogeneity of the magnetic
field and an accidental HV mismatch between the inner and outer field
cage (powered with two distinct HV supplies).
The distortions were studied in detail using cosmic-ray data obtained
with a special calibration run performed after the data taking
period. 
Appropriate distortion correction algorithms in the TPC reconstruction
software compensate for the voltage offset and for the inhomogeneities 
of the magnetic field.

Dynamic distortions which are caused by the build-up of ion-charge
density in the drift volume during the 400~\ms long beam spill are
observed in the chamber. 
Changes in the beam parameters (intensity, steering) cause
an increase or decrease in the dynamic distortions. 
While methods to correct the dynamic distortions are under development,
a pragmatic approach is chosen to limit the analysis to the early part of
the beam spill where the effects of dynamic distortions are still
small.
The time interval between spills is large enough to drain all charges in
the TPC related to the effect of the beam.
The combined effect of the distortions on the kinematic quantities
used in the analysis has been studied in detail, and only that part of
the data for which the systematic errors can be controlled with
physical benchmarks is used.
More than 30\% of the data can be retained.

\begin{figure}[tbp]
  \begin{center}
    \epsfig{figure=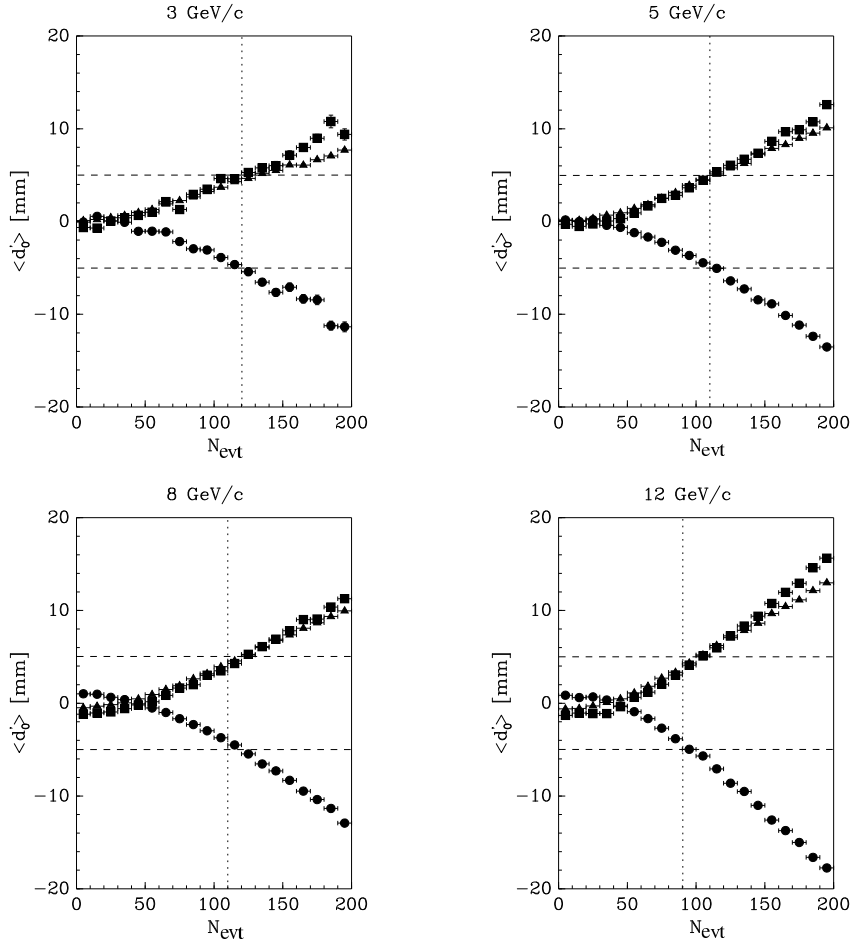,width=0.7\textwidth,angle=0}
  \end{center}
\caption{
Effect of dynamic distortions as a function of event
number in the spill for the four tantalum settings used in the
analysis emphasizing the first part of the spill (200 events).  
The symbols show the average extrapolated 
distance from the incoming beam particle trajectory for $\pi^-$
(filled circles), $\pi^+$ (filled squares), and protons
(filled triangles).
The momentum of the beam is indicated in the title of the panels.
Data with $\langle \dzeroprime \rangle < 5 \ \mm$ have been used in
the analysis. 
The dashed horizontal lines indicate the value at $\pm 5$ \mm to guide
 the eye, while the vertical dotted lines show the maximum value of
 \evtspill accepted in the analysis.
}
\label{fig:dynamic}
\end{figure}

The influence of the distortions can be monitored using the average
value of the extrapolated minimum distance of secondary tracks from the
incoming beam particle trajectory  $\langle \dzeroprime \rangle$.
In Fig.~\ref{fig:dynamic} this quantity
is plotted separately for positively and negatively 
charged pion tracks and protons as a function of the event number
within the spill for the four beam settings used. 
Due to the sign-convention for \dzeroprime, the distortions shift its
value in opposite directions for particles tracks of positive and
negative charge. 
It can clearly be seen that this distance increases with time. 
The effect also increases with beam momentum; this is expected from the
track multiplicity increase.
Also the beam intensity was higher for higher beam momenta.
As will be shown in the following, data taken under conditions where
the average \dzeroprime is smaller than 5~\mm can be
analysed reliably. 
For the analysis presented here, this results in a limit of 
$\approx 100$ events per spill, depending on the setting.
The performance of the chamber for this subset of the data was
studied using several methods, including the analysis of elastic
events in exposures of a liquid hydrogen target.
These results will be shown in subsequent sections.
The small mismatch extrapolated to \evtspill=~0 visible in the 8~\GeVc
 and 12~\GeVc data are due to residual static distortions.  Although the
 latter show a variation among different settings a common correction is
 applied.  The systematic error introduced in the momentum calibration
 by this approximation is estimated to be less than 1\%.

\section{Track reconstruction}
\label{sec:tracking}

The reconstruction of particle trajectories in the TPC is implemented
with a sequence of distinct steps.

After unpacking of the raw data, time-series of flash-ADC values
representing the charge collected on pads are combined into 
clusters on the basis of individual pad rows. 
Hits in neighbouring pads with time stamps that differ by
less than 600~\ns are included in the cluster.
Each cluster gets a weighted position in the $r\varphi$
direction along the pad row using the pad positions and in the $z$
direction using the time information. 
The reference time is defined on the rising
edge of the signal when the first pulse in a cluster goes over
threshold. 

The clusters are then assigned to tracks by a general purpose pattern
recognition algorithm.
This algorithm uses a general framework for track finding in
multi-dimensional space~\cite{uiterwijk}, in this case applied to a
3-D situation.  
The framework does not impose a preferred search direction. 
In an initial phase clusters are sorted into an $N$-dimensional
binary tree to prepare an efficient look-up of nearest neighbours. 
The algorithm then builds a network of all possible links
between the clusters.  
Links are acceptable if the distance between the clusters is small
(accepting hits on nearby pad rows, taking into account possible gaps in
the track hits). 
Then it builds a tree of connected clusters,
starting from `seeds'.  
As seed any of the linked pairs of hits is tried, taking first those on the
outer pad row and then the unused links on the next inner row, and so on.
Despite the magnetic field, the track
model approximates tracks locally as straight lines.  
When from a given link multiple continuations are possible a choice has
to be made which continuation is to be used to form the final track.
The branch of the tree which is retained as the best continuation of the
track is determined by examining pairs of fully grown branches and
selecting the better one.  
Since the tree is built recursively, it suffices to compare possible
continuations from a given link pair-wise.
The general framework allows the specific implementation to define the
criterion used to make this choice.
In the case of the trackfinding in the HARP TPC with its low occupancy
of hits the choice of the branch with the largest number of clusters
is sufficient.
Parameters in the framework which can be adjusted to the particular
situation are the minimum number of points for an accepted track, the
maximum curvature, the maximum distance between consecutive clusters
and the criterion to choose the best of two possible solutions for a
branch on a tree.

\subsection{Momentum measurement}
\label{sec:momentum:measurement}

Once clusters are assigned to a track, the track is fitted to a helix.
The fitting procedure is based on the algorithm developed by the ALEPH
Collaboration~\cite{ref:aleph} with slight modifications, {\em e.g.} 
the possibility to fit tracks which spiral for more than 2$\pi$~\cite{ref:morone}.
The fit consists of two consecutive steps: a circle-fit  
in the $x$--$y$ plane based on a least-square method~\cite{chernov} which 
defines three parameters, and a subsequent straight line fit in $z$--$s_{xy}$ 
plane\footnote{The $s_{xy}$ coordinate is defined as the arc 
length along the circle in the $x$-$y$ plane between a point and the 
impact point.} which defines two other parameters.
A helix is uniquely defined by these parameters.
The code uses the same sign conventions as in the TASSO and ALEPH 
software~\cite{ref:aleph} with a particle direction associated
to the motion along the helix itself.
Different classes of precision can be assigned to clusters along $r$
and $\phi$ depending on the number of hits that belong to a cluster
and depending on whether a cluster is near to a region of dead pads. 
This classification was developed from studies of the residuals
observed in the data and also quantified using simulated data.
As was done in the original ALEPH method, weights are applied to take
into account the differences in cluster quality, a method which is
applicable to errors of systematic nature. 

Tracks which are emerging from the target are refitted using the
position of the extrapolated beam particle as an extra point in the fit
with a weight similar to a TPC hit (`constrained fit').
Refitting the track parameters imposing the vertex constraint improves
the momentum resolution significantly at the cost of a moderate loss
of efficiency of a few percent. 
The energy-loss in the materials along the particle trajectory is not
taken into account in the fit\footnote{The constrained fit is
performed using the analytical helix track model.}.
However, in the analysis these effects are corrected for (see
Section~\ref{sec:corrections}) by applying the same procedure to the
data and the simulation. 

\begin{figure}[tbp]
  \begin{center}
    \epsfig{figure=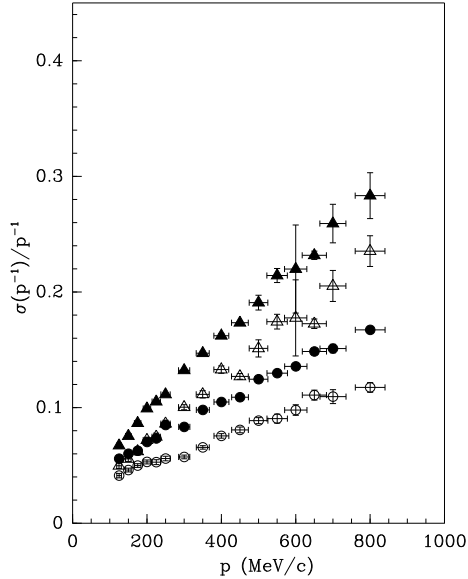,width=0.38\textwidth,angle=0}
  \end{center}
\caption{
  Simulation study of the resolution of the inverse of the momentum
  determination using the constraint of the extrapolated beam
  particle.  
  Results for charged pions emitted at large angles ($85^0$) are shown
  with triangles, while the circles represent the behaviour at smaller
  angles ($35^0$). 
  The resolution of the measured momentum is compared with the `true'
  momentum in the gas.
  All materials in the detector and the target are
  taken into account in the simulation.
  A fit with two Gaussians constrained to have the same mean has been
  performed to the distributions.
  Filled circles and triangles show the measurement of the RMS of the
  fitted function, while the open circles and triangles show the $\sigma$
  of the narrow Gaussian.  
}
\label{fig:p:sigma:rms}
\end{figure}

A study with the simulation program of the resolution of the inverse
of the momentum determination using the constraint of the extrapolated
beam particle is shown in Fig.~\ref{fig:p:sigma:rms}.
Results for particles emitted at large angles ($85^0$) are shown
together with the behaviour at smaller angles ($35^0$). 
The resolution of the measured momentum is compared with the `true'
momentum in the gas.
A fit to the distributions with two Gaussians constrained to have the
same mean has been performed.
The measurement of the RMS of the sum of the Gaussians is compared with the
$\sigma$ of the narrow Gaussian.  
The RMS is larger by 25\%--30\% than the $\sigma$ of the narrow
Gaussian, indicating the presence of non-Gaussian tails.
The difference between the two angles is expected from the fact that
the resolution is a function of \pt rather than $p$.
The tails of the distributions are fully taken into account in the
analysis. 
Although the track curvature is measured mainly in the gas, the
resolution extrapolates to \pt=~0 with a $\approx$2\% constant term. 
This is due to the use of the vertex point as constraint in the fit,
which adds the effect of multiple scattering in the inner field cage
and trigger counter, and to the use of a perfect helix as track
description neglecting inhomogeneities in the magnetic field which are
present in the simulation of the trajectories.
The experimental measurement of the resolution of the determination of
the momentum is consistent with the simulation and will be described
in a following section. 

\begin{figure}[tbp]
  \begin{center}
    \epsfig{figure=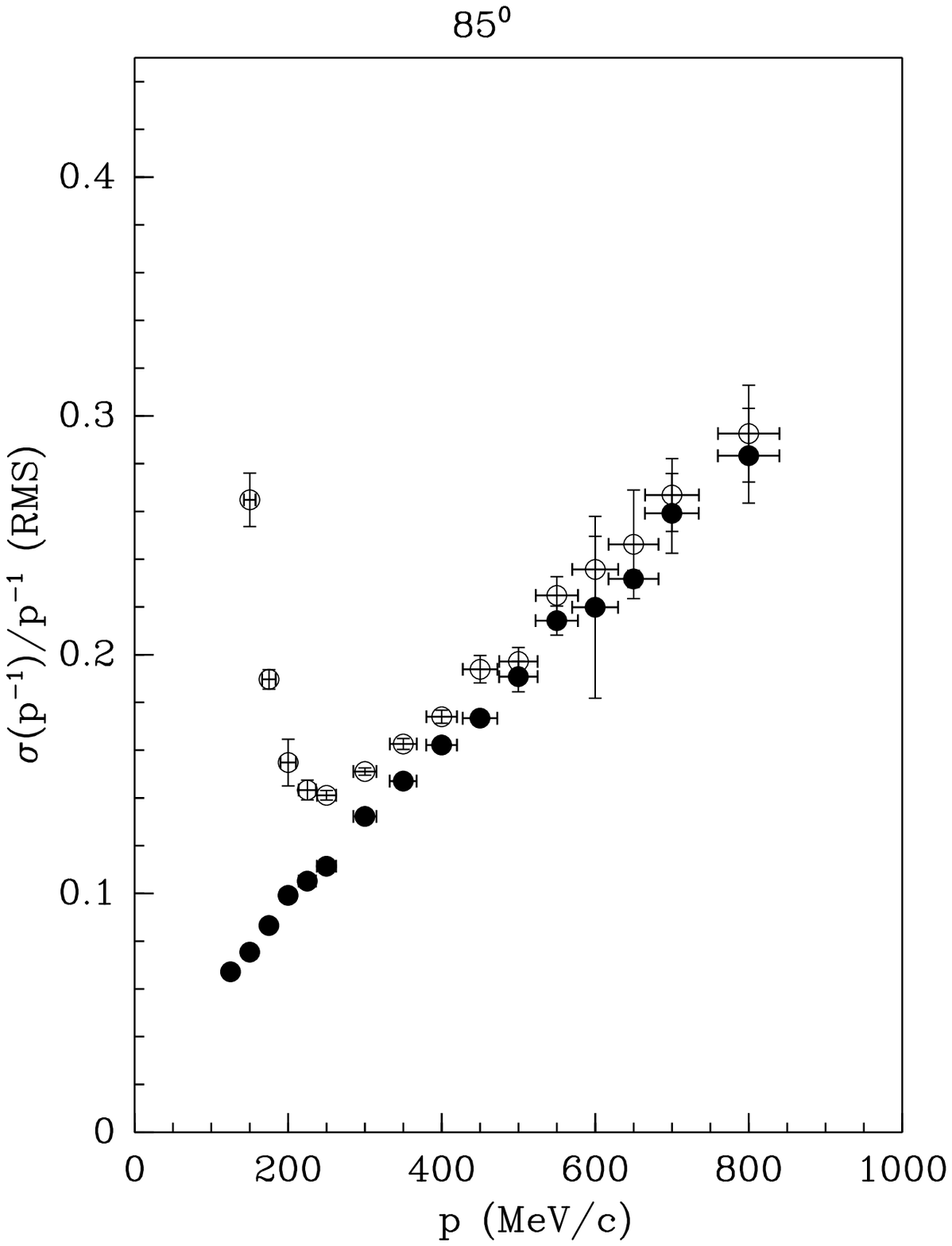,width=0.38\textwidth,angle=0}
    ~~
    \epsfig{figure=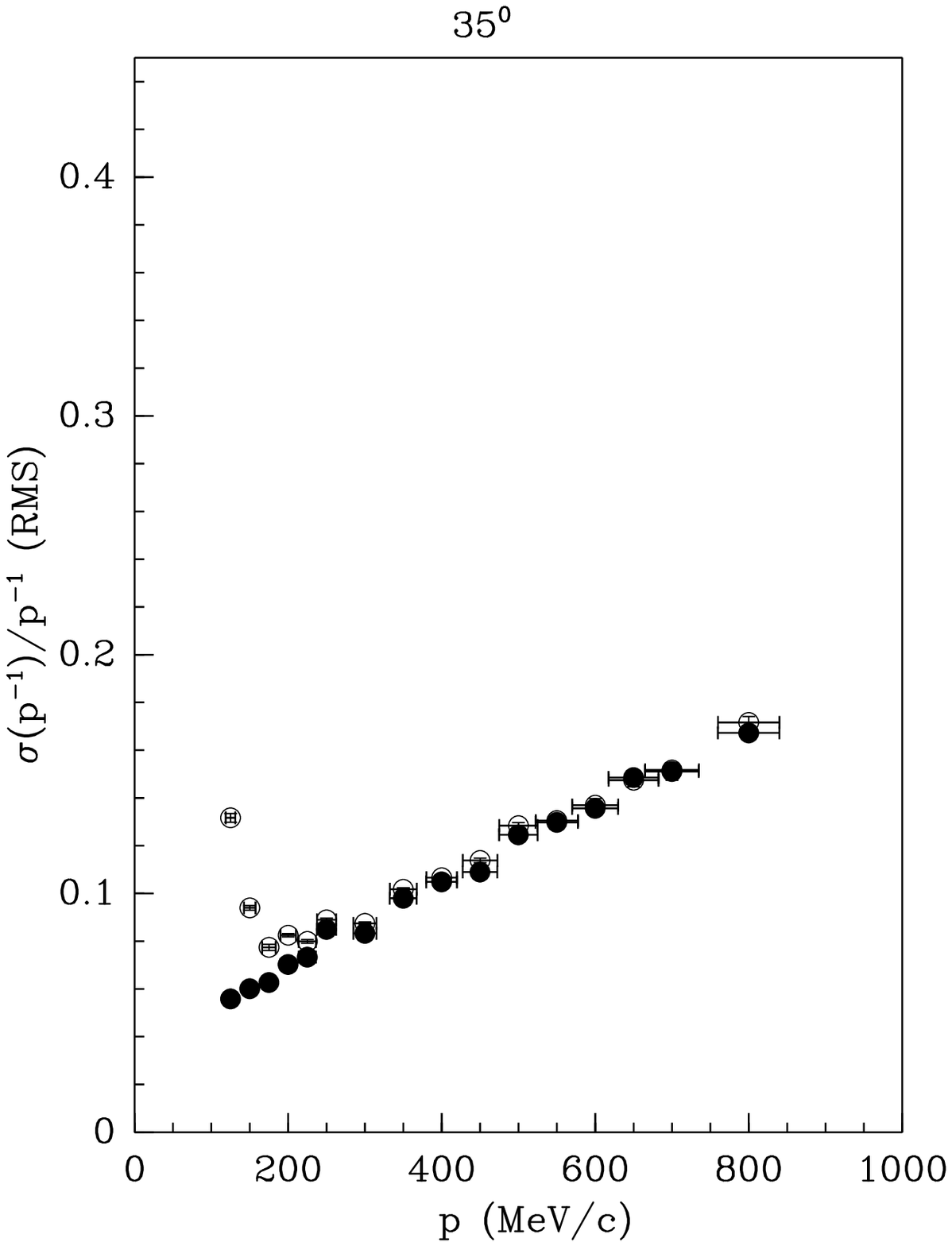,width=0.38\textwidth,angle=0}
  \end{center}
\caption{
 Simulation study of the resolution of the determination of the inverse
 of the momentum using the constraint of the extrapolated beam particle.  
 The left panel shows the results for charged pions emitted at large
  angles ($85^0$), while the right panel shows the behaviour at smaller
  angles ($35^0$).
  All materials in the detector and the target are
  taken into account.
  Filled circles show the resolution of the inverse of the momentum 
  determination with respect to the `true' momentum of the particle in
  the TPC gas volume for pions (measured by the RMS of the
  distribution).
  Open circles and boxes show the resolution with respect to the
  `true' momentum at the interaction vertex.  
  The effect of energy-loss in the material
  (target, trigger detector, IFC) is clearly visible.
}
\label{fig:p:resolution}
\end{figure}

Figure~\ref{fig:p:resolution} shows a simulation study of the
resolution of the inverse of the momentum using the constraint of the
extrapolated beam particle both with respect to the true momentum in the gas
and at the interaction vertex.  
The resolution of the momentum determination with respect to the
momentum at the interaction vertex suffers from the effect of
energy-loss in the material (target, trigger detector, IFC).
The large difference of the effect of the material between large and
small angles is due to the relatively large transverse dimensions of
the tantalum target (15~\mm radius) compared to the thickness of only
5.6~mm in the direction of the beam.

\subsection{Measurement of \dedx and time-of-flight}

The mean energy-loss per unit length for each particle trajectory is
calculated by an algorithm which evaluates the \dedx for each cluster
on the track associated to each curvilinear TPC pad row.
The d$x$ is calculated considering the segment of the helicoidal 
trajectory of the particle in that row,
the d$E$ is the total charge collected by the pad plane for
that cluster summing all ADC 
counts collected by the pads that belong to that cluster.
The resulting distribution of \dedx of the individual clusters is
expected to follow a Vavilov distribution.
To obtain the most reliable estimate for the peak (the most probable
value) of the Vavilov distribution, an algorithm has been developed
using a truncated mean.
The algorithm has been optimized by selecting all clusters of all
tracks in slices of momentum for preselected pion and proton samples,
respectively; hence this technique allowed a characterization of the
\dedx distributions to be made with extremely high statistics.
It is found that calculating an average value using 80\% of the clusters,
removing the 20\% with the highest \dedx, provides the best estimate of
the peak position. 
In Section~\ref{sec:pid} \dedx spectra as they are observed in
the data are shown.

The particle relativistic velocity $\beta$ is determined measuring the
time-of-flight (TOF) from its production point at the target
up to the system of resistive plate chamber (RPC) detectors arranged as
a barrel around the TPC.  
The path-length is determined using the trajectory measured in the
TPC. 
The time of production of the particle is measured using the time the
beam particle traverses the BTOF detectors and extrapolating it to the
interaction point.
The combined time resolution is 180~ps~\cite{ref:ieee:rpc}.
At present $\beta$ is not used for PID in the final analysis.
However, the PID capabilities with this TOF measurement are used to
select pure samples of pions and protons to measure the efficiency and
purity of the PID selection using \dedx.

\section{Data selection procedure}
\label{sec:selection}

The positive-particle beam used for this measurement contains mainly 
positrons, pions and protons, with small components of kaons and
deuterons and heavier ions.
Its composition depends on the selected beam momentum.
The analysis proceeds by first selecting a beam proton hitting the
target, not accompanied by other tracks. 
Then an event is required to be triggered by the ITC in order to be
retained. 
After the event selection the sample of tracks to be used for analysis
is defined.
The selection procedure is described below.

The beam time-of-flight system measures time over a distance
of 21.4~\m  which provides particle identification at low energy (up
to 5~\GeVc). 
At 3~\GeVc the time-of-flight measurement allows the selection of
pions from protons to be made at more than 5{$\sigma$},
the protons account for about 30$\%$ of beam at this momentum.
The fraction of protons increases with beam momentum.
At higher momenta protons are selected by rejecting particles with a
measured signal in either of the beam Cherenkov detectors. 
The selection of protons for the beam momenta with the Cherenkov
detectors has been described in detail in Ref~\cite{ref:alPaper}.
More details on the beam particle selection can be found in 
Ref.~\cite{ref:harpTech}.
Deuterons (and heavier ions) are removed by TOF measurements.  
A set of MWPCs is used to select events with only one
beam particle for which the trajectory extrapolates to the target.
An identical beam particle selection was performed for events
triggered with the incident-proton trigger in order to provide an
absolute normalization of the incoming protons.
This trigger selected every 64$^{th}$ beam particle coincidence
outside the dead-time of the data acquisition system.
The requirement of a trigger in the ITC keeps a sample of 
one million events for the analysis.
 
The beam particle has to be accepted by the criteria described above
and has to be identified as a proton.
In order to avoid large effects of the TPC dynamic distortions only the
first \evtspill events in each spill are retained.
Using calibration data sets, the deterioration of the performance of the
detector is determined as a function of the strength of the distortions
characterized by an average value of \dzeroprime for the same set of
events. 
As a practical solution and 
to simplify the analysis the `event number in spill'
\evtspill defines a 
measure of the time when the event occurred from the start of the
spill. 
This is a good measure of time since the readout time per event is
sufficiently constant (about 1~ms/event) and since the beam intensity
was so high that the DAQ was running close to saturation.
For each setting the \evtspill criterion was calibrated with
the behaviour of the average  $d_0'$, $\langle d_0' \rangle$.
The part of the spill accepted in the analysis is then defined by
determining for each data-taking condition for which value of \evtspill
the average value of \dzeroprime exceeds 5~\mm.
In practice, the value of \evtspill is close to 100 in all
settings analysed, compared to a typical total number of events per
spill of 300.

Cuts are defined to reject tracks from events which arrive randomly
in the 30~\micros drift time of the TPC secondaries from  
interactions of other beam particles (`overlays').
In addition, selection criteria are used which preferentially remove
tracks from secondary interactions ({\em i.e.} interactions of the
particles produced in the primary interaction).
The following selection was applied to retain well-measured particle
tracks with known efficiency and resolution.

Tracks are only considered if they contain at least twelve space
points out of a maximum of twenty.
This cut is applied to ensure a good measurement of the track
parameters and of the \dedx.
Furthermore, a quality requirement is applied on the fit to the
helix. 
The latter requirement introduces a very small loss of efficiency.

\begin{figure}[tbp]
  \begin{center}
    \epsfig{figure=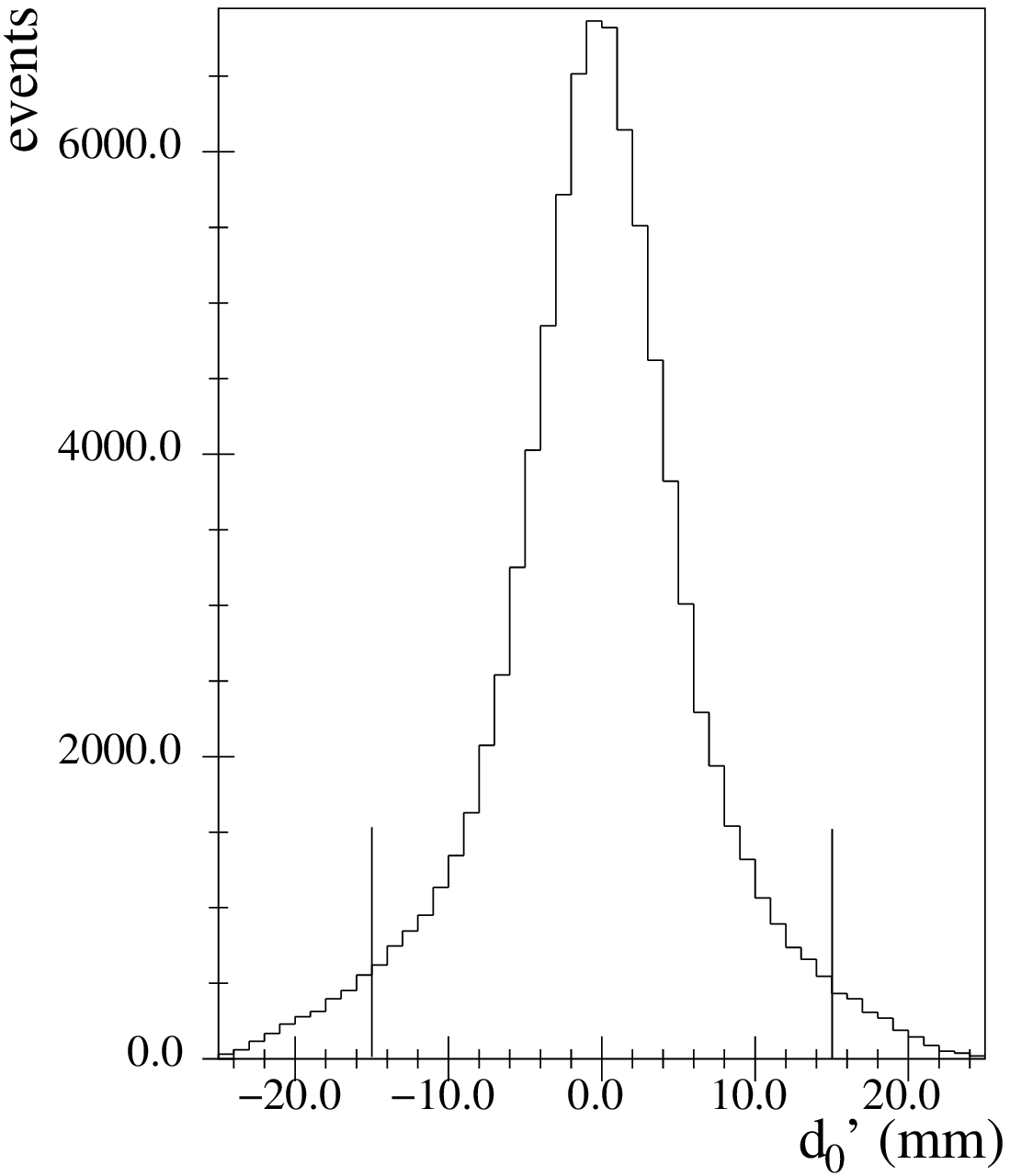,width=0.42\textwidth}
    ~~
    \epsfig{figure=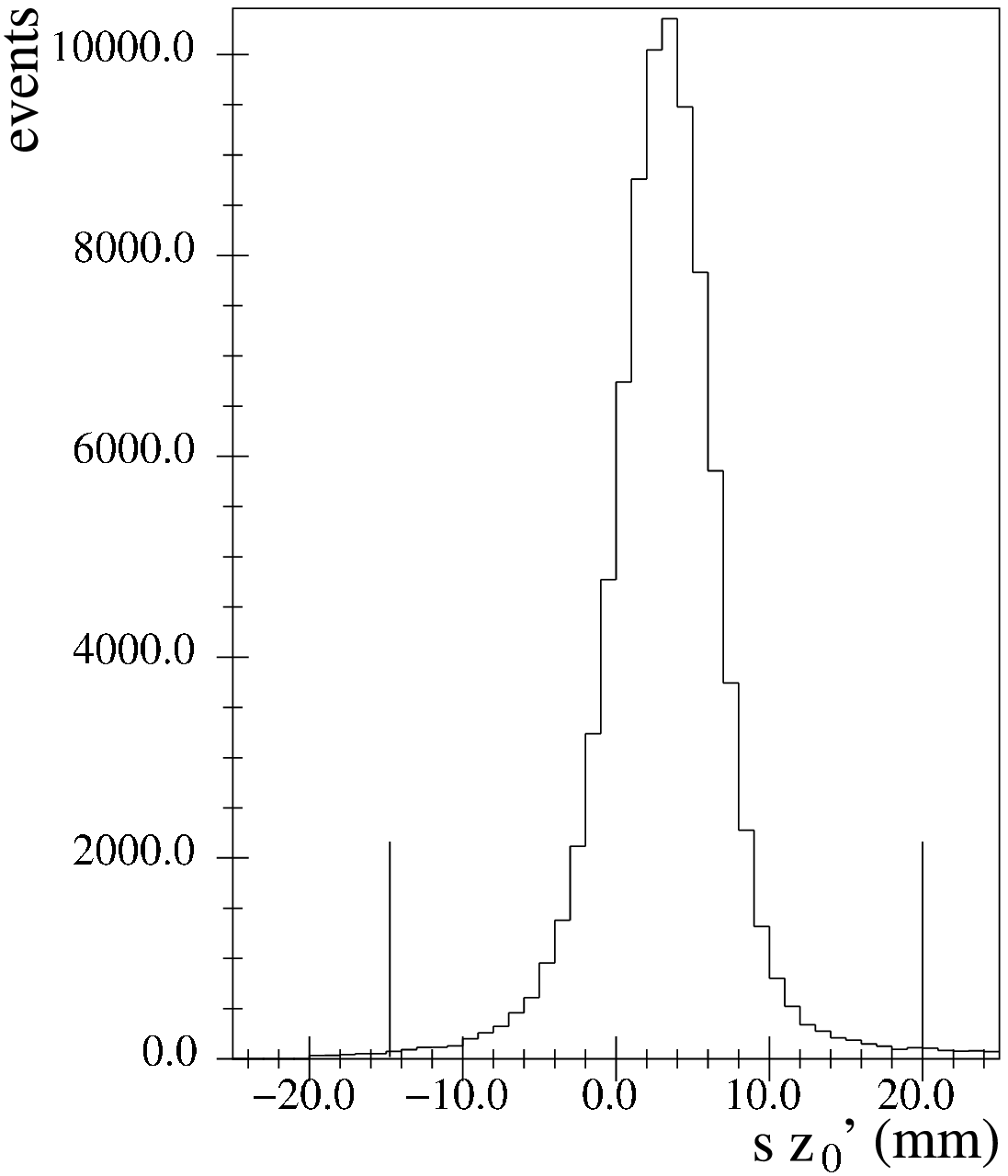,width=0.42\textwidth}
  \end{center}
\caption{The distribution of \dzeroprime (left panel) and $s$~\zzeroprime
  (right panel) taken 
  with an 8~\GeVc proton beam hitting a tantalum target for tracks with
  low \dedx.
  Cuts (indicated by the vertical bars) are applied at $| \dzeroprime
  | < 15 \ \mm$ and  $-14.4 \ \mm < s \ \zzeroprime  < 20.0 \
  \mm$. 
}
\label{fig:zeroprime}
\end{figure}

For tracks satisfying these conditions, a cut is made on \dzeroprime,
the distance 
of closest approach to the extrapolated trajectory of the incoming
beam particle in the plane perpendicular to the beam direction and
\zzeroprime, the 
$z$-coordinate where the distance of the secondary track and the beam
track is minimal.  
Figure~\ref{fig:zeroprime} shows the distribution of \dzeroprime and
$s$~\zzeroprime for the data taken with 8~\GeVc protons on a tantalum
target.
The variable $s$ is defined as $s = \sin \theta$, where $\theta$ is the 
angle of the particle measured with respect to the nominal beam axis. 
To avoid the bias due to the change of curvature which occurs for
highly ionizing protons traversing the ITC trigger counter and the
inner field cage, only outgoing tracks with low \dedx were used
for this figure. 
Cuts are applied at $| \dzeroprime | < 15 \ \mm$ and 
$-14.4 \ \mm < s \ \zzeroprime < 20.0 \ \mm$.
The $\sin \theta$ dependence in the cut has been introduced 
to take into account the angular dependence of the precision of the
extrapolation. 
The accepted $s \ \zzeroprime$ region is symmetric around the
centre of the target.
The target extends from $z=0$ to $z=5.6$~\mm in this coordinate system. 
The transverse coordinates of the interaction vertex are obtained
extrapolating the trajectory of the incoming beam particle measured by
the MWPCs. 
The longitudinal coordinate is taken from the position where the
fitted track is closest to the trajectory of the beam particle.

Finally, only tracks with momentum in the range between 100~\MeVc and
800~\MeVc are accepted. 
In addition, particles with transverse momentum below 55~\MeVc are
removed. 
This range meets the requirements of the data needed for the design of
the neutrino factory and is consistent with the acceptance and
resolution of the chamber. 
Table~\ref{tab:events} shows the number of events and tracks at
various stages of the selection.
To give an impression of the complexity of the events, one can define an
`average multiplicity' as the ratio of the number of tracks with at
least twelve hits in the TPC (regardless of their momentum, angle or
spatial position) and the number of events accepted by the selection
criteria with at least one such track. 
The average multiplicity obtained according to this definition is
reported in Table~\ref{tab:events}.
 
\begin{table}[tbp!] 
\caption{Total number of events and tracks used in the tantalum
  5\%~$\lambda_{\mathrm{I}}$ target data sets, and the number of
  protons on target as calculated from the pre-scaled trigger count.} 
\label{tab:events}
\begin{center}
\begin{tabular}{ l  r  r r r} \hline
\bf{Data set}&\bf{3 \bfGeVc}&\bf{5 \bfGeVc}&\bf{8 \bfGeVc}&\bf{12 \bfGeVc}\\ \hline
    Total events taken by the DAQ     & 2291133 & 2094286 & 2045628 & 886305 \\
    Protons on target (selected incident-proton$\times$64) & 1693376 & 3251136 & 6136960 & 3486016 \\
    Accepted protons with interaction triggers &  416131 & 447312 & 752377 & 436400 \\
    Accepted protons with Large Angle Int. (LAI)& 101509 & 218293 & 442624 & 269927 \\
    Maximum \evtspill                & 120 & 110 &  110 & 90 \\
    LAI in accepted part of the spill &  38281 & 72229 & 137033 & 82994 \\
    Fraction of triggers used        & 38\% & 33\% &  31\% & 31\%\\
    LA tracks with $\ge 12$ hits         &  68340 & 188754 & 464308 & 346856 \\
    Average multiplicity                 &  2.3 &  3.5  & 4.7 & 6.0 \\
    Accepted momentum (vertex constraint)&  50985 & 138261 & 338598 & 242114 \\
    From target and in kinematic region  &  34430 & 93220 & 214339 & 148012 \\
    Negative particles                   &  3836 & 14485 & 42159 & 33095 \\
    Positive particles                   & 30594 & 78735 & 172180 & 114917 \\
    \bf{$\bfpim$ selected with PID}        & {3526} & {13163} & {37706} & {29105} \\
    \bf{$\bfpip$ selected with PID}        & {4706} & {15791} & {42296} & {31407} \\
\end{tabular}
\end{center}
\end{table}

\section{Performance of the detector}
\label{sec:performance}

The present measurement concentrates on the production of particles at
large angles from the beam direction as measured in the TPC.  
To calibrate the performance of the TPC one would ideally enter
particles of known momentum and type into the sensitive volume of the
chamber. 
To achieve this either
the chamber would have to be rotated or moved to another position or
the beam would have to be  steered far from its normal trajectory.
Neither option was available so that other methods had to be
employed to characterize the performance of the chamber. 
Cosmic-ray tracks and the elastic events in the data taken
with hydrogen targets were used to characterize the TPC.
Additional constraints were obtained making use of the characteristic
momentum dependence of the \dedx for particle tracks in the TPC.

The measured quantities used in the analysis are the 
momentum, scattering angle with respect to the beam particle and
particle identification.
Therefore, the performance of the detector needs to be characterized
for these quantities and for the efficiency to reconstruct the tracks
as a function of these quantities.
In addition, the ability of the simulation to reproduce these has 
to be studied.
The resolutions, measurement biases and efficiencies need to be known
as function of the important kinematic variables.

To investigate which fraction of the data can be used in the presence
of dynamic distortions the behaviour of the quantities relevant for
the analysis has been studied as a function of the strength of these
distortions. 
As discussed above, the average $d_0'$ of the tracks produced by the
beam in the target is used as a parameter to characterize the strength of the
dynamic distortions.

It will be demonstrated that each important
reconstructed quantity and its behaviour as a function of time in spill
can be characterized using constraints from the data themselves.
The absolute scale of the momentum is determined making use of the
kinematics of elastic scattering.
Its resolution is measured with cosmic-ray tracks with consistency
checks based on \dedx and elastic scattering.
The evolution with the effect of dynamic distortions is measured with 
elastic scattering and \dedx constraints, while the analysis of \dedx
sets a limit on any possible charge-asymmetry in the momentum
measurement.  
Similarly, the resolution of the measurement of the scattering angle is
obtained with cosmic-ray data, supported by consistency checks from
elastic scattering.
The absolute measurement of the angle and its sensitivity to dynamic
distortions is constrained by the kinematics of elastic scattering. 
In the analysis, PID is based on the measurement of \dedx.
The robustness of the \dedx measurement was observed with elastic
scattering and with minimum-ionizing particles using the fact that the
\dedx is independent of the momentum measurement for these particles. 
The efficiency and purity of the identification of the particle type
was measured using an independent selection based on time-of-flight.
Finally, an absolute measurement of the efficiency and its evolution
as a function of strength of the dynamic distortions was obtained with
elastic scattering.
The most important points will be elaborated below.

\subsection{Study with cosmic-ray events}
\label{sec:cosmics}

Cosmic-ray data were taken during and outside the beam data taking
periods. 
During beam periods, cosmic-ray triggers were collected between
the beam spills.
Additional cosmic-ray exposures were performed close to the data
taking periods with the beam to ensure that the detector conditions remained the
same. 
For these data the outer barrel RPCs were used to provide a trigger
for the cosmic-rays.
In the year following data taking (2003) an extensive
cosmic-ray exposure was performed providing a dedicated calibration. 
In particular, the trigger was provided by a scintillator rod positioned 
in the inner field cage to obtain tracks following the same
trajectory as secondary tracks during beam exposures.
The rod was placed at the nominal target position with transverse
dimensions similar to the beam spot size, but more extended in $z$
than the usual targets.

\begin{figure}[tbp]
  \begin{center}
    \epsfig{figure=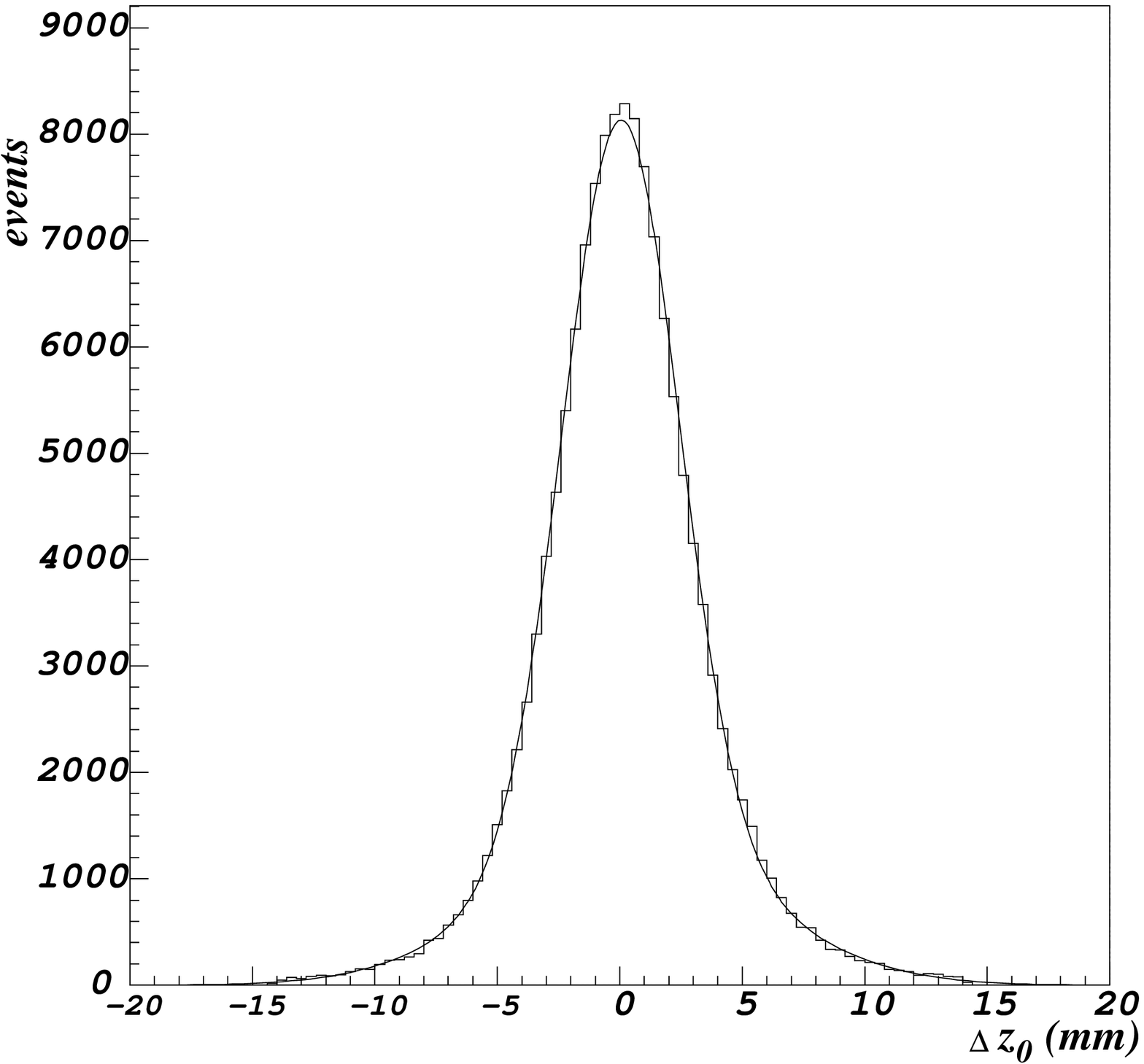,width=0.46\textwidth,angle=0}
    \epsfig{figure=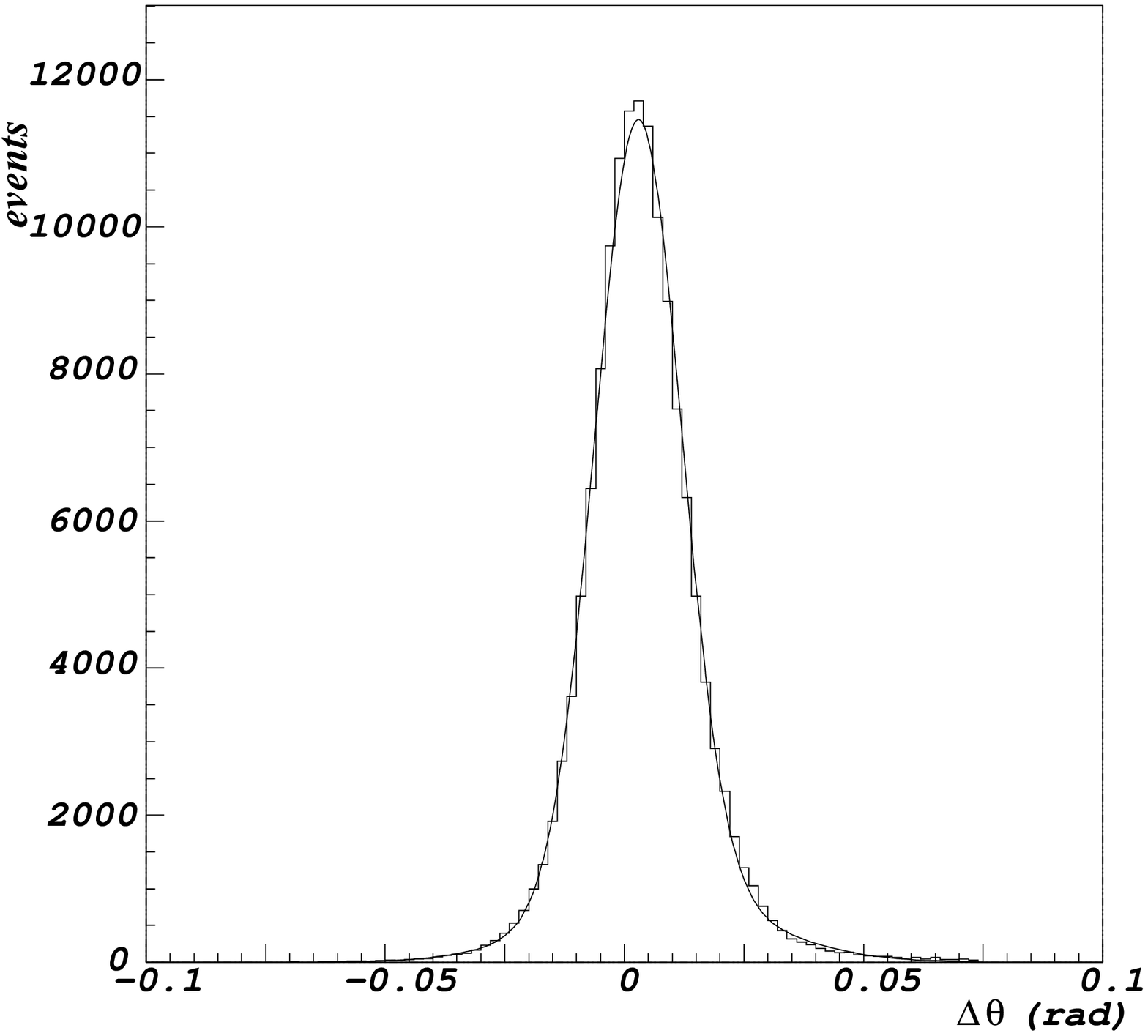,width=0.475\textwidth,angle=0}
  \end{center}
\caption{
  Left: Measurement with cosmic-ray tracks of the \zzeroprime resolution of
  the TPC.  
  The distribution has an RMS of 3.4~\mm.
  Right: measurement with cosmic-ray tracks of the \tht resolution of
  the TPC. 
  The distribution has an RMS of 12~\mrad.
} 
\label{fig:cos-zz}
\end{figure}

The resolution of the extrapolated track position at the target was 
measured by splitting the cosmic-ray track into two halves and 
taking the distance between the two extrapolated trajectories.
The difference in the extrapolation in the direction of the
beam measures the \zzeroprime resolution.
Figure~\ref{fig:cos-zz}~(left) shows the result of this measurement.
The distribution has an RMS of 3.4~\mm.
The resolution in the angle of secondary tracks with respect to the
beam direction \tht can be measured again by comparing the two track
segments. 
The resolution measured in this way is shown  in the right panel of
Fig.~\ref{fig:cos-zz}. 
The distribution has an RMS of 12~\mrad.
The mean value is non-zero, reflecting a small systematic uncertainty in
the measurement of $\theta$ of the order of 5~\mrad.
This bias is caused by the limited precision of the equalization constants
of the pad pulse heights which can introduce an $r$-dependent systematics
in the determination of the $z$-position of hits.
In the absence of interactions by beam particles  
the equalization constants determined for the cosmic-ray data-taking
periods  are expected to be less precise than for the normal data-taking
periods. 
This effect has been found to induce a negligible uncertainty in the
analysis.

\begin{figure}[tb]
 \begin{center}
  \begin{minipage}[b]{0.66\textwidth}{
   \begin{turn}{90}\mbox{~~~~~~~~~~~~~~~~~~~~~~~~~~~~~~~~~~~~~~~~~~~~~~~~~~~~~~~~$\sigma(p_\mathrm{T}^{-1})/p_\mathrm{T}^{-1}$}
   \end{turn}
   \epsfig{figure=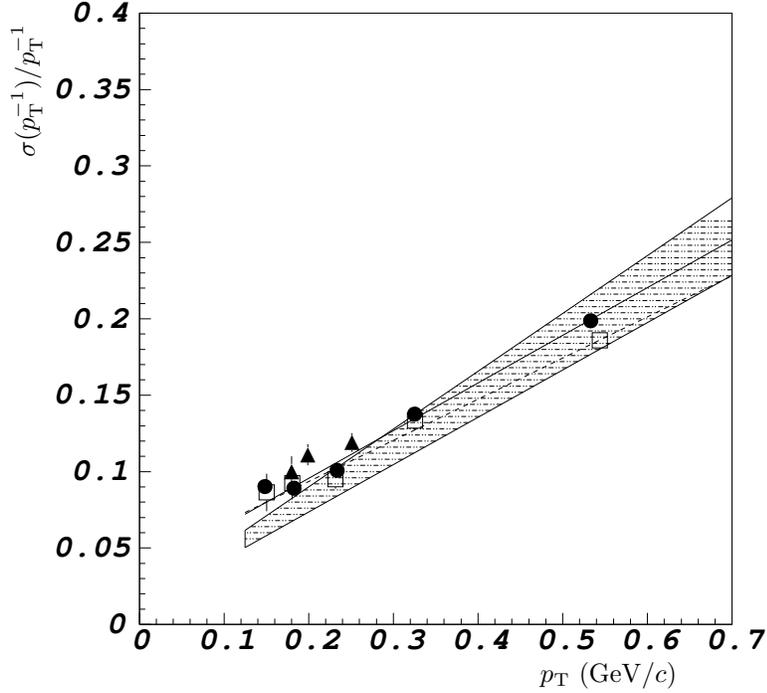,width=0.9\textwidth}
   \\
   \hfill \mbox{~~~~~~~~~~~~~~~~~~~~~~~~~~~~~~~~~~~~~~~~~~~~~~~~~~~~~~~~~~~~$p_\mathrm{T} \ (\GeVc)$}
   }
  \end{minipage}
  \end{center}
\caption[] {Momentum resolution in the TPC.  The filled circles (open
  boxes) and 
  the drawn (dashed) straight line refer to the cosmic-ray data taken in
  2003 (2002).  The filled triangles are the upper limits (`over-estimates')
  obtained from the \dedx selection.  The shaded area refers to
  a straight-line fit to the Monte Carlo calculations.
}
\label{fig:cos-p}
\end{figure}

To measure the transverse momentum resolution three estimates of the
transverse momentum of the cosmic-ray track are obtained:
the transverse momentum measured from the curvature of the two halves of
the track separately and the transverse momentum from the curvature of
the complete track.
The relative resolution is then obtained from the distribution of the
difference of the inverse of the transverse momenta of the two
half-tracks divided by the inverse of the overall track transverse
momentum.
The sigma of the Gaussian fit (divided by the square-root of two) is
plotted as a function of the transverse momentum of the overall track.
The result of this analysis is shown in Fig.~\ref{fig:cos-p}.
Since the resolution is expected to be Gaussian in the curvature
($1/p$), the resolution is shown in this quantity.
The resolution measured with cosmic-rays is compared with the
over-estimates which can be
obtained by selecting a small slice of the 
steep part of the dependence of the \dedx on the momentum\footnote{The
momentum resolution is over-estimated since the effects of the size of
the \dedx slice and the \dedx resolution are not corrected for.}.
Subdividing the data-sample into different bins of $\theta$ a fixed
\dedx slice (corresponding to a given momentum) can be used to
determine several points at different \pt.
The resolution expected from the simulation using the point--to--point
resolution measured with \dedx in the data is consistent with the
cosmic-ray measurement. 
We do observe, however, a slightly larger constant term in the
cosmic-ray data as predicted by the simulation.
In the 2003 data the cosmic-ray tracks were triggered by a scintillator
on the TPC axis.  
The amount of material of the scintillator is similar to the
amount of material in the inner field cage and trigger detector.
Hence, the constant
term is expected to be slightly larger for the cosmic-rays than for
the MC.   In the 2002 data, no such trigger detector was available.
Therefore the tracks do, in majority, not pass the nominal axis of the
TPC, such that they see effectively a larger amount of the inner field
cage material.

\subsection{Study with elastic scattering data}
\label{sec:elastics}

Elastic scattering interactions of protons and pions on hydrogen
provide events where the kinematics are fully determined by the
direction of the forward scattered beam particle.
The kinematic properties of the elastic scattering reaction
were exploited to provide a known `beam' of protons pointing into the
TPC sensitive volume. 
Data were taken with liquid hydrogen targets at beam momenta from
3~\GeVc to 15~\GeVc. 
A good fraction of forward scattered protons or pions in the elastic
scattering reaction enter into the acceptance of the forward spectrometer.
The full kinematics of the event can be constrained by a precise
measurement of the direction of the forward scattered beam particle. 
In particular, the direction and momentum of the recoil proton are
precisely predicted.
Selecting events with one and only one track in the forward direction and
requiring that the measured momentum and angle are consistent with an
elastic reaction already provides an enriched sample of elastic
events. 
By requiring that only one barrel RPC hit is recorded at the
position predicted for an elastic event 
(the precision of the prediction from the forward spectrometer is
within the RPC pad size)
and within a time window consistent with a proton time-of-flight 
a sample of recoil protons with known momentum vector of a purity of
about 99\% is obtained. 
At beam momenta in the range 3~\GeVc--8~\GeVc the kinematics are such
that these protons point into the TPC with angles of $\approx
70^{\circ}$ with respect to the beam direction.
Once a clean sample of elastic-scattering events is isolated the
efficiency of the track finding and fitting procedure can be
measured and an estimate of the resolution and biases of the
measurement of momentum and angle can be obtained.
The correlation of the forward scattering angle and
recoil proton momentum is such that an unavoidable threshold in recoil
proton momentum ($\approx 350 \ \MeVc$) translates into a minimum
angle for the scattered particle. 
The threshold is relatively high due to the need to detect the proton
also in the barrel RPC system outside the outer field cage of the TPC.
This requirement can be removed only in cases where a small amount of
background can be tolerated.
Due to the geometry of the rectangular aperture of the dipole magnet
of the forward spectrometer
only two small horizontal sectors of the TPC can be populated with
recoil protons above threshold momentum in the 3~\GeVc beam.
In the 5~\GeVc beam the situation is much better and all azimuthal
angles can be populated, although not yet homogeneously.
In the 8~\GeVc beam the population is homogeneous in $\phi$, but the
error propagation of the measurement of the forward scattering angle
into the prediction of momentum and angle of the recoil proton becomes
less favourable.
Summing up all these arguments, the 8~\GeVc beam is most suitable for the
determination of average efficiency, the 5~\GeVc beam is still useful
for efficiency measurements and provides a good sampling of the
resolution of the detector, while the  3~\GeVc beam can be used
to study the resolution with the most favourable situation for the
prediction.  
The numbers of selected elastic events total about 15,000 for the
8~\GeVc data sample, and 5,000 for the 5~\GeVc and 3~\GeVc data
samples each. 

\begin{figure}[tbp]
  \begin{center}
    \epsfig{figure=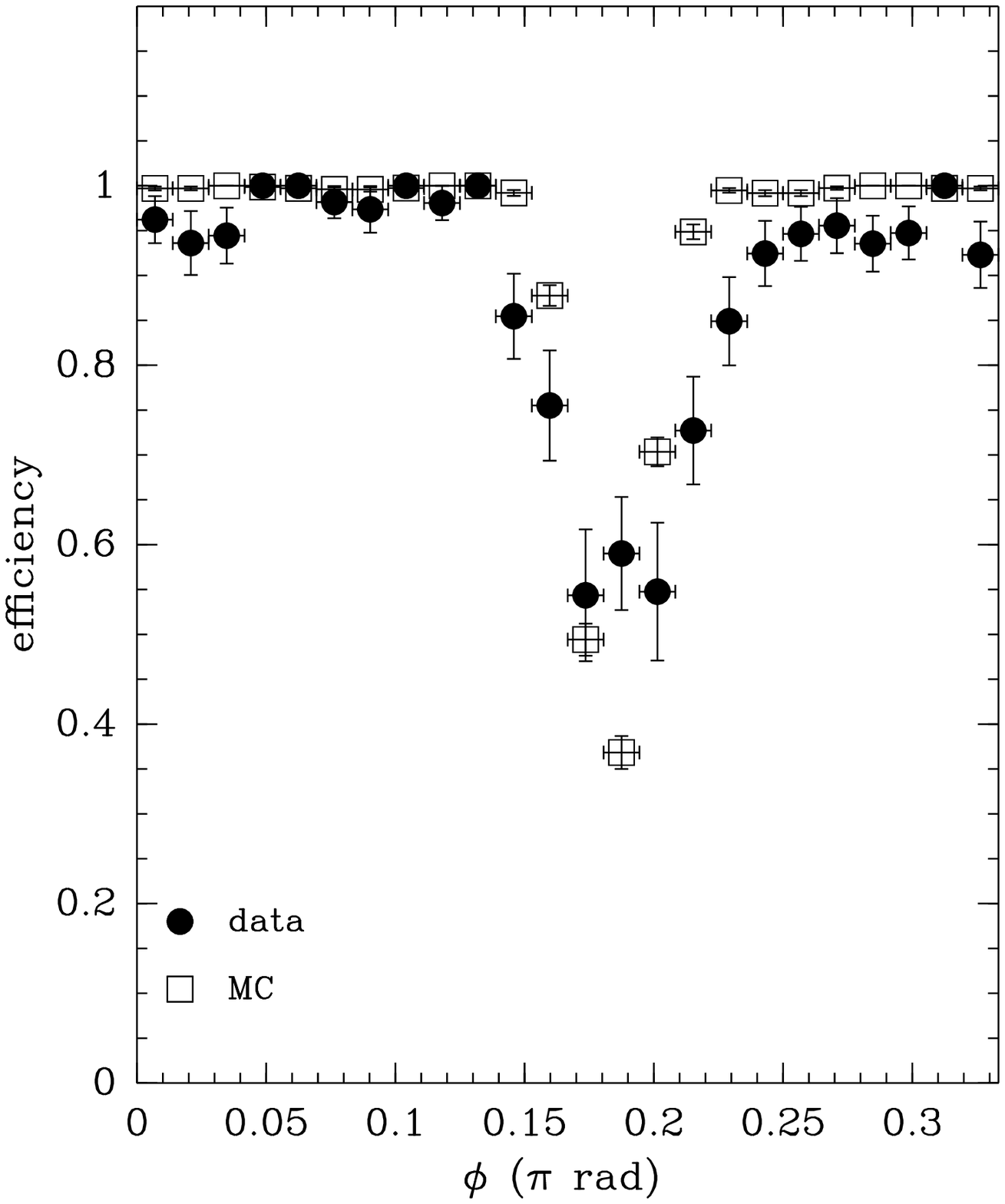,width=0.4\textwidth,angle=0}
    ~~
    \epsfig{figure=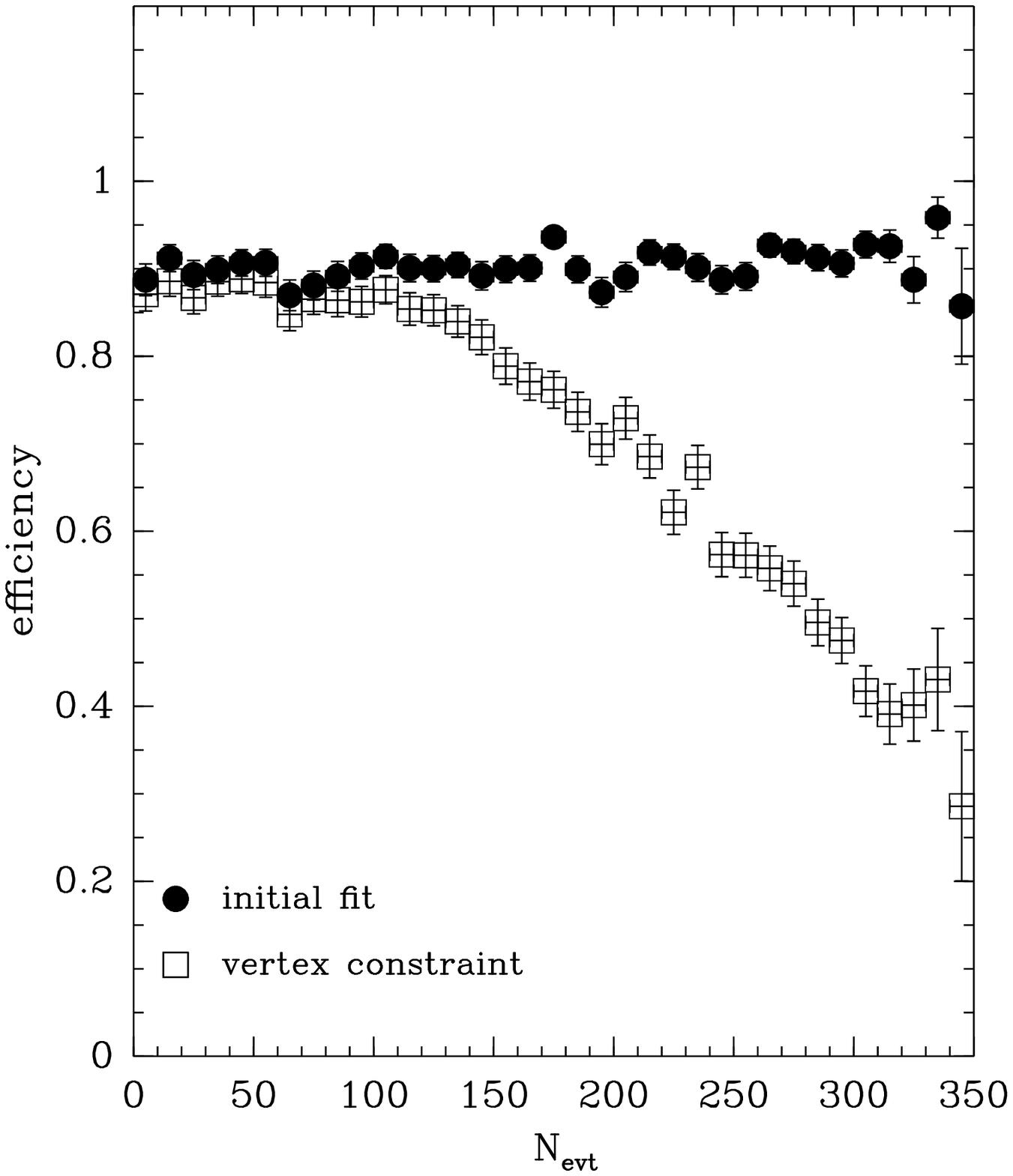,width=0.41\textwidth,angle=0}
  \end{center}
\caption{
Left panel: 
the track finding efficiency as a function of $\phi$ within the
sectors of the pad-plane of the TPC for 8~\GeVc elastic scattering
data measured with elastic events (first 80 events in the spill). 
The filled circles show the efficiency for recognizing tracks including
the fit to a helix in the data, the open squares show the simulated
efficiency.
The integral of the efficiency is well reproduced, although the
details near the spokes are different.  
Right panel:
Efficiency for the pattern recognition and momentum
  reconstruction for elastically produced recoil protons as a function of
  event number in spill.  Closed circles: trajectory fit without vertex
  constraint;  open squares: trajectory fit with vertex
  constraint.  The efficiency includes the effect of the cut on 
  \dzeroprime.
  The efficiency for the reconstruction and fit
  using the vertex constraint remains constant within $\approx$~1\% up
  to a distortion corresponding 
to $\langle \dzeroprime \rangle = 6 \ \mm$. 
}
\label{fig:elas:eff}
\end{figure}

Based on the 8~\GeVc data the track reconstruction efficiency was
determined to be $91\%\pm 1\%$ compared with an efficiency of  93\%
calculated with the simulation.
In the 5~\GeVc beam the efficiency is the same as that for
8~\GeVc data. 
In the data a $\approx 1\%$ loss of efficiency can be attributed to
channels with intermittent connection problems, an effect not
simulated in the Monte Carlo (MC).  
The inefficiency is dominated by the effect of the `spokes', the place
where the wires of the wire-planes are fixed as shown in
Fig.~\ref{fig:elas:eff}~(left). 
The integral of the efficiency is well reproduced, although the
details near the spoke are different. 
This is due to the smearing effect in the measurement of $\phi$ under
the influence of dynamic distortions.
Since the analysis is performed integrating over $\phi$ this has to
first order no effect.
The good agreement of the measurements of the absolute efficiency with
the simulation justifies the use of the simulation to determine the
efficiency to measure pions. 
The systematic error is estimated by changing the effective cut on the
number of points to accept tracks.

Figure~\ref{fig:elas:eff}~(right) displays the results of this analysis for
the reconstruction with and without vertex constraint.
The efficiency of the reconstruction without vertex constraint is
insensitive to distortions ({\em i.e.} the track will be found and 
measured), while the momentum reconstruction using the vertex
constraint keeps a constant efficiency up to
$\evtspill \approx 90$, corresponding to 
$\langle d_0' \rangle \approx 6 \ \mm$ for this data set.
The loss of efficiency for the constrained fit is due to the need to
apply a cut in \dzeroprime to ensure that the track originates from the
vertex. 
It can therefore be concluded that the efficiency for the reconstruction and fit
using the vertex constraint remains constant within $\approx$~1\% up
to a distortion corresponding to $\langle \dzeroprime \rangle = 6 \
\mm$.

\begin{figure}[tbp]
  \begin{center}
    \epsfig{figure=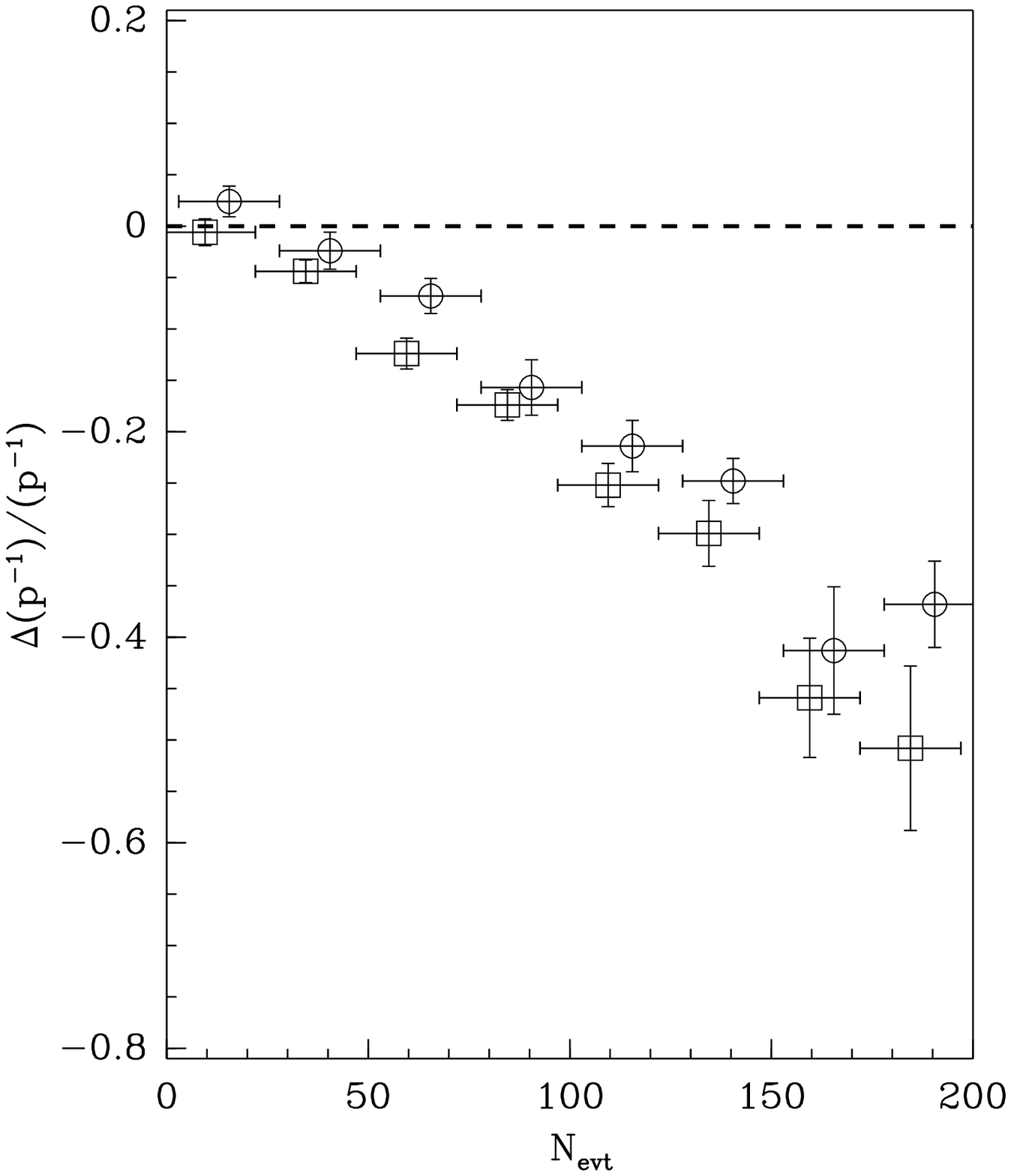,width=0.413\textwidth,angle=0}
    \epsfig{figure=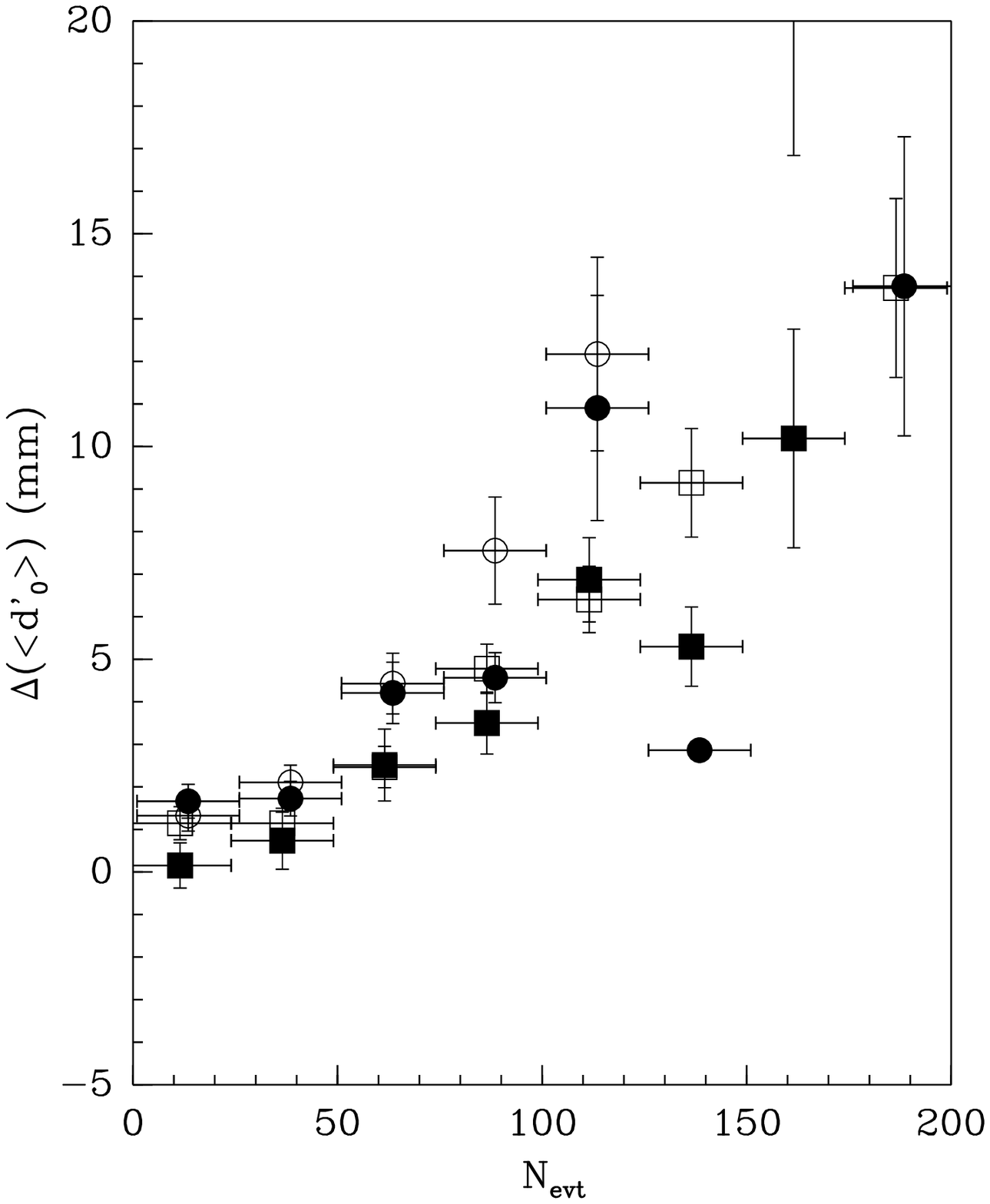,width=0.4\textwidth,angle=0}
  \end{center}
\caption{
Left panel:
The shift in average momentum for  elastic scattering data (3~\GeVc:
open squares,  5~\GeVc: open circles) measured with elastic 
events as a function of the value of \evtspill.
The momentum estimator from the fit not constrained by the impact
point of the incoming beam particle is used here.
Right panel:
The shift in average \dzeroprime as a function of the event number in
spill for  elastic scattering data (3~\GeVc: filled and open boxes,
5~\GeVc: filled and open circles) measured with elastic 
events as a function of the value of \evtspill.
The open symbols show the data for predicted momenta below 450~\MeVc and the
filled symbols for predicted momenta above 450~\MeVc.
} 
\label{fig:el-p-bias-spill}
\end{figure}

It was verified with the data that the value of \tht is not modified
by the dynamic distortions. 
However, the momentum estimated with the fit not using the impact
point of the incoming beam particle and the  value of \dzeroprime is
biased as a function of event in spill due to 
the effect of these distortions as shown in
Fig.~\ref{fig:el-p-bias-spill}.
The results of this analysis justify the use of only a limited
number of events in each spill in order not to introduce large
uncertainties due to distortions.
The analysis of the elastic scattering events sets very
stringent constraints on the maximum effect of distortions of all
kinds on the measurements of kinematic quantities with the TPC.
Therefore, solid estimates for the magnitude of the systematic error
sources are obtained.  
For the hydrogen runs a higher beam intensity was used roughly
equalizing the interaction rate between these runs and the runs with a
tantalum target.
In fact, the dynamic distortion effects were in strength similar to the
12~\GeVc Ta runs. 
Since the analysis takes into account this variation by applying a cut
at a different value of  \evtspill the measurements are representative for
all datasets\footnote{The 
value of ``maximum \evtspill'' reported in Table~\ref{tab:events}, can
be used to estimate a value of equal strength of dynamic distortions for
different datasets.}.

\begin{figure}[tbp]
  \begin{center}
    \epsfig{figure=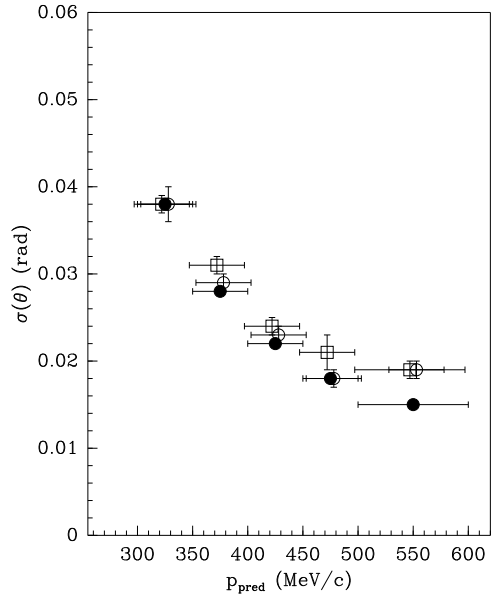,width=0.4\textwidth,angle=0}
  \end{center}
\caption{
The resolution in \tht for  elastic scattering (3~\GeVc: open boxes,
5~\GeVc: open circles) data measured with elastic 
events as a function of the momentum predicted by the forward
scattered track compared to a simulation of the same sample of events
at 5~\GeVc (filled circles). 
} 
\label{fig:el-th-mom}
\end{figure}

The resolution in the measurement of the polar angle \tht is shown in
Fig.~\ref{fig:el-th-mom} as a function of the predicted momentum 
of the proton when it enters the gas.
The comparison of the experimental result with the simulation shows
good agreement. 
For low-momentum protons ($p < 500 \ \MeVc$) the resolution is
dominated by multiple scattering. 

\begin{figure}[tbp]
  \begin{center}
    \epsfig{figure=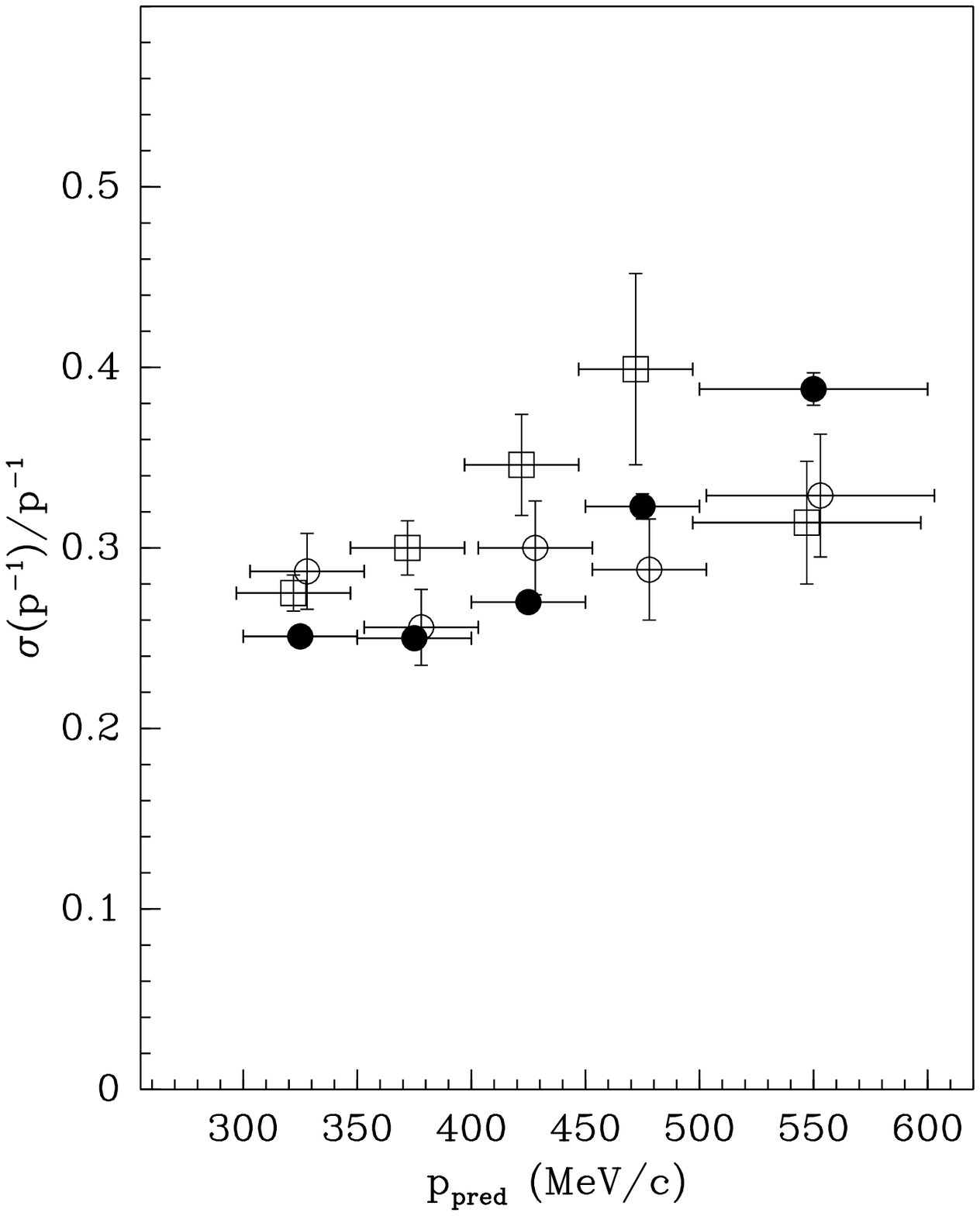,width=0.4\textwidth,angle=0}
    ~~
    \epsfig{figure=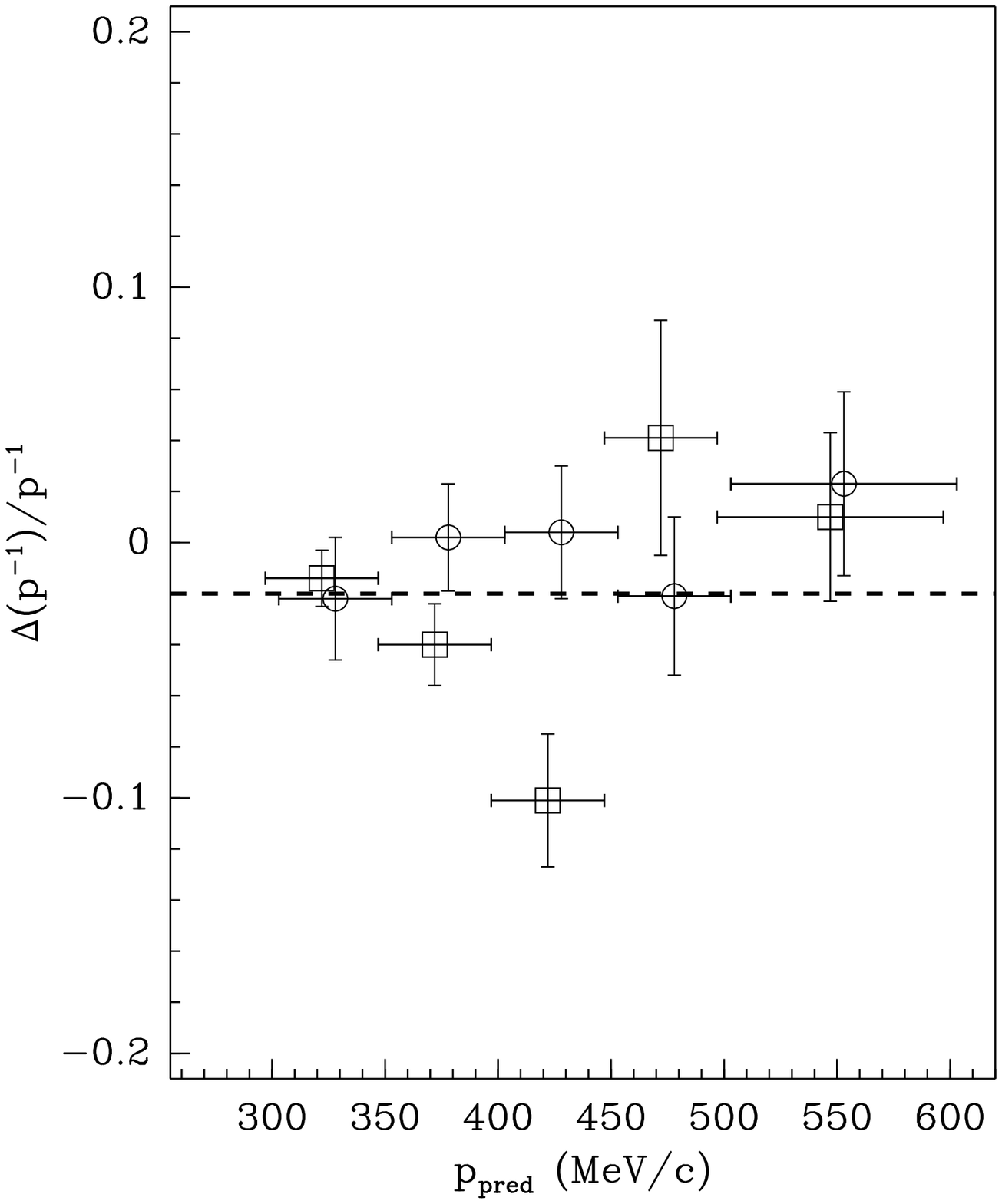,width=0.412\textwidth,angle=0}
  \end{center}
\caption{
Left panel:
The momentum resolution of the fit without vertex constraint for
elastic scattering data (3~\GeVc: open squares,  5~\GeVc: open circles)
measured with elastic events as a function of the momentum predicted
by the forward scattered track.
The resolution is dominated by the effect of energy-loss and multiple
scattering and is consistent with the measurement with cosmic-ray
tracks. 
The filled circles show a full simulation of the elastic events using
a realistic elastic cross-section model and detector description.
The agreement between data and simulation is good.
Right panel:
the momentum bias of the fit without vertex constraint measured with
 elastic scattering data (3~\GeVc: open squares,  5~\GeVc: 
open circles) with elastic 
events as a function of the momentum predicted by the forward
scattered track.
In the absence of a clear trend, the average of the points constrains
the bias to be smaller than 3\%.
For these comparisons only the first 50 events in the spill are
used since the unconstrained fit is sensitive to dynamic distortions
beyond this value. 
} 
\label{fig:el-p-mom-bias}
\end{figure}

Since the energy loss in the material of the cryogenic target, trigger
counter, and inner field cage is large for protons in the energy range
covered by elastic scattering, there is a significant change of
curvature of the trajectory of these protons in that region of the
detector. 
This effect could introduce a bias in the measurement of the momentum
using the vertex constraint for these low-momentum protons.
Therefore, it is more significant to study the behaviour of
the momentum measurement for protons without making use of the vertex
constraint. 
For pions, it was checked independently that the constrained fit is
unbiased with respect to the unconstrained fit for tracks
reconstructed in the data and the simulated data. 
The momentum measured for recoil protons in elastic scattering events
using the fit without vertex constraint is 
compared with the prediction based on the forward scattering angle
including a correction for energy-loss in the liquid hydrogen target
and the material surrounding the target (including the trigger counter
and inner field cage).
The comparison is made in the variable $1/p$.
The measurement of the momentum resolution of the fit without vertex
constraint is shown as a function of 
momentum in Fig.~\ref{fig:el-p-mom-bias}~(left).
Although the resolution is consistent with the measurement with
cosmic-ray tracks, this is not a very strong constraint since it is
dominated by the effect of energy-loss and multiple scattering. 
The simulation predicts for protons a resolution of $\approx 30\%$ in
the range from 300~\MeVc to 600~\MeVc.
The  momentum bias using the fit without vertex constraint is shown as
function of predicted momentum in Fig.~\ref{fig:el-p-mom-bias}~(right).
The average of the bias is ($2\pm1$)\%.
In the absence of a clear trend one concludes that the bias is less
than 3\%. 
From the precision in knowledge of the absolute beam momentum and the
precision in the measurement of the kinematical quantities of the
forward scattered track one cannot expect a precision better than
2\% in this cross-check.
For this comparison only the first 50 events in the spill are used in
order to avoid the effect of dynamic distortions in the unconstrained
fit as shown in Fig.~\ref{fig:el-p-bias-spill}.

\begin{figure}[tbp]
  \begin{center}
    \epsfig{figure=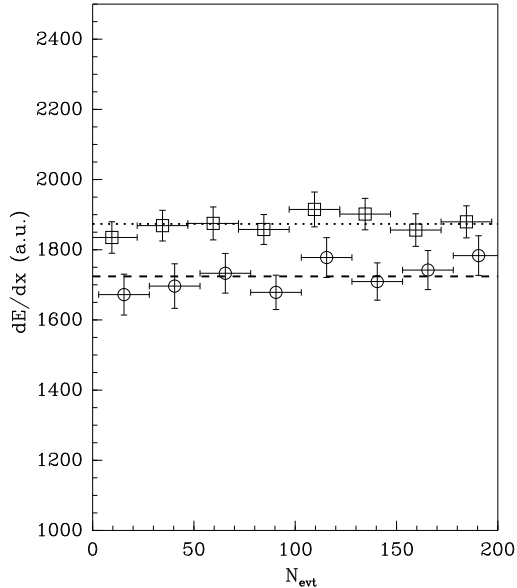,width=0.42\textwidth,angle=0}
  \end{center}
\caption{
The average value of \dedx as a function of the event number in
spill for  elastic scattering data (3~\GeVc: open boxes,  5~\GeVc:
open circles) measured with elastic 
events.
The dotted and dashed lines show the average value for 3~\GeVc and
5~\GeVc, respectively.
} 
\label{fig:el-dedx-bias-spill}
\end{figure}

Since the behaviour of the fit constrained with the impact
point of the incoming beam particle cannot be studied very well using
low momentum protons, the effect of distortions on this estimator is
studied using other physical benchmarks.
These will be described in the following section.
The fact that the measurement of \dedx is insensitive to the distortions
will be used in these studies.
The robustness of this quantity can be observed in
Fig.~\ref{fig:el-dedx-bias-spill}.
The average \dedx is shown as a function of event number in spill for
the sample of elastic events selected using the forward spectrometer.
The definition of the sample is independent of measurements in the
TPC.
Both in the 3~\GeVc and 5~\GeVc beam this quantity is stable.
The higher average \dedx in the  3~\GeVc beam is caused by the
lower average momentum of the protons.

\subsection{Systematic checks of the momentum measurement}
\label{section:tpc:rpc:shift}

\begin{figure}[tb]
\centering
\includegraphics[width=0.4\textwidth]{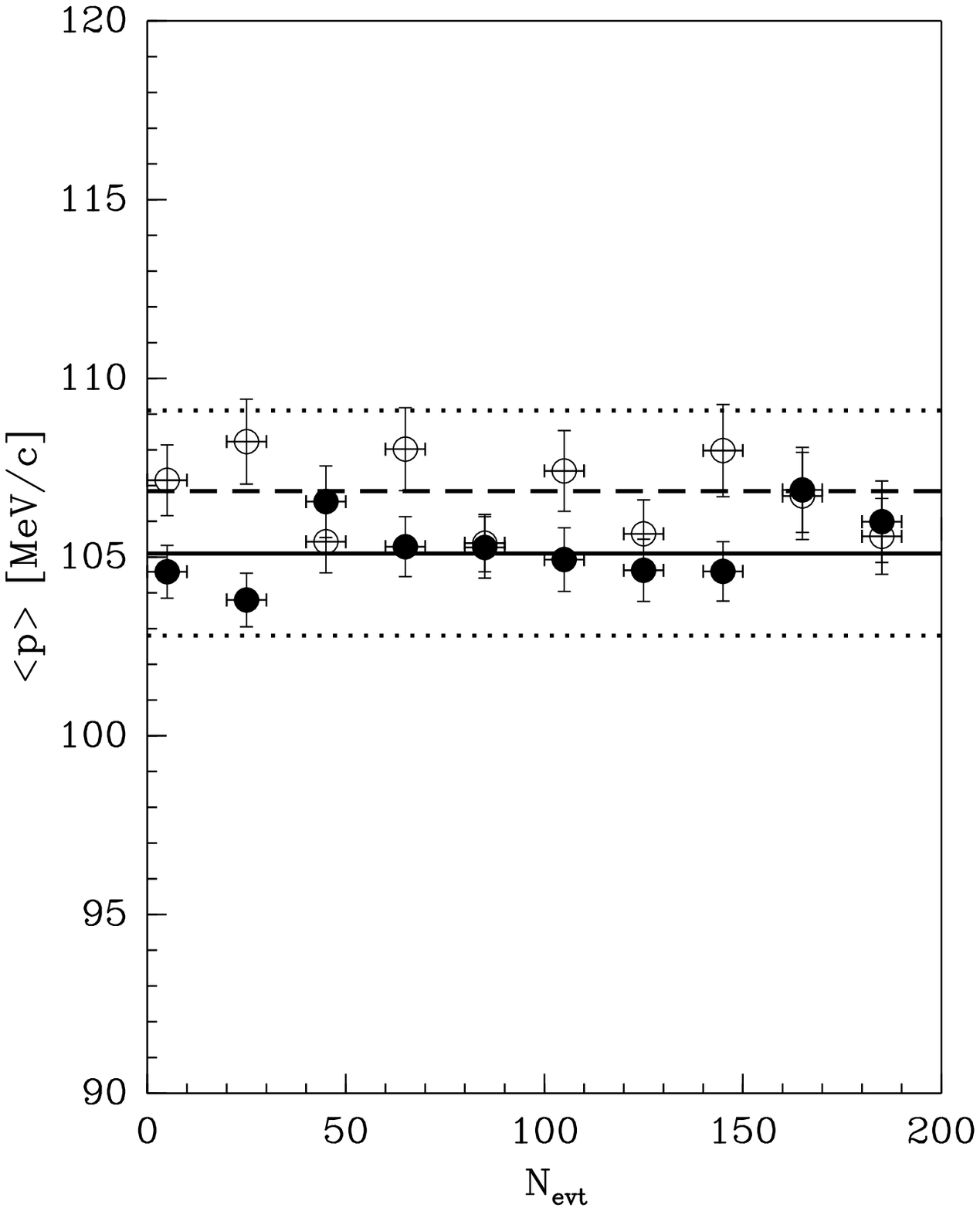}
~~
\includegraphics[width=0.4\textwidth]{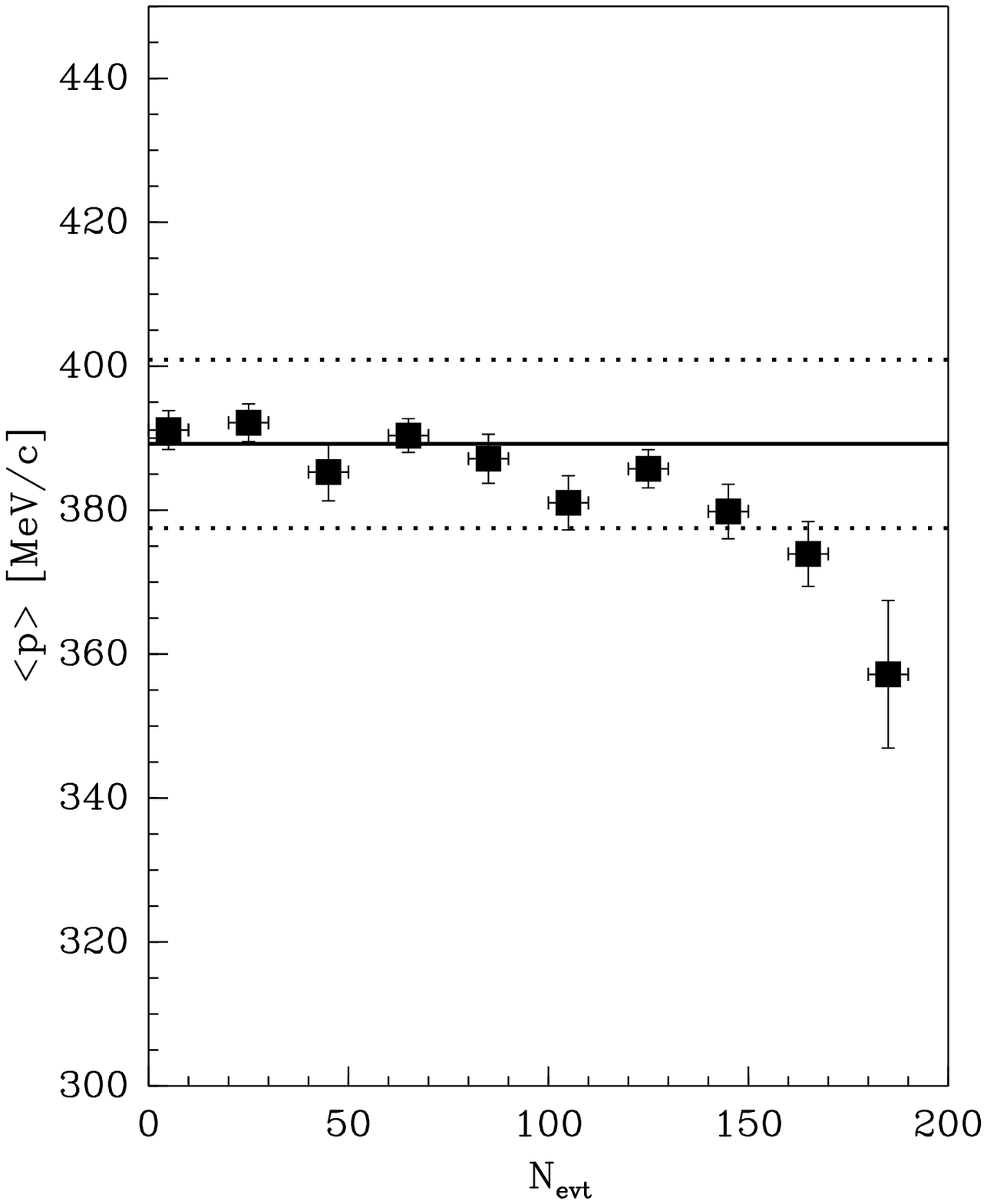}
\caption{Comparison of the average reconstructed momentum as a
  function of event number in spill: left, for charged 
  pions selected using \dedx, open circles are for $\pi^-$, closed
  circles are   for $\pi^+$;   right, for protons using a higher
  value of \dedx for the selection.
  In the left panel the straight lines indicate the average values for
 the first 100 events:
  dashed line for $\pi^-$ and solid line for $\pi^+$.
  In the right plot the solid line shows the average for protons for
  the first 100 events.
 The dotted lines in both panels show the $\pm 3\%$ variation around the
 averages. 
 For both panels a selection 1.0~rad~$<\theta <$~1.5~rad was applied to
 ensure a limited range of $p_{\mathrm{T}}$.
 For the protons in the right panel the requirement that they have a hit
 in both barrel RPC layers was applied to set a momentum threshold.
}
\label{fig:dedx:pions}
\end{figure}

In Fig.~\ref{fig:dedx:pions} the sensitivity of the momentum measurement
to dynamic distortions is shown.   
Particles were selected in narrow bands of \dedx in regions where \dedx
depends strongly on momentum.  
To select a sample with the highest possible momentum, the protons in
the right panel of Fig.~\ref{fig:dedx:pions} were required to reach the
RPC system (low momentum protons would be absorbed before reaching the
RPCs) 
in
addition to the requirement $\dedx > 4.8 \ \mathrm{MIP}$. 
A further selection 1.0~rad~$<\theta <$~1.5~rad ensures a limited
range of $p_{\mathrm{T}}$.
The same angular selection was applied for the tracks shown in the left
panel, together with a selection 
$2.3 \ \mathrm{MIP} < \dedx< 2.8 \ \mathrm{MIP}$.
Owing to the combined selection of a \dedx interval and momentum
interval, the sample of tracks in the left panel of
Fig.~\ref{fig:dedx:pions} is a pure pion sample and in the right panel
of the same figure a pure proton sample.
The analysis was performed for the combined data set taken with
3~\GeVc, 5~\GeVc, 8~\GeVc and 12~\GeVc beams on Be, C, Cu, Sn, Ta and Pb
targets. 
As can be seen in Fig.~\ref{fig:dedx:pions}
the momentum measurement using the vertex constraint is robust with
respect to the dynamic distortions within a few percent for values of
$\langle d_0' \rangle$ smaller than 5~mm.
This robustness is contrary to the effect observed with the fit not
using the vertex constraint which is much more sensitive to
distortions as shown in Section~\ref{sec:elastics}. 
The average momentum obtained from a Gaussian fit to the momentum
distribution shows that the average momentum stays constant within a
few percent up to $\evtspill = 200$ at $p_{\mathrm{T}} \approx 95 \ \MeVc$
(pions)  and up to 
$\evtspill = 100$ at $p_{\mathrm{T}} \approx 350 \ \MeVc$ (protons),
respectively. 
For this data set $\langle d_0' \rangle$ is $\approx 5 \ \mm$ at
$\evtspill = 100$ and roughly twice as large at $\evtspill = 200$.
The $p_{\mathrm{T}}$-range covered by this cross-check represents a large range
of the kinematic domain used in the analysis. 
It is expected that the transverse momentum measurement of lower \pt
tracks ($p_{\mathrm{T}} \approx 95 \ \MeVc$) is less affected by dynamic
distortions than that of higher \pt tracks 
($p_{\mathrm{T}} \approx 350 \ \MeVc$) since an equal  shift in
position of the clusters induces a smaller fractional change in
curvature for tracks with large curvature.

The measurement of the angle of the particles' trajectory with respect
to the beam direction $\theta$ remains constant within a few mrad up to 
$\langle d_0' \rangle = 10 \ \mm$.

To check asymmetries of the momentum reconstruction between $\pi^+$
and $\pi^-$ one can inspect the results of the analysis where tracks emitted
almost perpendicular (1.0 rad  $<\theta <$ 1.5 rad) to the beam
direction with a \dedx of about three times the value of a minimum
ionizing particle were selected.
Using the fact that the $\pi^+$ and $\pi^-$ spectra are expected to be
similar at these angles and that the \dedx selection keeps only
pions in a narrow momentum region, one can constrain reconstruction
asymmetries. 
For this selection we find an average pion momentum of 105~\MeVc with
an asymmetry of 1\%, as shown in Fig.~\ref{fig:dedx:pions}.
This is negligible compared to other systematic errors in the
analysis (see Section~\ref{sec:elastics}).

\subsection{Efficiency}
\label{sec:tracking-eff}

\begin{figure}[tbp]
  \begin{center}
    \epsfig{figure=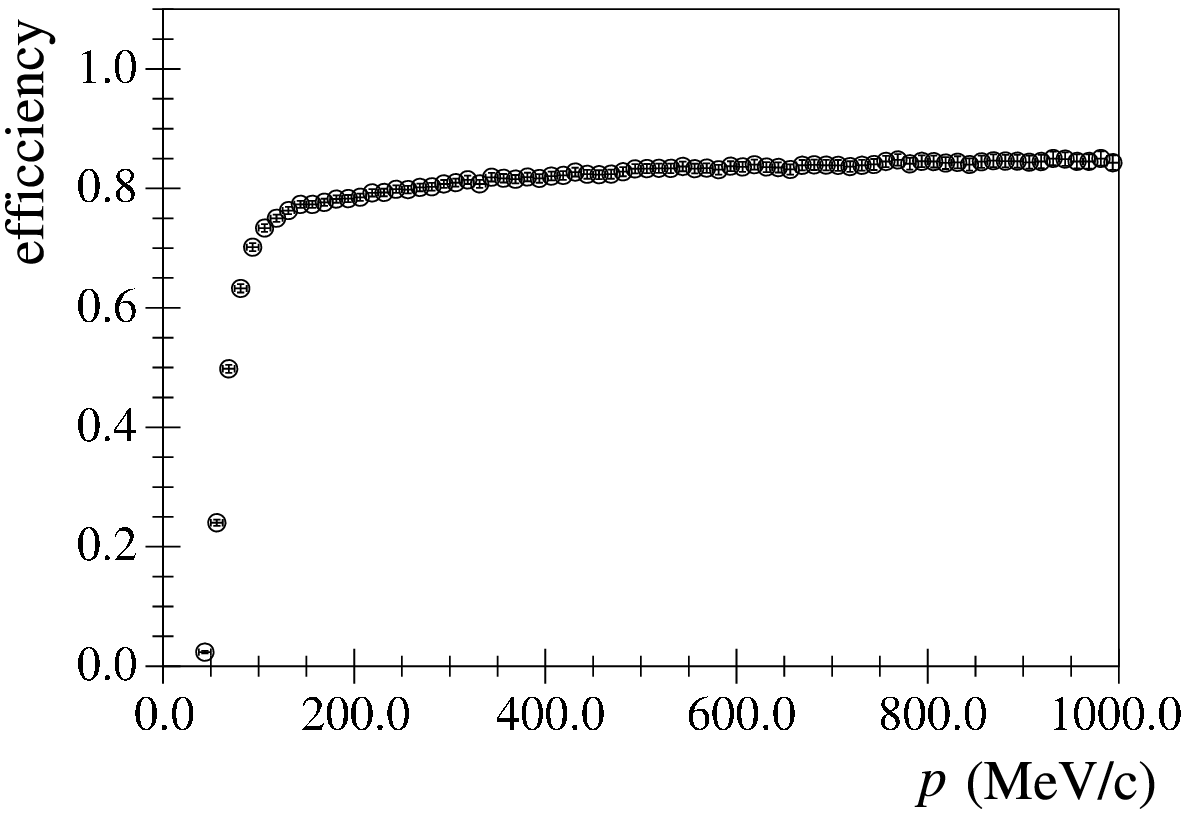,width=0.48\textwidth,bburx=350,bbury=245}
    \epsfig{figure=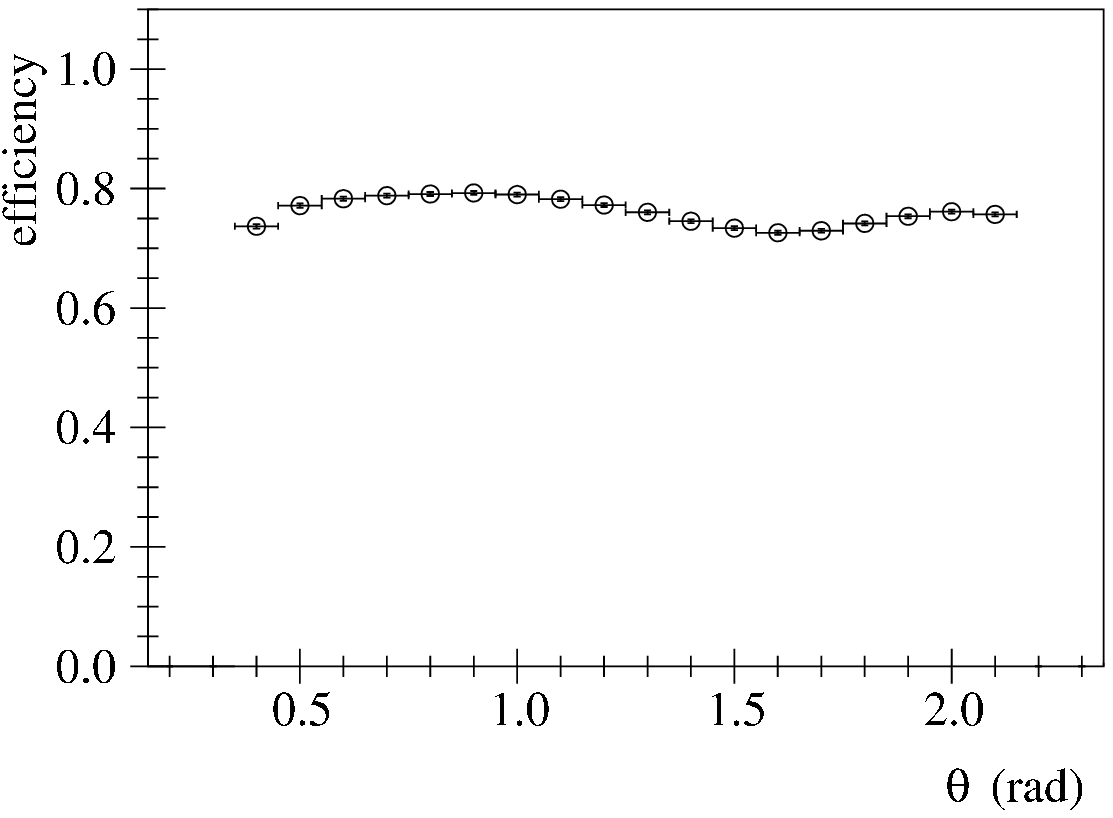,width=0.48\textwidth,bburx=350,bbury=245}
  \end{center}
\caption{
Left: the efficiency as a function of total momentum at their production point for pions.
Right: the efficiency as a function of \tht for pions.
}
\label{fig:eff-pt}
\label{fig:eff-th}
\end{figure}

Having verified the ability of the Monte Carlo program to simulate the
efficiency for protons (Section~\ref{sec:elastics}), the simulation is
then used for pions. 
The efficiency calculation was done by
simulating single \pip and \pim in bins of \tht and $p$.
The map of dead channels in the TPC was applied corresponding to the
data set to be corrected.
Thus a different simulation was run for each of the momentum settings.
The same cuts used for the data were applied to the reconstructed MC
tracks.  
Figures~\ref{fig:eff-pt}~(left) and \ref{fig:eff-th}~(right) show the
efficiency for pions as a function of $p$ and \tht, respectively.
The variable on the abscissa in Fig.~\ref{fig:eff-pt}~(left) is the
momentum of the pion in the gas of the TPC, hence after energy loss in
the target and the material around the inner field cage.
The result confirms that the efficiency is strongly limited at
low momentum ($p \leq$ 75 \ \MeVc for pions) due to  the energy-loss
in the materials surrounding the target and inside the target itself.  
Consequently, the measurement will be limited to pions with momentum 
at their production point above 100~\MeVc.
The dip at $\theta = 1.57 \ \rad$ ($0.5 \ \pi \ \rad$) in
Fig.~\ref{fig:eff-th}~(right) is due to the absorption and energy loss
in the target. 
The amount of material represented by the target with its 30~mm diameter
is large for tracks traversing it at 90 degrees with respect to the beam
direction. 
This effect dominates over the increase in material seen by tracks which
traverse the inner field cage at small angles with respect to the beam,
but which traverse only on average half of the 5.6~mm thickness of the
target in the beam direction.

\subsection{Particle identification}
\label{sec:pid}

\begin{figure}[tbp]
  \begin{center}
   \begin{minipage}[b]{0.49\textwidth}{
    \begin{turn}{90}\mbox{~~~~~~~~~~~~~~~~~~~~~~~~~\dedx (a.u.)}
    \end{turn}
    \epsfig{figure=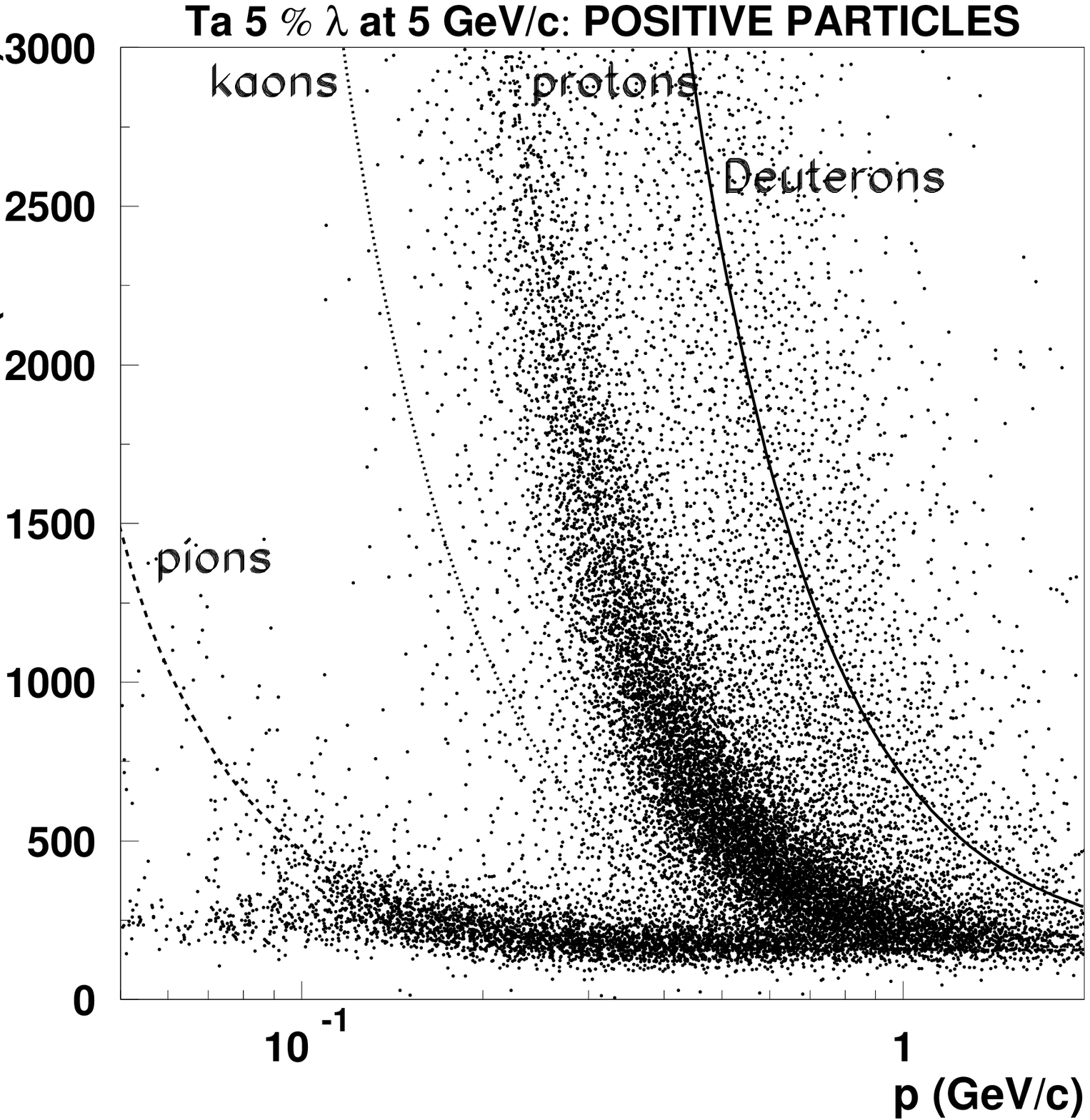,width=0.9\textwidth}
    }
    \end{minipage}
   \begin{minipage}[b]{0.49\textwidth}{
    \begin{turn}{90}\mbox{~~~~~~~~~~~~~~~~~~~~~~~~~\dedx (a.u.)}
    \end{turn}
    \epsfig{figure=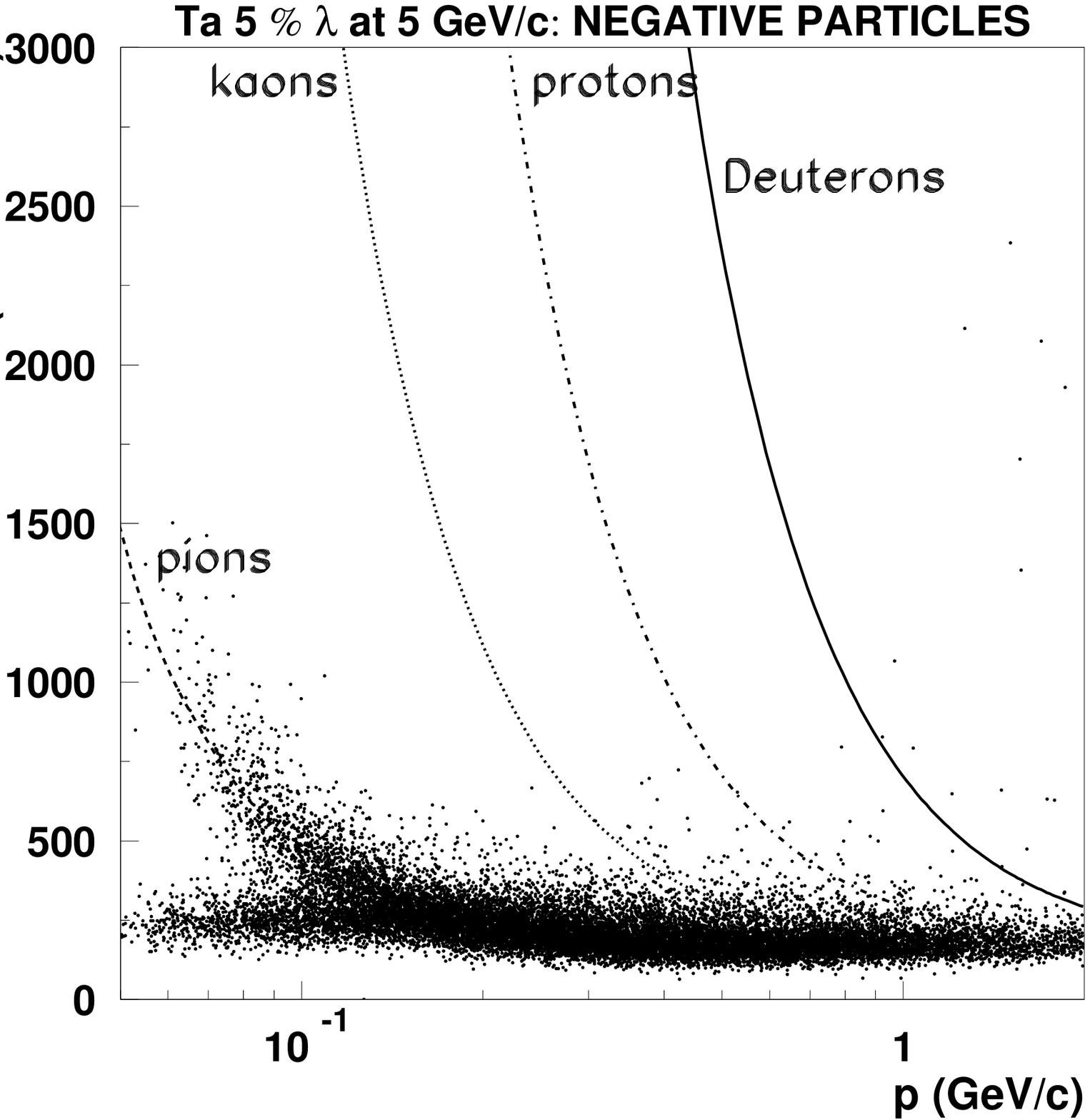,width=0.9\textwidth}
    }
    \end{minipage}
    \vspace{9pt}
   \begin{minipage}[b]{0.49\textwidth}{
    \begin{turn}{90}\mbox{~~~~~~~~~~~~~~~~~~~~~~~~~\dedx (a.u.)}
    \end{turn}
    \epsfig{figure=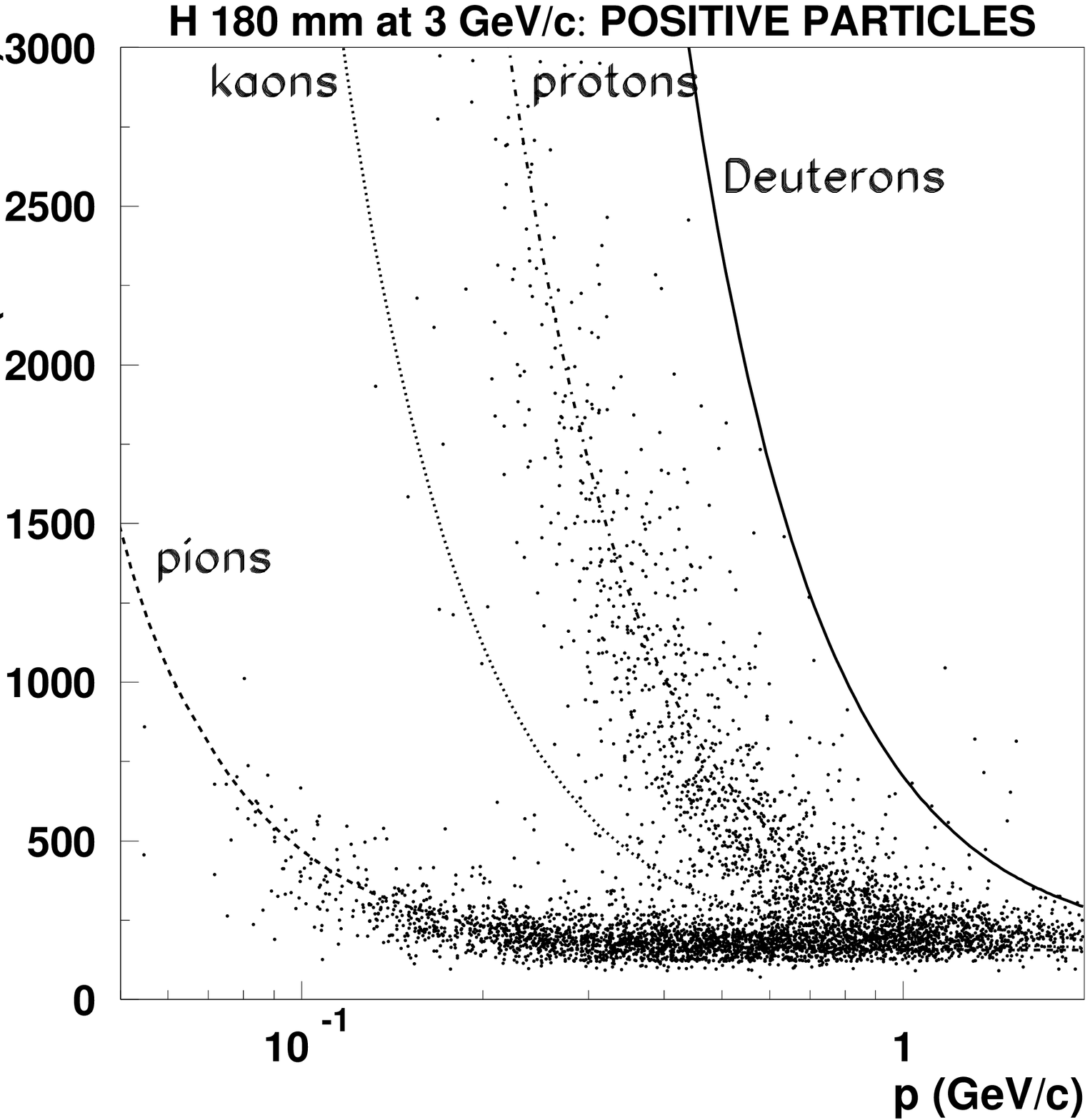,width=0.9\textwidth}
    }
    \end{minipage}
   \begin{minipage}[b]{0.49\textwidth}{
    \begin{turn}{90}\mbox{~~~~~~~~~~~~~~~~~~~~~~~~~\dedx (a.u.)}
    \end{turn}
    \epsfig{figure=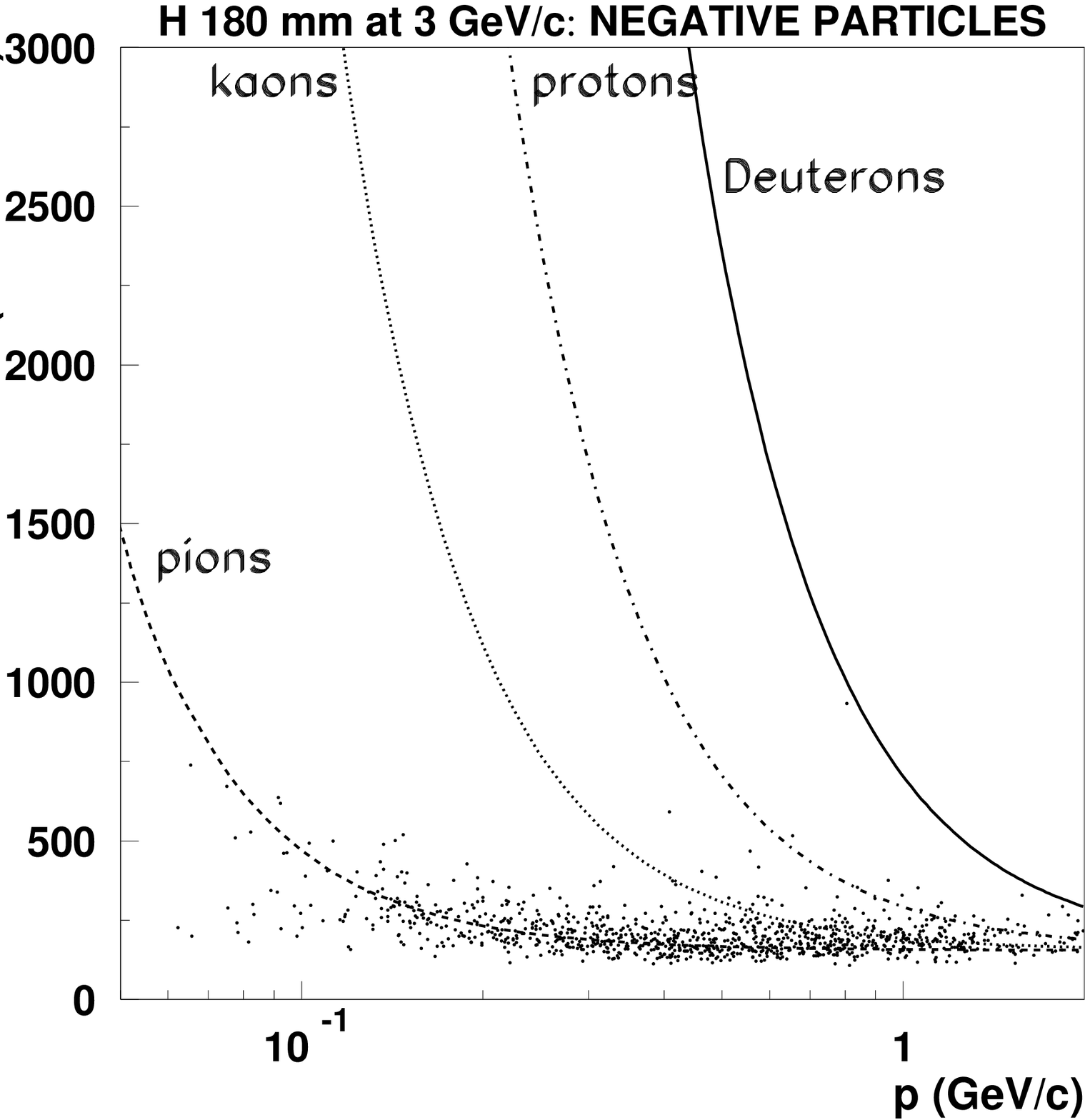,width=0.9\textwidth}
    }
    \end{minipage}
  \end{center}
\caption{d$E$/d$x$ (in arbitrary units) versus momentum (in \GeVc). 
  Top: p--tantalum data in the
  5~\GeVc beam; Bottom: p--hydrogen data in the
  3~\GeVc beam; Left: for positive tracks; Right: for negative
  tracks. 
  The lines are simple $\beta^{-2}$ curves and merely indicate the
 regions  populated by the various particle types.  
  The fact that the band marked `deuterons' is not present in the
  hydrogen data clearly shows that the population in this band in the
  tantalum data is not an artefact of the momentum reconstruction but
  deuteron production.
}
\label{fig:dedx}
\end{figure}

\begin{figure}[tbp]
\centering
\mbox{
  \epsfig{file=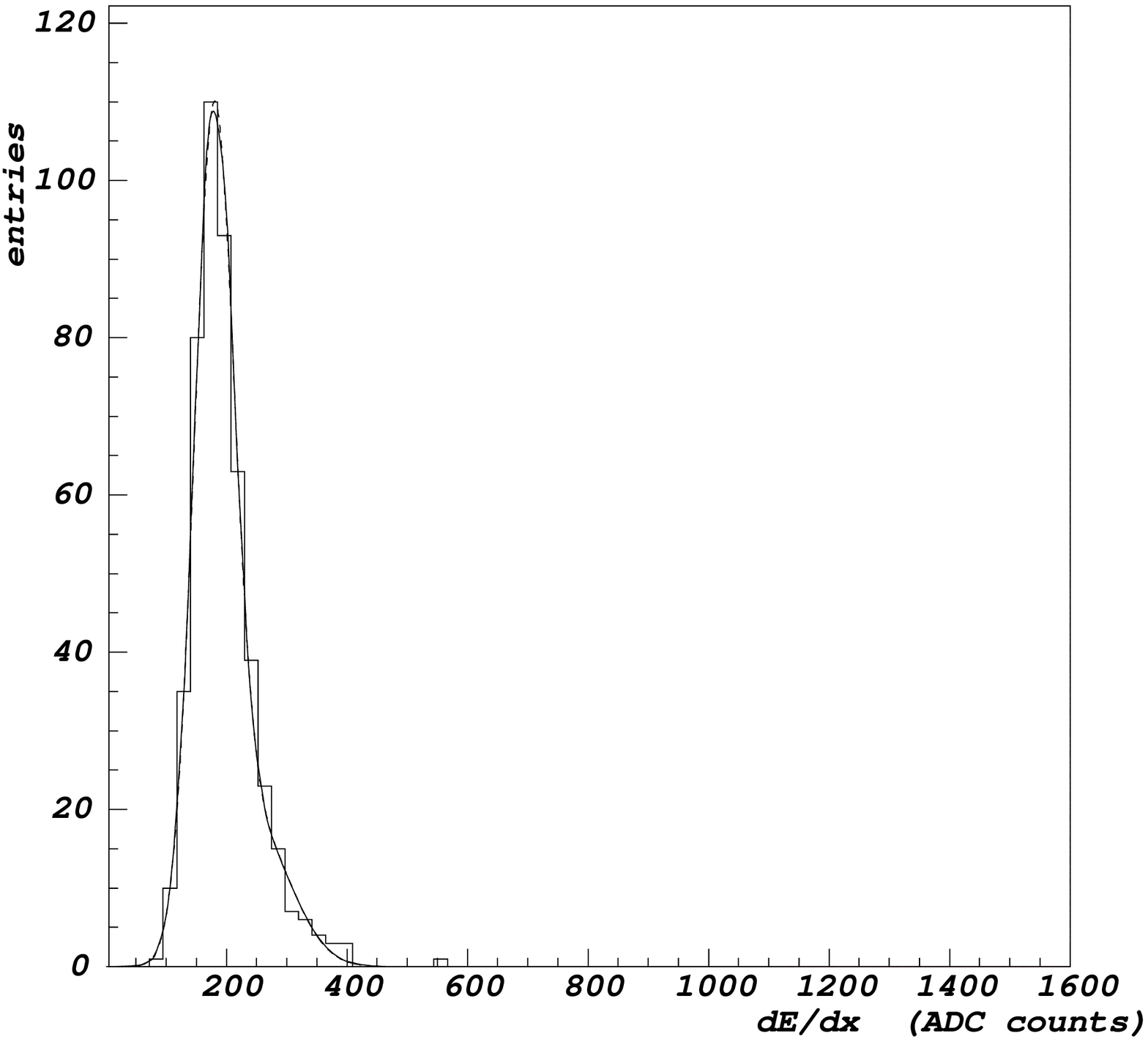,width=0.40\textwidth}
  \epsfig{file=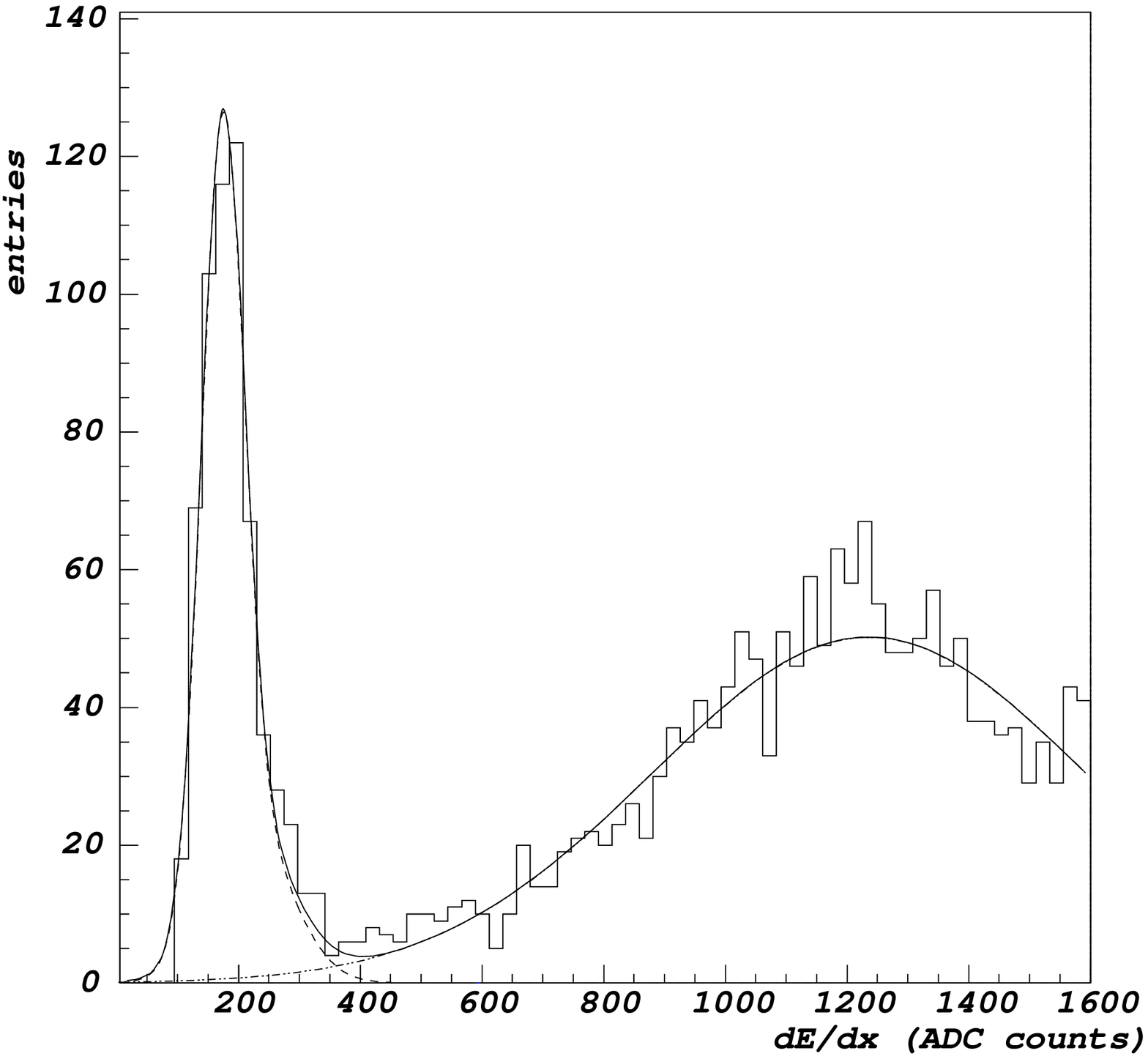,width=0.40\textwidth}
}
\caption{\dedx spectra for negative particles (on the left) and
  positive (on the right) with momentum between 300~\MeVc and
  350~\MeVc. The curves show the Landau distributions fitted to the
  data.  The protons are clearly visible in the distribution for
  positive particles at high \dedx and absent for the negatively
  charged particles.} 
\label{fig:alternative:dedx300}
\end{figure}

\begin{figure}[tbp]
\begin{center}
\mbox{
  \epsfig{file=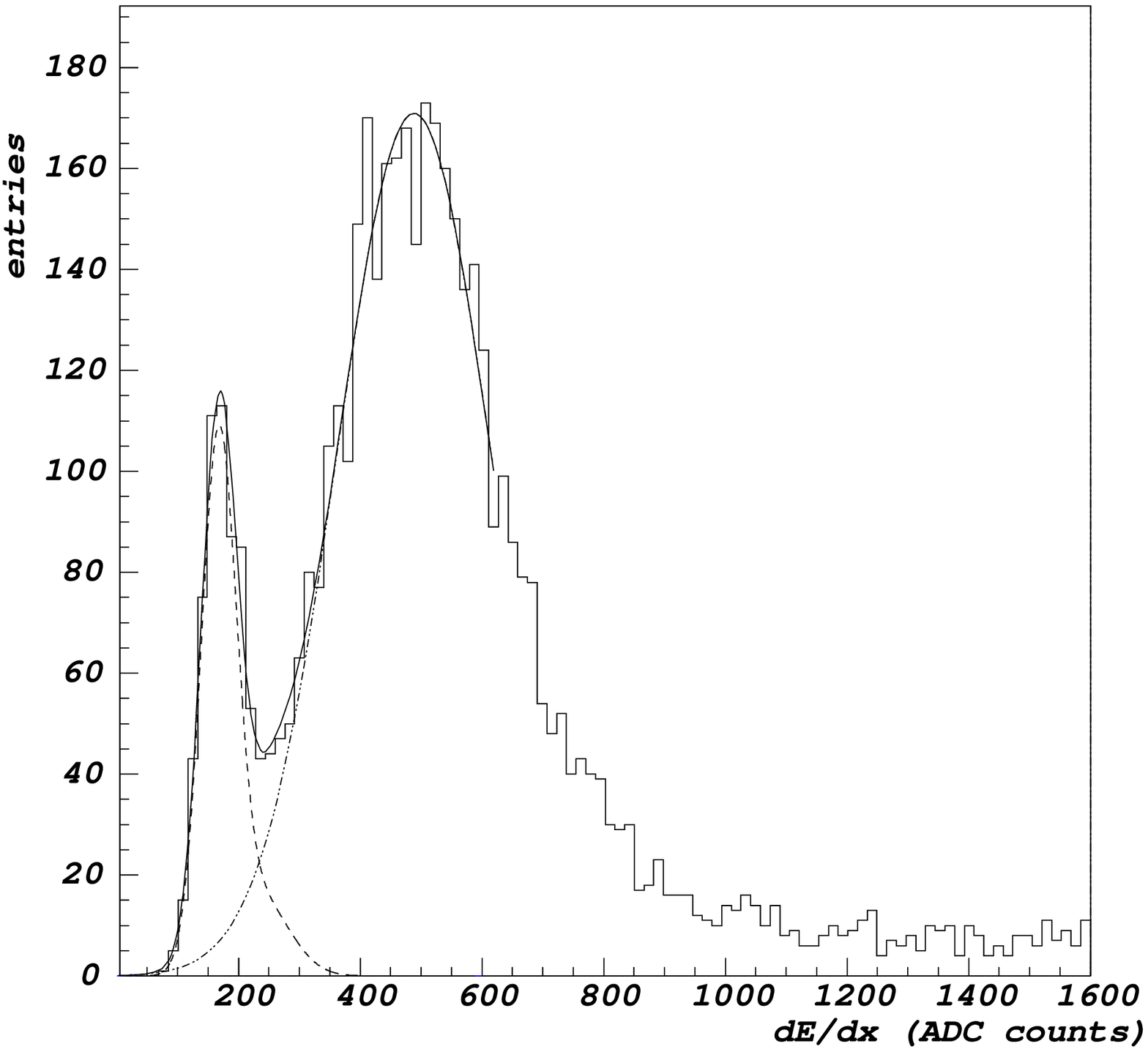,width=0.40\textwidth}
  \epsfig{file=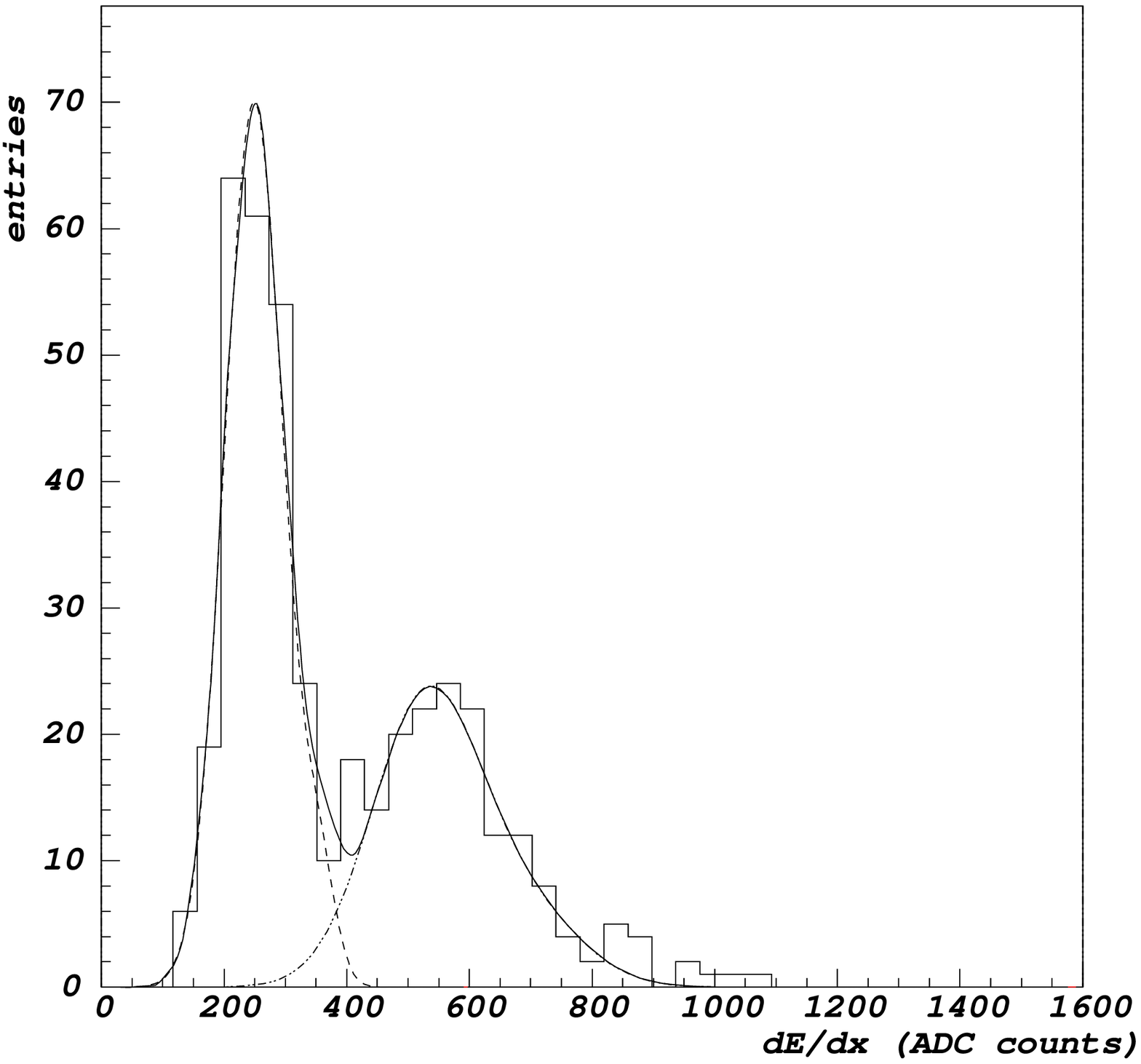,width=0.40\textwidth}
}
\caption{Left panel: \dedx spectra for 
  positive particles with momentum between 500~\MeVc and
  600~\MeVc.  The curves show the Landau distributions fitted to the
  data.  The distributions of pions and protons are distinct but not
  cleanly separated.
  Right panel: The d$E$/d$x$ spectrum for negative particles with momentum
  between 75~\MeVc and 100~\MeVc.  The curves indicate the fits to the
  two components using two Landau distributions.  The distribution of
  the electrons with low \dedx is clearly visible to the left of the
  highly ionizing negative pions.}
\label{fig:alternative:dedx75}
\end{center}
\end{figure}

The particle identification in the large-angle region mainly uses 
the \dedx information provided by  the TPC. 
The measurement of \dedx is shown as a function of momentum in
Fig.~\ref{fig:dedx}.  
The electron, pion and proton populations are well separated at most
momentum values. 
As an example, the distributions in various momentum ranges are shown
in Fig.~\ref{fig:alternative:dedx300} and
\ref{fig:alternative:dedx75}.
These figures show the separation between electrons and pions in the
low momentum region, and the pion--proton separation at intermediate
and higher momenta.
Fits with two Landau distributions (corresponding to the different
particle types) are also shown in the figures.
In this analysis simple momentum dependent cuts are used to
separate the different populations. 
The pions are identified by removing electrons and protons.
The kaon population is negligible.
The cuts were optimized to maximize the purity of the pion sample,
accepting a lower efficiency in the selection.

The measurement of the velocity $\beta$ of secondary particles by the
time-of-flight determination with the RPC detectors using the BTOF as
starting-time reference provides complementary particle identification.
It allows the efficiency and purity of the PID algorithm using \dedx
to be studied for a large subset of the TPC tracks.  
Combining the samples taken with the different beam momenta used in
this analysis a statistical accuracy of the order of 0.2\% can be
obtained in the PID efficiency determination. 

The choice to use \dedx as principal PID estimator is motivated by two
facts. 
The first argument is given by the fact that \dedx is obtained as a
property of the same points which constitute the TPC track, while the
TOF is obtained by matching the track to an external device.
It is observed that the background in the matching is not negligible. 
Converted photons from \piz production can hit the same -- rather large
-- RPC pad as the one pointed to by the track.  
This background depends on the position in the RPC barrel where the pad
is located and is different for every momentum setting.
Thus a different background subtraction would have to be determined for
each momentum--target dataset.
The second argument is the increased complexity of the analysis which
would be introduced by having to combine two PID detectors of which the
response is highly non-Gaussian.
The probability density functions of both the response of the \dedx and
of the TOF would have to be determined as function of all relevant
parameters. 
The gain in efficiency one would obtain with such a procedure would be
rather limited and would not balance the additional systematics
introduced. 
On the contrary, the availability of an independent PID device makes it
possible to determine the efficiency and purity of the selection with
the main device in a straightforward manner, without the need to know
the efficiency of the independent auxiliary PID device.

The measurement of $\beta$ allows an almost independent selection of a
very pure proton sample to be made in the momentum range
300~\MeVc--800~\MeVc  with a purity better than 99.8\%. 
The purity of the sample was checked using negative particles and
verifying that no particles identified as anti-protons are present.
While a proton sample was obtained using interactions of incoming
protons, a pure pion sample was prepared by using negative pions
selected by TOF produced by incident positive pions.
The behaviour of positive pions was also checked for momenta below
500~\MeVc (where they can be selected without proton contamination)
and was found to be equal to that of negative pions.

\begin{figure}[tbp]
  \begin{center}
    \epsfig{figure=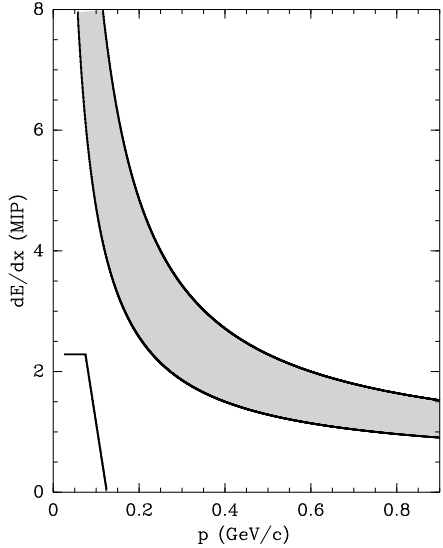,width=0.36\textwidth}
  \end{center}
\caption{Criteria used for the PID selection using the \dedx (expressed
 in MIP) as a
 function of the measured momentum of the particle.  Low momentum
 electrons and positrons are rejected when their \dedx is below the
 lower left curve.  The remaining paricles are classified as protons if
 their \dedx is above the gray band, as pions if they are below the gray
 band and rejected when they lie inside the gray band.
 The value of the MIP is calibrated for each setting.
}
\label{fig:dedx:pidcuts}
\end{figure}

The cuts were defined favouring purity over efficiency and are shown
graphically in Fig.~\ref{fig:dedx:pidcuts}.  
Protons are selected by requiring a high \dedx, while at higher momenta
pions are selected with low \dedx.  
To ensure purity of both samples there are `unidentified' particles
between the two samples.
At low momenta electrons are rejected by selecting low \dedx, while pions
are accepted with a higher \dedx.
This separation is not pure above 125~\MeVc, so an electron
subtraction is needed in the analysis.

\begin{figure}[tbp]
  \begin{center}
    \epsfig{figure=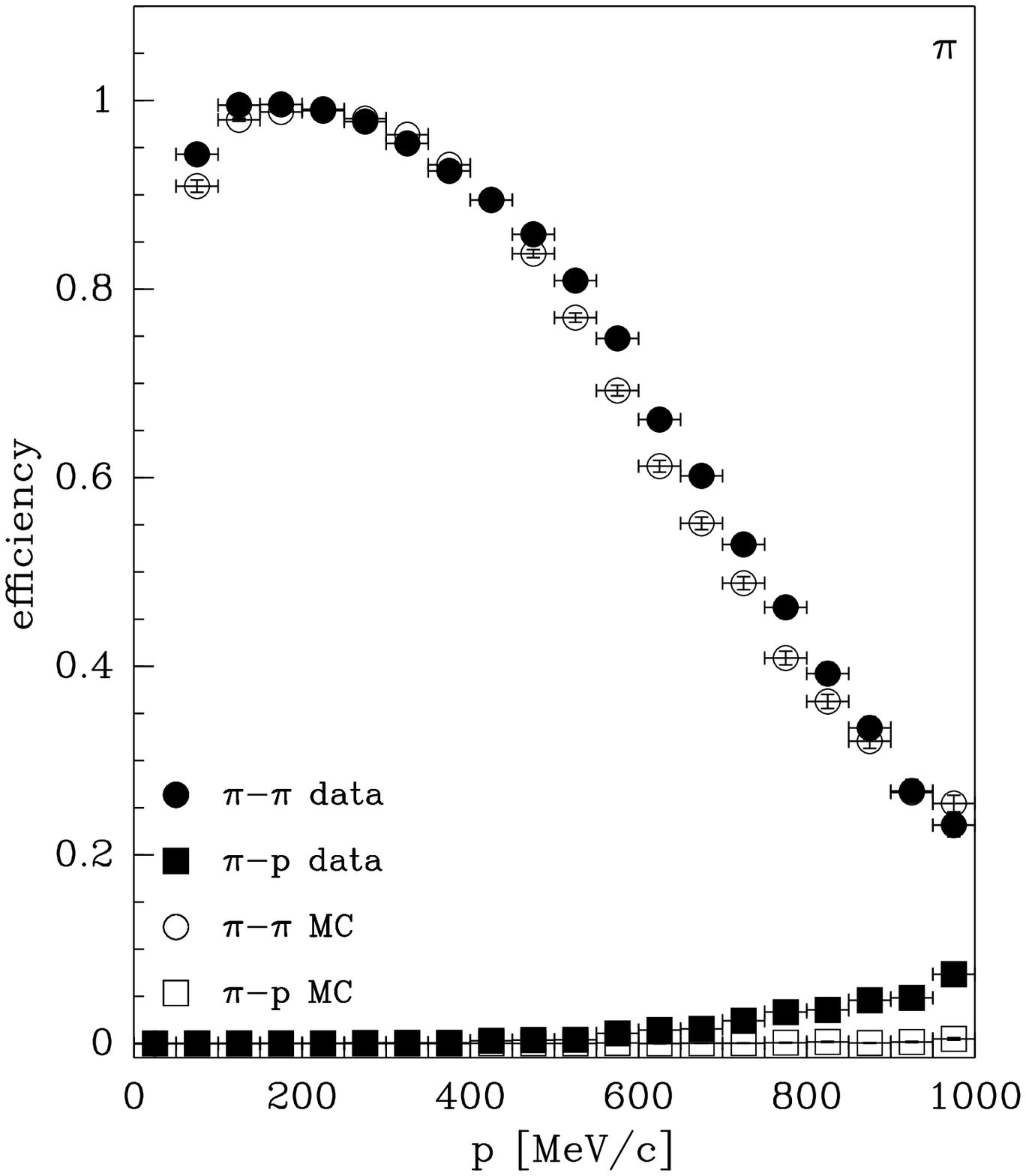,width=0.36\textwidth}
    ~~
    \epsfig{figure=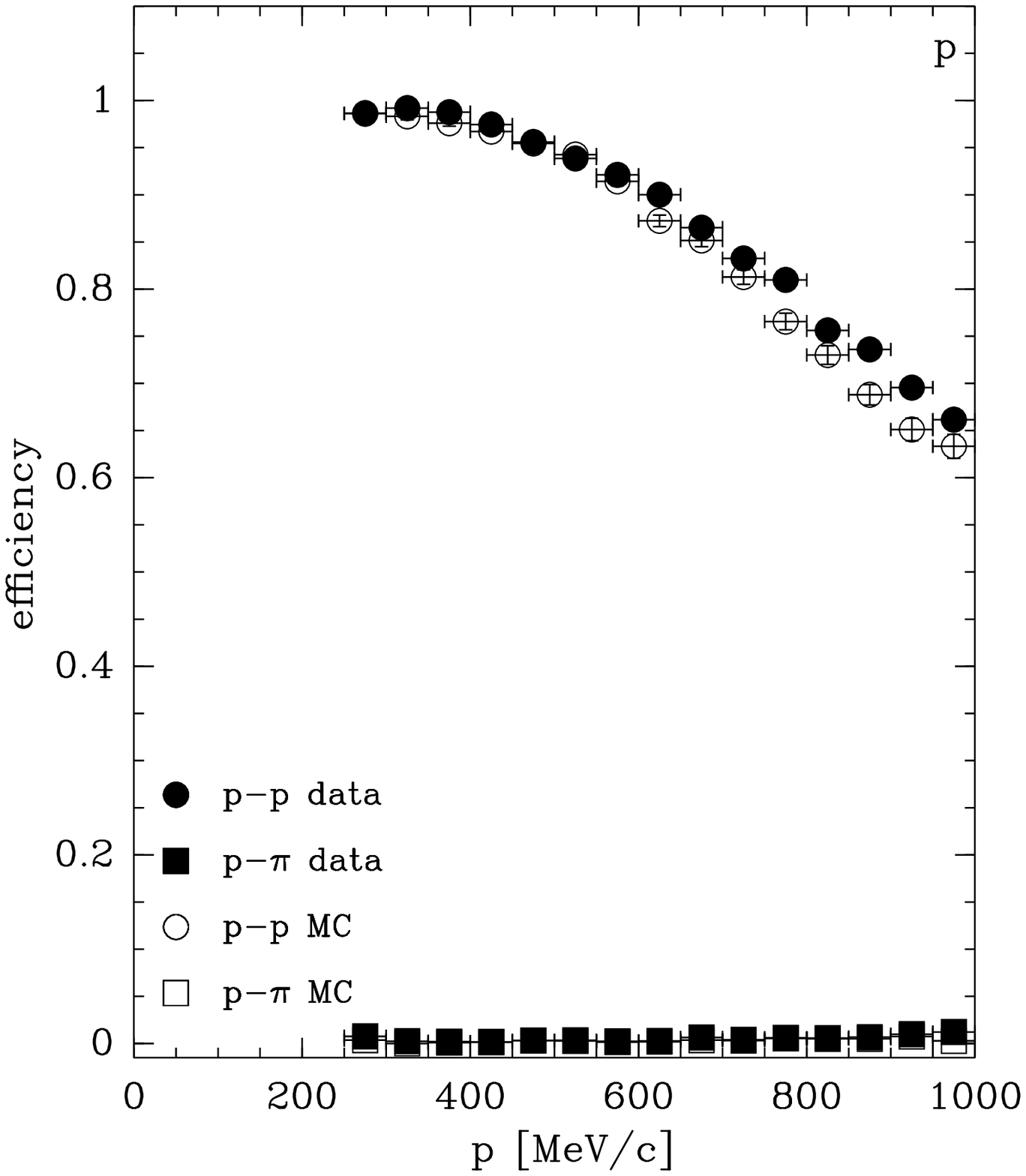,width=0.36\textwidth}
  \end{center}
\caption{Performance of the PID using the \dedx as a function of the
  measured momentum of the particle.  The particles are selected using
  TOF.  Left: for negative pions produced in a positive pion beam;
  Right: for protons produced in a proton beam.
  The filled (open) circles show the efficiency measured with the data
  (Monte Carlo), the filled (open) squares represent the fraction of
  particles misidentified as anti-protons (left) and pions (right) in the
  data (Monte Carlo).
}
\label{fig:dedx:pid}
\end{figure}

The result of this analysis in terms of efficiency and 
of the fraction misidentified particles  
is shown in Fig.~\ref{fig:dedx:pid}.
For the pions, the drop in efficiency toward higher momenta is caused by
the need to make a hard cut to remove protons. 
The migration of pions and protons into the wrong sample is kept below
the percent level in the momentum range of this analysis ($p <
800~\MeVc$). 
This is important for the measurement of the  \pip production rate 
since the proton production rate is five to ten times larger in some
of the bins.
The small differences in efficiency (up to $\approx5$\%)  which are
visible between the data and the simulation are dealt with in the
analysis by an {\em ad hoc} correction to the cross-sections.
It is checked that the angular dependence of the PID efficiency and
purity are negligible. 

\begin{figure}[tbp]
\begin{center}
\epsfig{figure=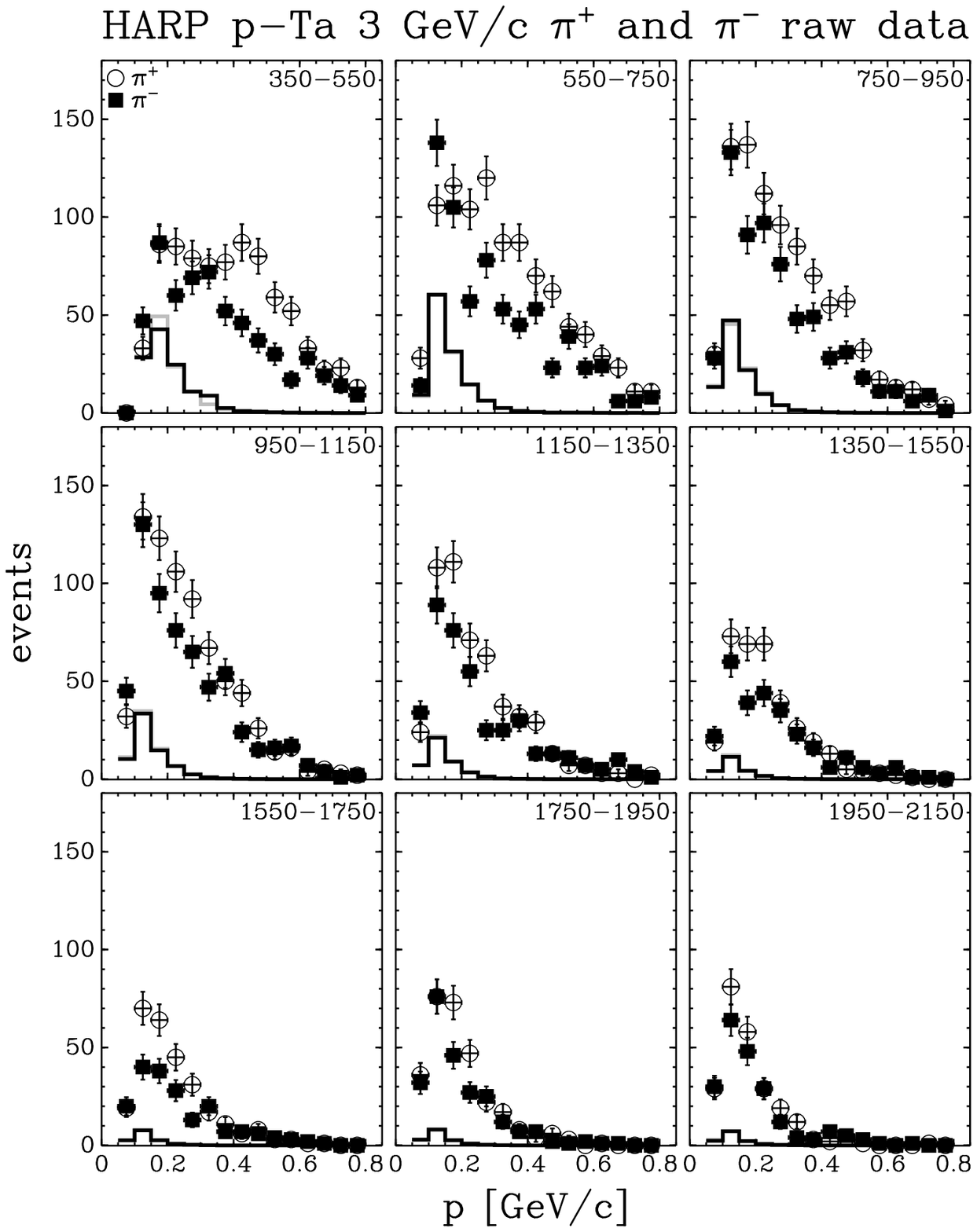,width=0.48\textwidth}
\epsfig{figure=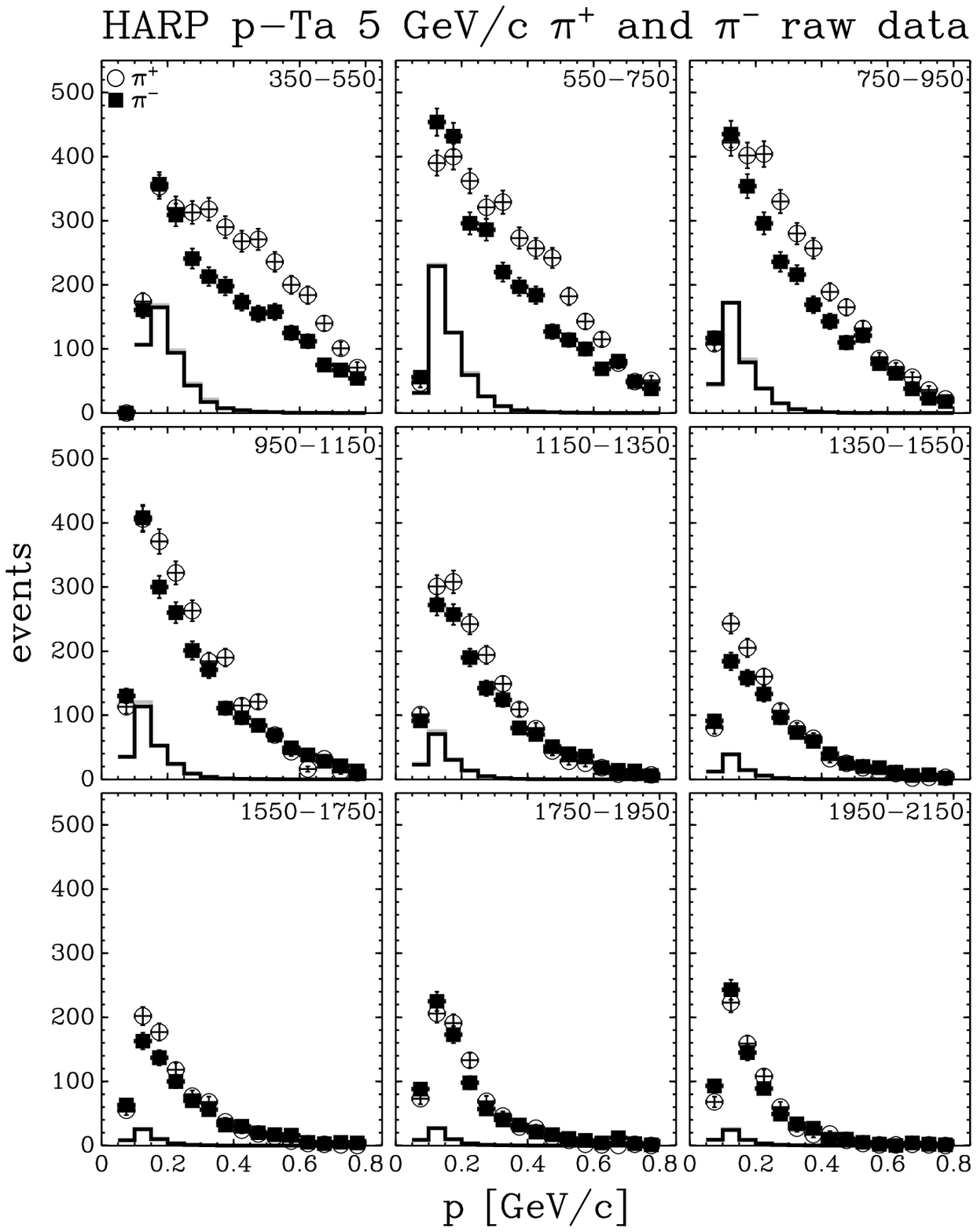,width=0.48\textwidth}
\vspace{9pt}
\epsfig{figure=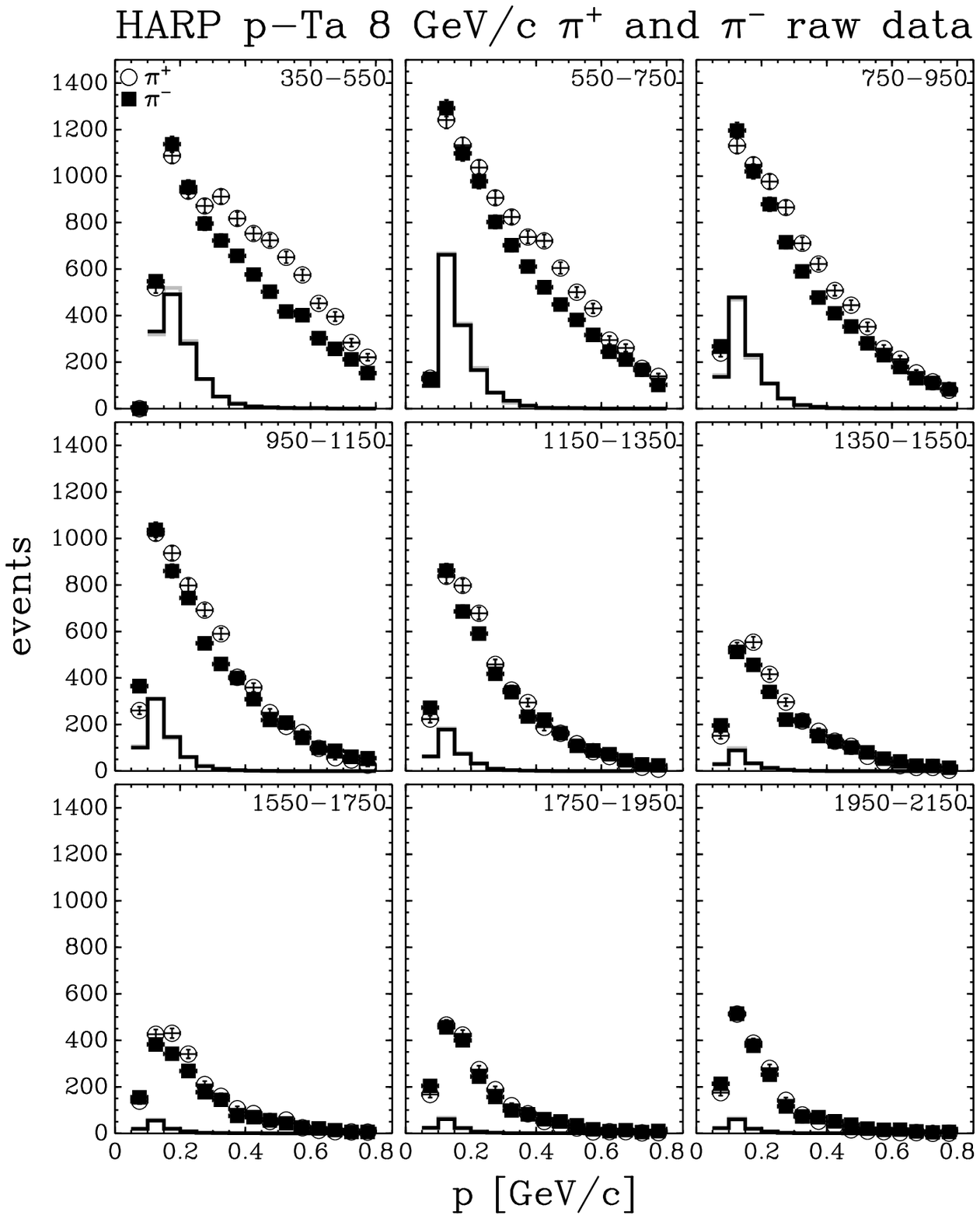,width=0.48\textwidth}
\epsfig{figure=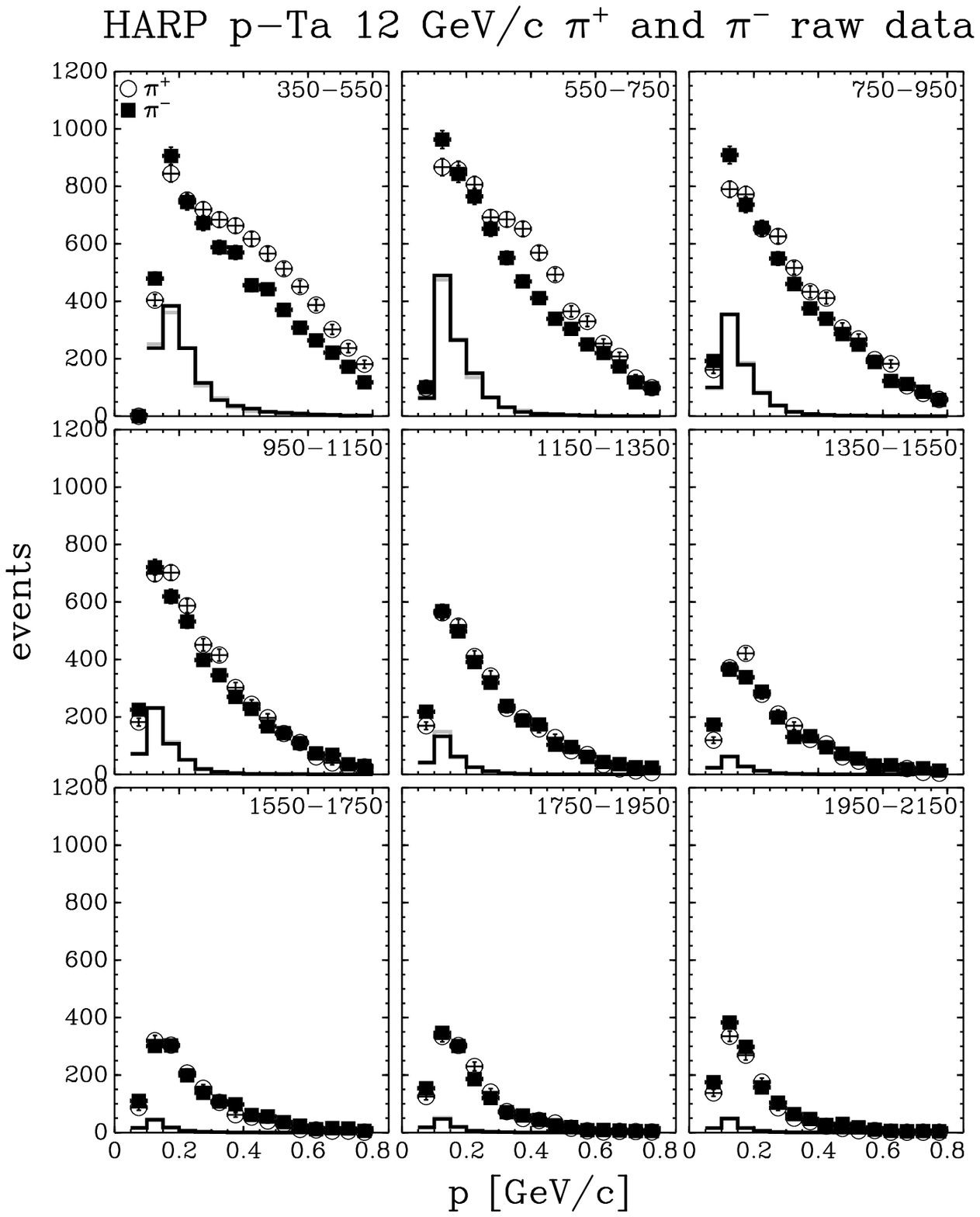,width=0.48\textwidth}
\end{center}
\caption{Distribution for positive (open circles) and
  negative pions (filled squares) using the PID algorithm based on \dedx
  as a function of momentum and in different angular bins (indicated in
 mrad in the panels) taken
  with 3~\GeVc, 5~\GeVc, 8~\GeVc and 12~\GeVc proton beam hitting a
  tantalum target.
  The histograms show the distributions calculated for the \piz
  subtraction (see Section~\ref{sec:xsec}).  
  The shaded (black) lines show the positrons (electrons) to be
  subtracted from the \pip (\pim) spectra.  The two sets of lines coincide
  almost everywhere as expected from the production mechanism and are
 therefore almost indistinguishable in the figure.
}
\label{fig:raw-pions}
\end{figure}

With the cuts as described above, the momentum distributions of pions
are obtained in angular bins (indicated in mrad in the panels) as shown
in Fig.~\ref{fig:raw-pions}. 
The distributions in this figure are not corrected for efficiencies
and backgrounds.  

\subsection{Simulation program}
\label{sec:simulation}

The experiment simulation is based on the GEANT4
toolkit~\cite{ref:geant4}. 
The materials in the beam line and the detector are accurately
reproduced in this simulation, as well as the relevant features of the
detector response and the digitization process.
The simulation starts from a beam proton 4.5~m upstream of the
tantalum target. 
The characteristics of the proton beam are reproduced from the
measurements with the MWPC for each momentum setting of the beam
separately.  
The response of the relevant beam detectors is simulated in terms of
measurements of time, position and direction, so that the
reconstruction of simulated events gives realistic results.

The most important detectors to simulate for this analysis are the TPC,
the RPC system and the trigger counters.
In the TPC all stochastic processes in the energy deposition of the
particles along their trajectories are reproduced, including the
electron drift, the signal formation on the pad plane, the response of
the electronics and the digitization process.
Important details such as the individual behaviour of each single
electronics channel in terms of pulse shape and signal amplitude are
taken into account.
For each different setting (beam--target combination) the precise
knowledge of dead pads and equalization constants as observed in the
data are reproduced.
The RPCs are simulated using their actual geometrical details,
and the response is reproduced from the overall performance
observed in the data.

All relevant physical processes are simulated using the GEANT4 tools,
including multiple scattering, energy-loss, absorption and
re-interactions. 

\section{Analysis procedure}
\label{sec:xsec}

The double-differential cross-section for the production of a particle of 
type $\alpha$ can be expressed in the laboratory system as:

\begin{equation}
{\frac{{\mathrm{d}^2 \sigma_{\alpha}}}{{\mathrm{d}p_i \mathrm{d}\theta_j }}} =
\frac{1}{{N_{\mathrm{pot}} }}\frac{A}{{N_A \rho t}}
M_{ij\alpha i'j' \alpha'}^{-1} \cdot
{N_{i'j'}^{\alpha'}} 
\ ,
\label{eq:cross}
\end{equation}

where $\frac{{\mathrm{d}^2 \sigma_{\alpha}}}{{\mathrm{d}p_i \mathrm{d}\theta_j }}$
is expressed in bins of true momentum ($p_i$), angle ($\theta_j$) and
particle type ($\alpha$).
The summation over reconstructed indices $i'j'\alpha'$ is implied in
the equation.
The terms on the right-hand side of the equation are as follows.

The so called `raw yield' $N_{i'j'}^{\alpha'}$ 
is the number of particles of observed type $\alpha'$ in bins of reconstructed
momentum ($p_{i'}$) and  angle ($\theta_{j'}$). 
These particles must satisfy the event, track and PID 
selection criteria.

The matrix $ M_{ij\alpha i'j' \alpha'}^{-1}$ 
corrects for the  efficiency and resolution of the detector. 
It unfolds the true variables $ij\alpha$ from the reconstructed
variables $i'j'\alpha'$ and corrects  
the observed number of particles to take into account effects such as 
trigger efficiency, reconstruction efficiency, acceptance, absorption,
pion decay, tertiary production, 
PID efficiency, PID misidentification and electron background. 
The method used to correct for the various effects will be described in
more detail in the following section.

The factor  $\frac{A}{{N_A \rho t}}$ 
is the inverse of the number of target nuclei per unit area
($A$ is the atomic mass,
$N_A$ is the Avogadro number, $\rho$ and $t$ are the target density
and thickness)\footnote{We do not make a correction for the attenuation
of the proton beam in the target, so that strictly speaking the
cross-sections are valid for a $\lambda_{\mathrm{I}}=5\%$ target.}. 

The result is normalized to the number of incident protons on target
$N_{\mathrm{pot}}$. 

Although, owing to the stringent PID selection,  the background from
misidentified protons in the pion sample is small, the pion and proton
raw yields ($N_{i'j'}^{\alpha'}$, for 
$\alpha'=\pim, \pip, \mathrm{p}$) have been measured simultaneously. 
This makes it possible to correct for the small remaining proton
background in the pion data without prior assumptions concerning the
proton production cross-section.

\subsection{Correction for resolution, energy-loss, efficiency and backgrounds}
\label{sec:corrections}
Various techniques are described in the literature to obtain the
matrix   $ M_{ij\alpha i'j' \alpha'}^{-1}$.
In this analysis an unfolding technique is used.
It performs  
a simultaneous unfolding of $p$, $\theta$ and PID, with a correction
matrix $M^{-1}$ computed using the Monte Carlo simulation.

A Bayesian technique, described in Ref.~\cite{dagostini} is used
to calculate the unfolding matrix.
The central assumption of the method is that the 
probability density function in the (`true') physical parameters (`physical
distribution') can be approximated by a histogram with bins
of sufficiently small width.
A population in the physical distribution of events in a
given cell $ij\alpha$ generates a distribution in the measured variables,
$M_{ij\alpha i'j'\alpha'}$, 
where the indices $ij\alpha$ indicate the binning in the physical angular, 
momentum and PID variables, respectively, and  $i'j'\alpha'$ the
binning in the measured variables. 
Thus the observed distribution in the measurements can be
represented by a linear superposition of such populations.
The task of the unfolding procedure consists then of finding the
number of events in the physical bins for which the predicted
superposition in the measurement space gives the best description of
the data.
The application of this unfolding method is described in Ref.~\cite{ref:grossheim}.

In order to predict the population of the migration matrix element 
$M_{ij\alpha i'j'\alpha'}$, the resolution, efficiency
and acceptance of the detector are obtained from the Monte Carlo.
This is a reasonable approach, since the Monte Carlo
simulation describes these quantities correctly
(see Section~\ref{sec:performance}). 
Where some deviations
from the control samples measured from the data are found, 
the data are used to introduce (small) {\em ad hoc} corrections to the
Monte Carlo.

A central point in the unfolding method is the construction of the
`migration matrix' $M_{ij\alpha i'j'\alpha'}$, that is the matrix which
describes the 
distribution of the measurements ($p_{\mathrm{m}}$, $\theta_{\mathrm{m}}$,
and $A_{\mathrm{m}}$, where $A$ represents the integer PID variable) given
a bin in the corresponding physical (`true') variables ($p_{\mathrm{p}}$,
$\theta_{\mathrm{p}}$ and $A_{\mathrm{p}}$).   
In this analysis the entries in this matrix are obtained with the use of a
`single particle Monte Carlo'.
This type of Monte Carlo consists of generating a single particle per
event in the target of a given particle type at a given $p$ and \tht
into the full  detector simulation.
The effect of this particle measured in the detector is ideally a
single particle reconstructed with the same kinematic variables and
properly identified.
However, all known complications are simulated in the Monte Carlo.
In particular, for each of the individual beam momentum settings
(corresponding to a period of data taking of about one calendar day)
the calibrations of the TPC obtained for these particular runs as well
as the characteristics of the incoming beam were used in the Monte
Carlo.  
Especially important is the effect on the efficiency of variations in
the map of dead channels.

The efficiency and the effect of cuts are taken into account by keeping
track of the number of 
generated particles and by entering the measured particle into the
migration matrix only when it has been reconstructed.
This procedure is equivalent to a multiplicative bin--by--bin efficiency
correction. 
The systematic uncertainty in the efficiency is estimated from the
variation observed with the elastic scattering data and the difference
of the efficiency observed for the data and the simulation for the
protons. 

Each point (or bin) in the 3-dimensional phase space ($p_{\mathrm{p}}$,
$\theta_{\mathrm{p}}$ and $A_{\mathrm{p}}$) generates a distribution
in the measured variables.
The corresponding distributions in the measured variables are then the
result of the smearing according to the resolution of the measurements.
For this reason the number of bins in the measured variables is larger
than in the `true' variables, in order not to lose the information
provided by the resolution of the measurements.
The unfolding matrix is obtained using equidistant bins in the true
variables.
The final binning is then defined taking into account the resolution
of the detector and the statistics of the data sample. 
During this re-binning procedure the full information of the covariance
matrix is propagated. 
The Monte Carlo description of the momentum resolution, although
checked with cosmic-ray tracks and elastic scattering data, may not be
perfect. 
Possible discrepancies up to 10\% of the resolution are taken into
account in the systematic error. 
The value of the uncertainty is obtained from the analysis of elastic
scattering and cosmic-ray data.

Using the unfolding approach, possible known biases in the measurements
are taken into account automatically as long as they are described by
the Monte Carlo.
For example the energy-loss of particles inside the target and
material around the inner field cage translate into an average shift
of the measured momentum distribution compared to the physical
momentum. 
Known biases are therefore treated in the same way as resolution
effects. 
Uncertainties in the description of the energy-loss and a potential
bias in the momentum determination are estimated to be of the order of
3\% using the elastic scattering analysis. 
This variation has been applied in the estimation of the corresponding
systematic error.

Also the effects of imperfect PID are treated by representing the
distribution of the measured PID of a single particle type over all
possible bins in the migration matrix.
This procedure allows the background of protons in the pion sample to
be subtracted without {\em a priori} assumptions about the proton spectrum.
The effects of a possible difference of the Monte Carlo description of
the efficiency and purity of the PID are estimated by varying 
the cuts differentially for the data and the simulation within the
limits estimated with the analysis described in
Section~\ref{sec:pid}. 
The performance of the PID is correlated with the momentum and angular
measurements, hence the importance of the choice to perform the
unfolding simultaneously in these three variables, $p$, $\theta$ and
$A$. 

The absorption and decay of particles is simulated by the Monte Carlo.
The generated single particle can re-interact and produce background
particles by the hadronic or electromagnetic processes.
These processes are simulated and can give rise to additional
particles reconstructed in the TPC in the same event.
In such cases also the additional measurements are entered into the
migration matrix.
Thus the complete set of observed effects of a single particle
generated inside the target are taken into account.
Uncertainties in the absorption of secondaries in the
material of and close to the IFC of the TPC are taken into account by
a variation of 10\% of this effect in the simulation. 
The uncertainty in the production of background due to tertiary
particles is larger. 
A 30\% variation of the secondary production was applied.
The value of the variation was estimated from a comparison of the
results for the cross-sections in the energy regime of this experiment
with the predictions of the model used in the simulation.
The uncertainty estimate is reasonable since the secondary
interactions are in majority produced by protons 
with a momentum around 1~\GeVc where one expects hadronic models to be
more reliable than in the energy range of the present measurements. 

A different approach is needed for backgrounds generated by
other secondary particles, such as \piz's produced in hadronic
interactions of the incident beam particle.
The assumption is made that the \piz spectrum is similar to the
spectrum of charged pions.
Initial \pim and \pip spectra are obtained in an analysis without \piz
subtraction. 
The \pim spectra are then used in the MC for the \piz distributions. 
A full simulation of the production and decay into $\gamma$'s with
subsequent conversion in the detector materials is used to predict the
background electron and positron tracks.
Most of these tracks have a momentum below the threshold for this
analysis or low enough to be recognized by \dedx.
The tracks with a PID below the expected value for pions can be
rejected as background.
In the region below 120~\MeVc a large fraction of the electrons can be
unambiguously identified.
These tracks are used as relative normalization between
data and MC.
The remaining background is then estimated from the distributions of
the simulated electron and positron tracks which are accepted as pion
tracks with the same criteria as used to select the data.
These normalized distributions are subtracted from the data
before the unfolding procedure is applied.
Uncertainties in the assumption of the \piz spectrum are taken into
account by an alternative assumption that their spectrum follows the
average of the \pim and \pip distribution.
An additional systematic error of 10\% is assigned to the
normalization of the \piz subtraction using the identified electrons
and positrons.
At low momenta and small angles the \piz subtraction introduces the
largest systematic uncertainty. 
It is in principle possible to reject more electrons and positrons by
constructing a combined PID estimator based on \dedx and TOF.
To obtain a reliable result, the complete \dedx
and $\beta$ distributions need to be described including their
correlations. 
In addition, the measurement of the TOF introduces an inefficiency and
it has tails coming from background hits.
Indeed, such an analysis was performed and gave consistent results.
However, its systematic errors are more difficult to estimate. 

The absolute normalization of the result is calculated in the first
instance relative to the number of incident beam particles accepted by
the selection. 
After unfolding, the factor  $\frac{A}{{N_A \rho t}}$ is applied.
Especially at lower momenta, beam particles may miss the target even
if their trajectory measured in the MWPCs extrapolates to the target.  
The effects of such a `targeting efficiency' were estimated
counting secondaries produced in the forward direction and measured in
the forward spectrometer as a function of impact radius measured from
the centre of the target and found to be smaller than 1\%.
The measured variation in the target thickness is used as an estimate
of an additional uncertainty in the absolute normalization (less than 1\%).
The target thickness uncertainty cancels in the comparison of
data with different incoming beam momenta, while the uncertainty in
the efficiency to hit the target introduces an error into this
comparison. 
The beam normalization using down-scaled incident-proton triggers with the
same beam particle selection introduces for all settings a statistical
uncertainty significantly less than 1\%
\footnote{The statistical error corresponding to down-scaled triggers
  is smaller than the square-root of the number of collected triggers
  because the sampling is not random.}.  
The combination of above mentioned uncertainties are smaller than 2\%
  for all beam momentum settings.

The background due to interactions of the primary
protons outside the target (called `Empty target background') is
measured using data taken without the target mounted in the target
holder. 
Owing to the selection criteria which only accept events from the
target region and the good definition of the interaction point this
background is negligible ($< 10^{-5}$).

The use of a simulation where only one secondary particle is generated
in the target neglects the possible influence of particles on the
measurement of the trajectories of each other.
Owing to the relatively low multiplicity which is spread over a large
solid angle this simplification does not introduce a significant
error. 

The effects of these uncertainties on the final results are estimated
by repeating the analysis with the relevant input modified within the
estimated uncertainty intervals.
In many cases this procedure requires the construction of a set of
different migration matrices.
The correlations of the variations between the cross-section bins are
evaluated and expressed in the covariance matrix.
Each systematic error source is represented by its own covariance
matrix. 
The sum of these matrices describes the total systematic error.

\section{Results}
\label{sec:results}

\begin{figure}[tbp]
\begin{center}
\epsfig{figure=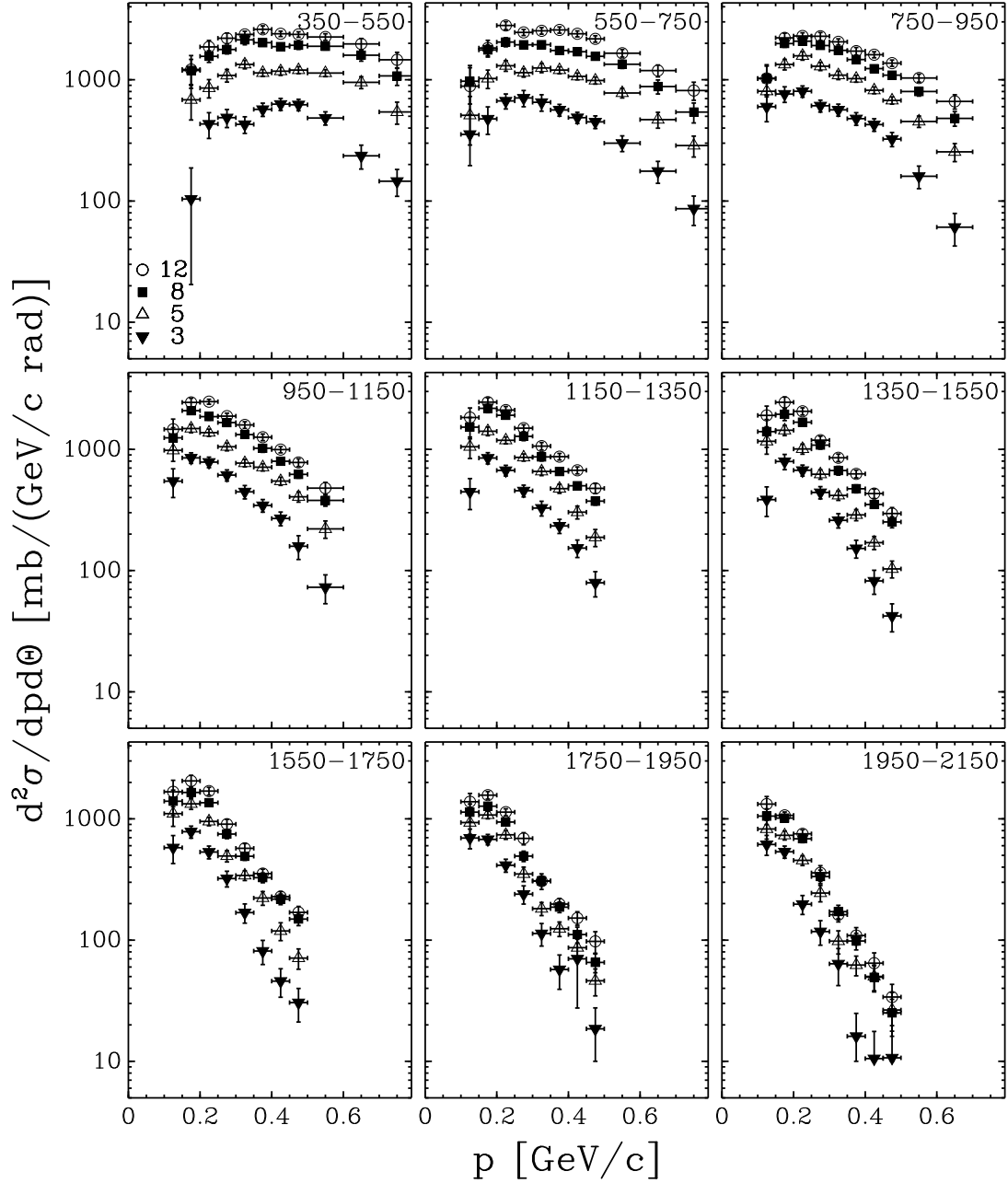,width=0.9\textwidth} 
\caption{
Double-differential cross-sections for \pip production in
p--Ta interactions as a function of momentum displayed in different
angular bins (shown in \mrad in the panels).
The results are given for all incident beam momenta (filled triangles:
3~\GeVc; open triangles: 5~\GeVc; filled rectangles: 8~\GeVc; open
circles: 12~\GeVc). 
The error bars take into account the correlations of the systematic
 uncertainties. 
}
\label{fig:xs-p-th-pbeam-plus}
\end{center}
\end{figure}

\begin{figure}[tbp]
\begin{center}
\epsfig{figure=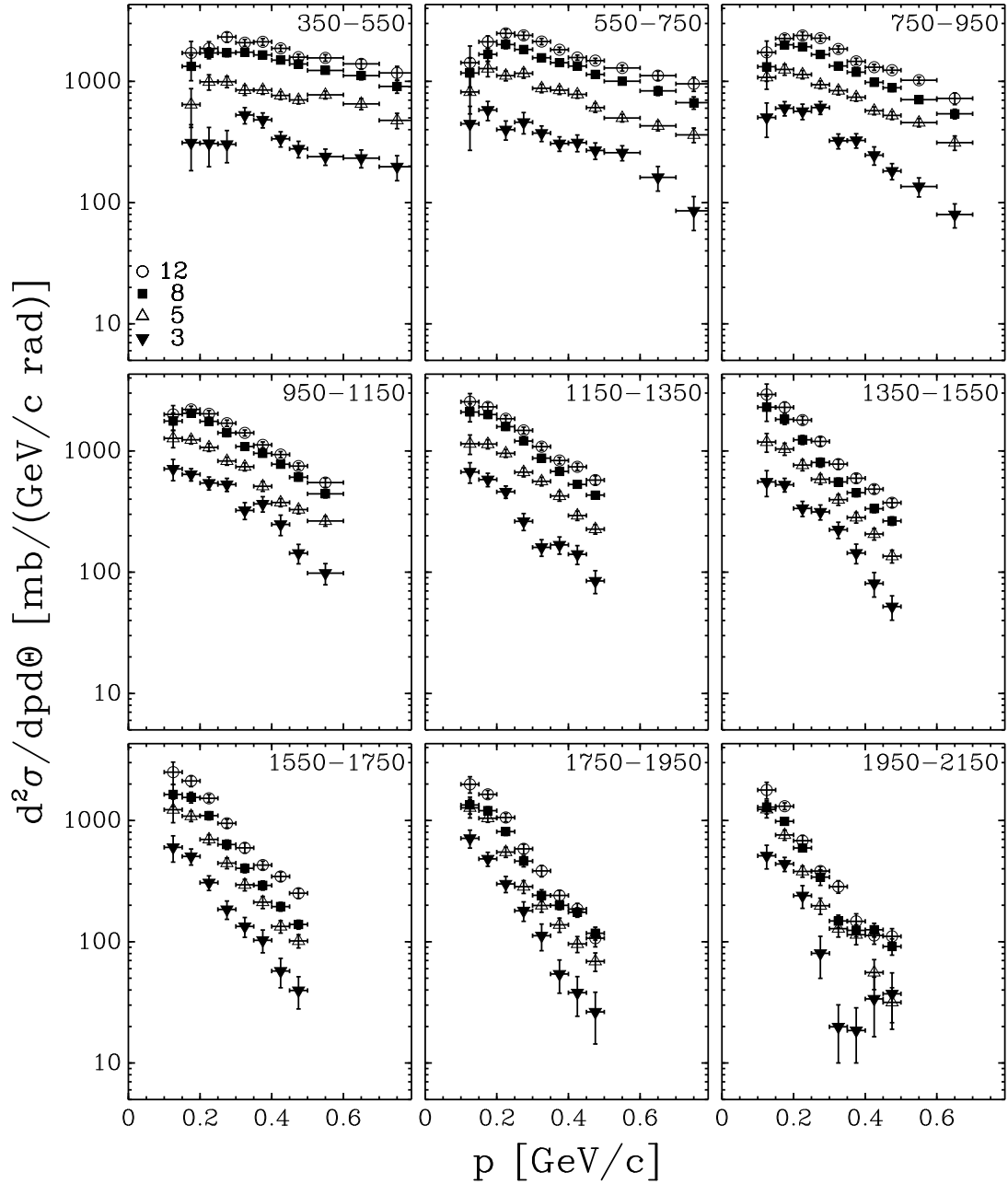,width=0.9\textwidth} 
\caption{
Double-differential cross-sections for \pim production in
p--Ta interactions as a function of momentum displayed in different
angular bins (shown in \mrad in the panels).
The results are given for all incident beam momenta (filled triangles:
3~\GeVc; open triangles: 5~\GeVc; filled rectangles: 8~\GeVc; open
circles: 12~\GeVc).
The error bars take into account the correlations of the systematic
 uncertainties. 
}
\label{fig:xs-p-th-pbeam-minus}
\end{center}
\end{figure}

Figures~\ref{fig:xs-p-th-pbeam-plus} and \ref{fig:xs-p-th-pbeam-minus}
show 
the measurement of the double-differential cross-section for
the production of 
positively (Fig.~\ref{fig:xs-p-th-pbeam-plus}) and negatively
(Fig.~\ref{fig:xs-p-th-pbeam-minus}) charged pions in the laboratory
system as a function of 
the momentum and the polar angle (shown in \mrad in the panels) for each
incident beam momentum.  
The error bars represent the combined statistical and systematic
error.
Correlations cannot be shown in the figures.
The errors shown are the
square-roots of the diagonal elements in the covariance matrix.
Tables with the results of this analysis are also given in Appendix A.
A discussion of the error evaluation is given below. 
The overall scale error (2\%) is not shown.
The measurements for the different beam momenta are overlaid in the
same figure.

\begin{figure}[tbp]
  \epsfig{figure=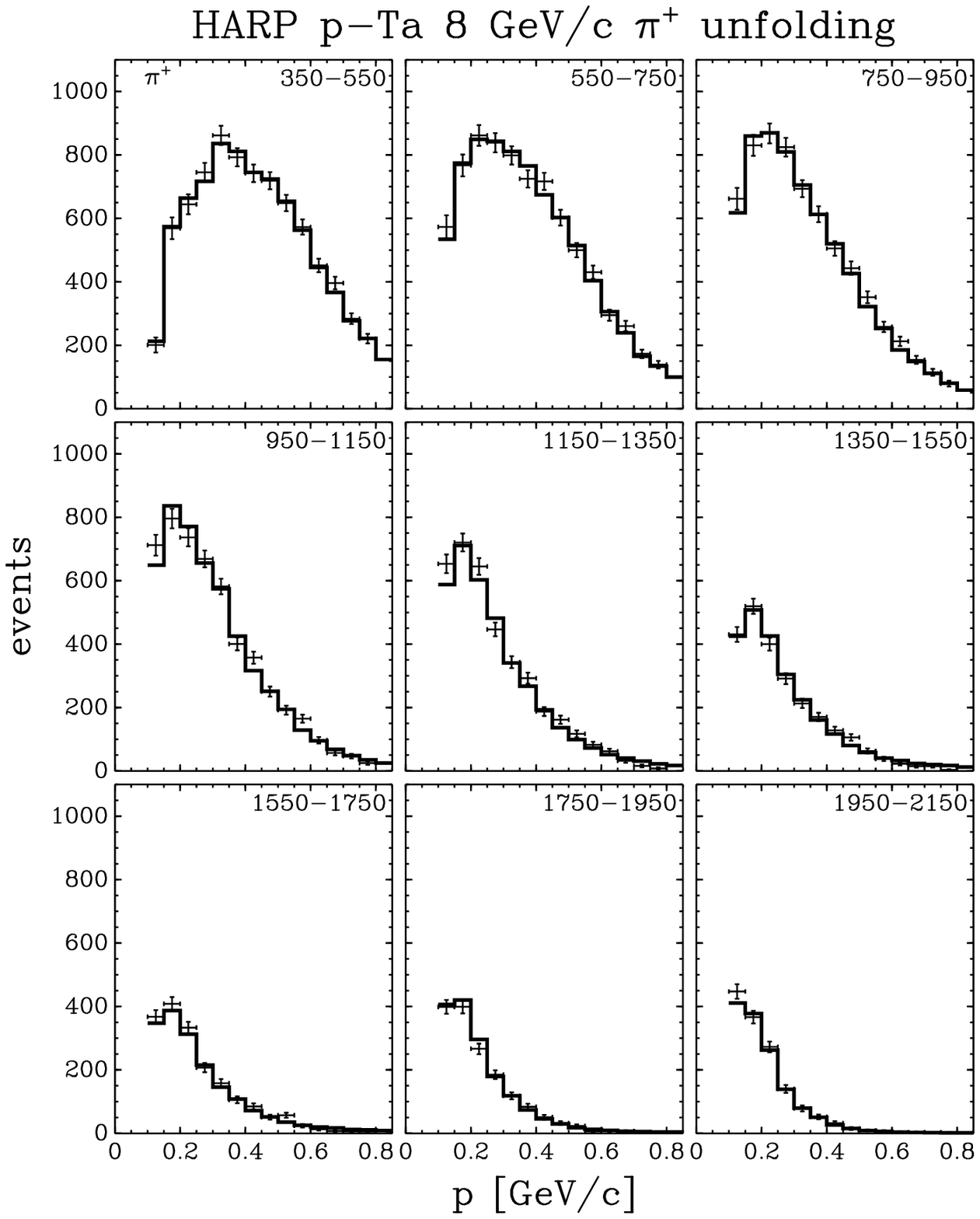,width=0.47\textwidth}
  ~
  \epsfig{figure=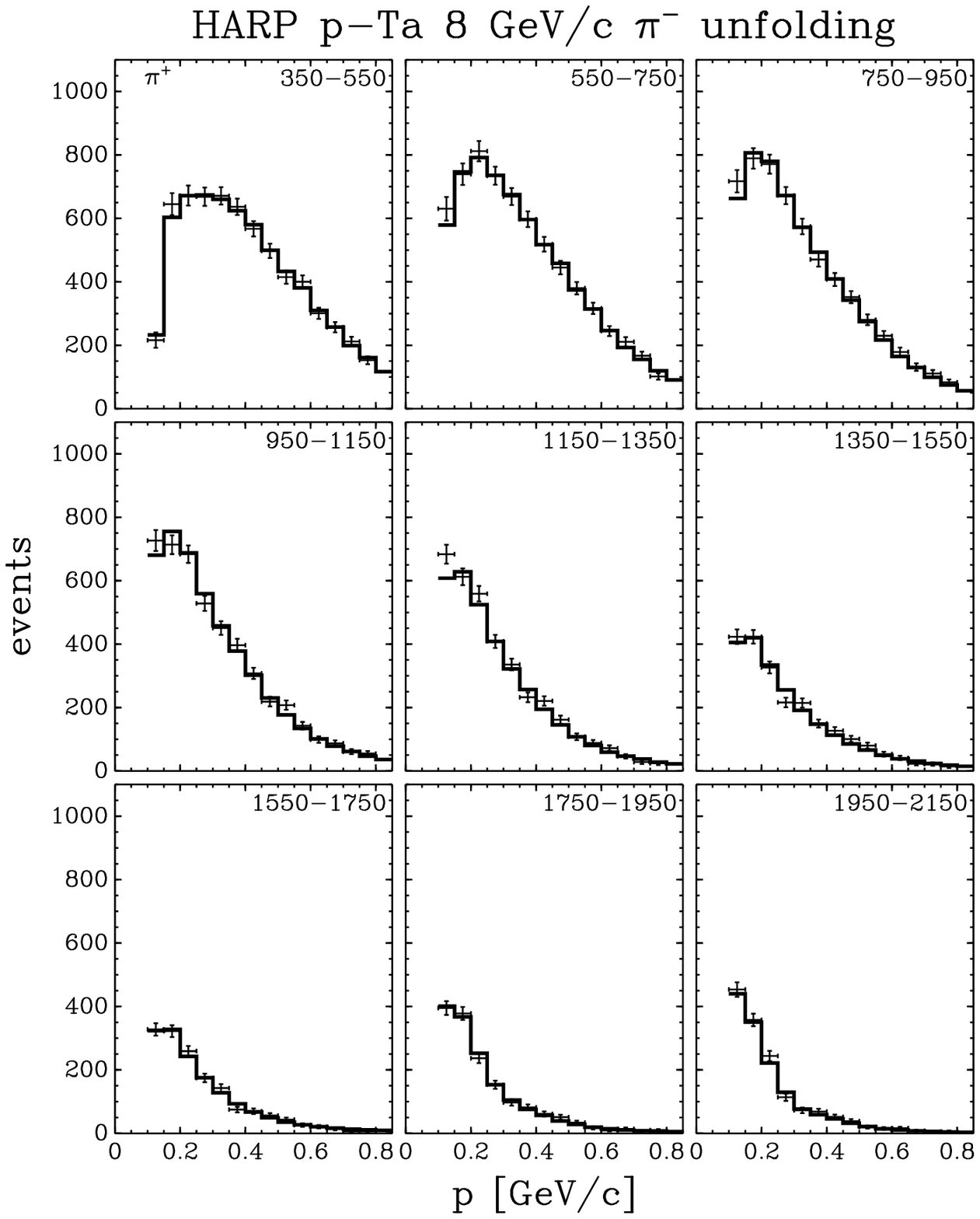,width=0.47\textwidth}
\caption{
Comparison of the \piz subtracted raw data in the 8~\GeVc beam (data points) with the
 prediction in the  measured variables  corresponding to the result of
 the unfolding (histogram).  Left panel: \pip; right panel: \pim. 
 The error bars represent the statistical error of the background
 subtracted data. 
 In the unfolding fit a binning twice as fine as shown here is used
 (both in angle and momentum).  For the sake of clarity these bins are
 summed four-by-four with the appropriate error propagation to obtain
 the spectra as shown here.
}
\label{fig:unfolding}
\end{figure}

The result of the unfolding procedure is the physical cross-section
(represented as a histogram) which provides the best fit to the measured
data, taking into account background, efficiency and resolution.  
The quality of the fit can be judged from Fig.~\ref{fig:unfolding},
where the raw data after \piz subtraction are compared to the
description corresponding to the unfolding result for the 8~\GeVc data.  
This data set has the highest statistics and therefore represents the
most stringent test.

\begin{figure}[tbp]
  \epsfig{figure=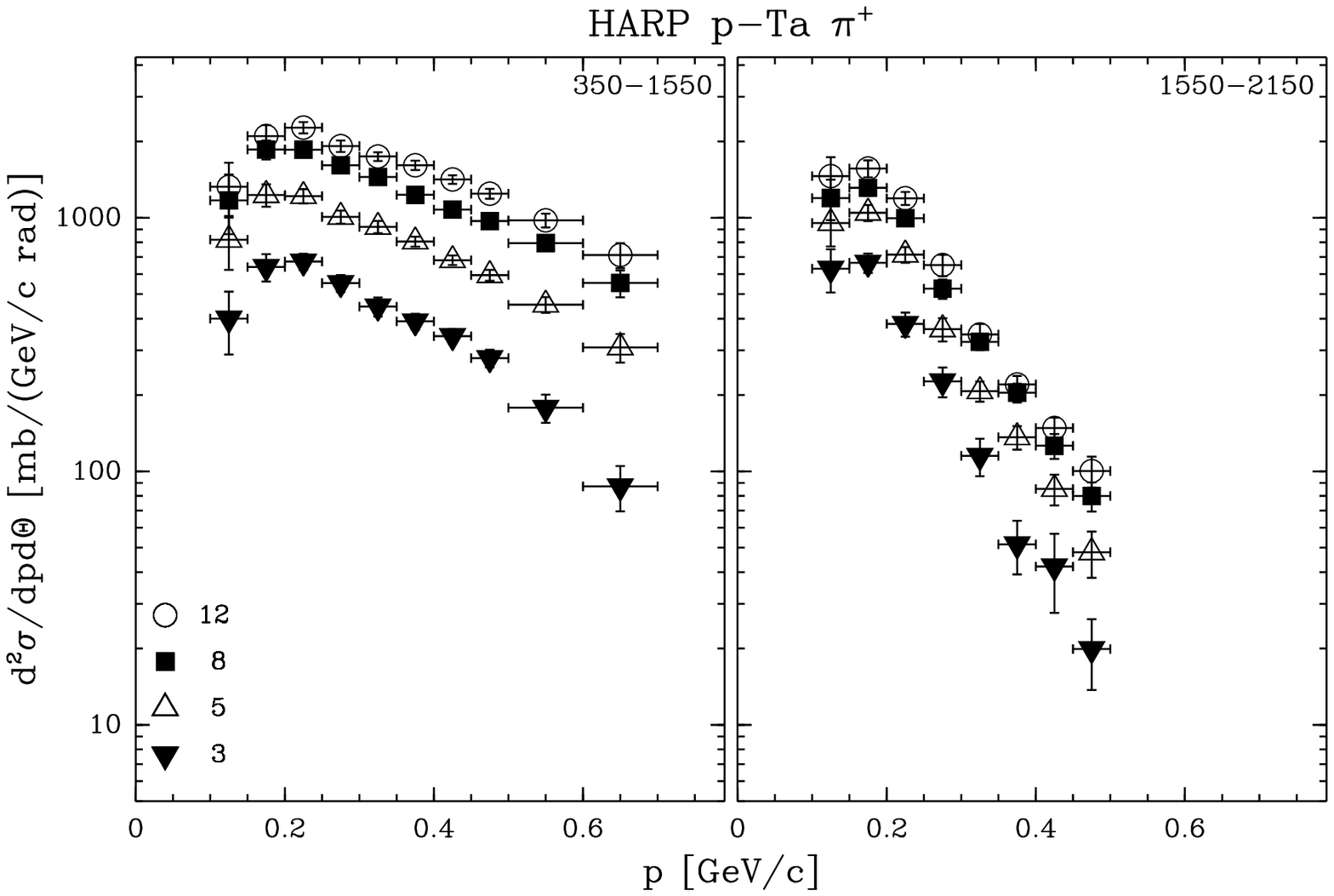,width=0.9\textwidth}
  ~
  \epsfig{figure=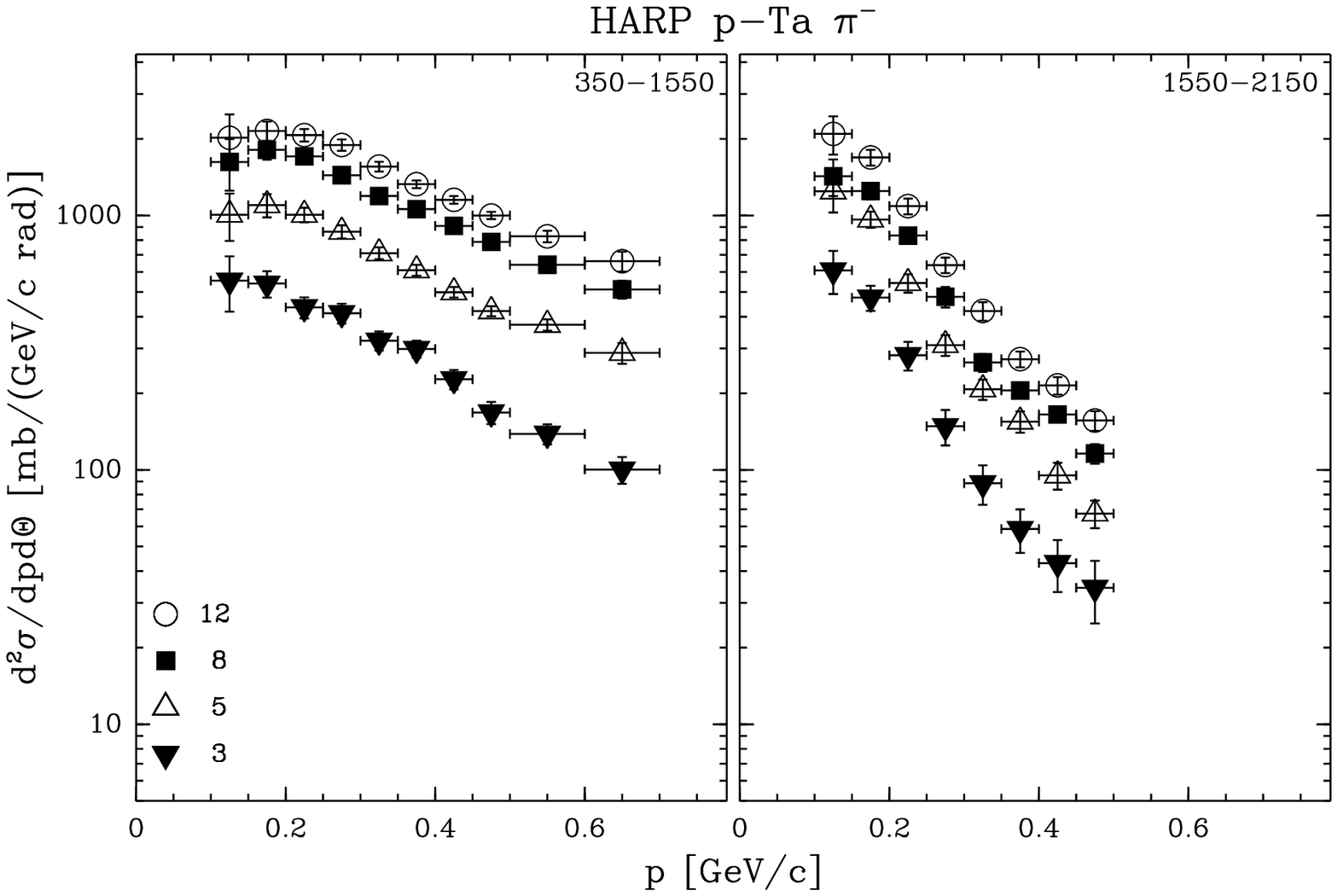,width=0.9\textwidth}
\caption{
Double-differential cross-sections for \pip (top panel) and \pim (bottom
panel) production in 
p--Ta interactions as a function of momentum averaged over the
angular region covered by this experiment (shown in mrad).
Left: forward production (350~\mrad  $\le \theta <$ 1550~\mrad); 
Right: backward production (1550~\mrad  $\le \theta <$ 2150~\mrad).
The results are given for all incident beam momenta (filled triangles:
3~\GeVc; open triangles: 5~\GeVc; filled rectangles: 8~\GeVc; open
circles: 12~\GeVc).
}
\label{fig:xs-p-pbeam}
\end{figure}

To better visualize the dependence on the incoming beam momentum, the
same data integrated over the angular range (separately for the
forward going and backward going tracks) covered by the analysis are shown
separately for \pip and \pim in Fig.~\ref{fig:xs-p-pbeam}.
The spectrum of pions produced in the backward direction falls much
more steeply than that of the pions produced in the forward direction.

\begin{figure}[tbp]
\begin{center}
  \epsfig{figure=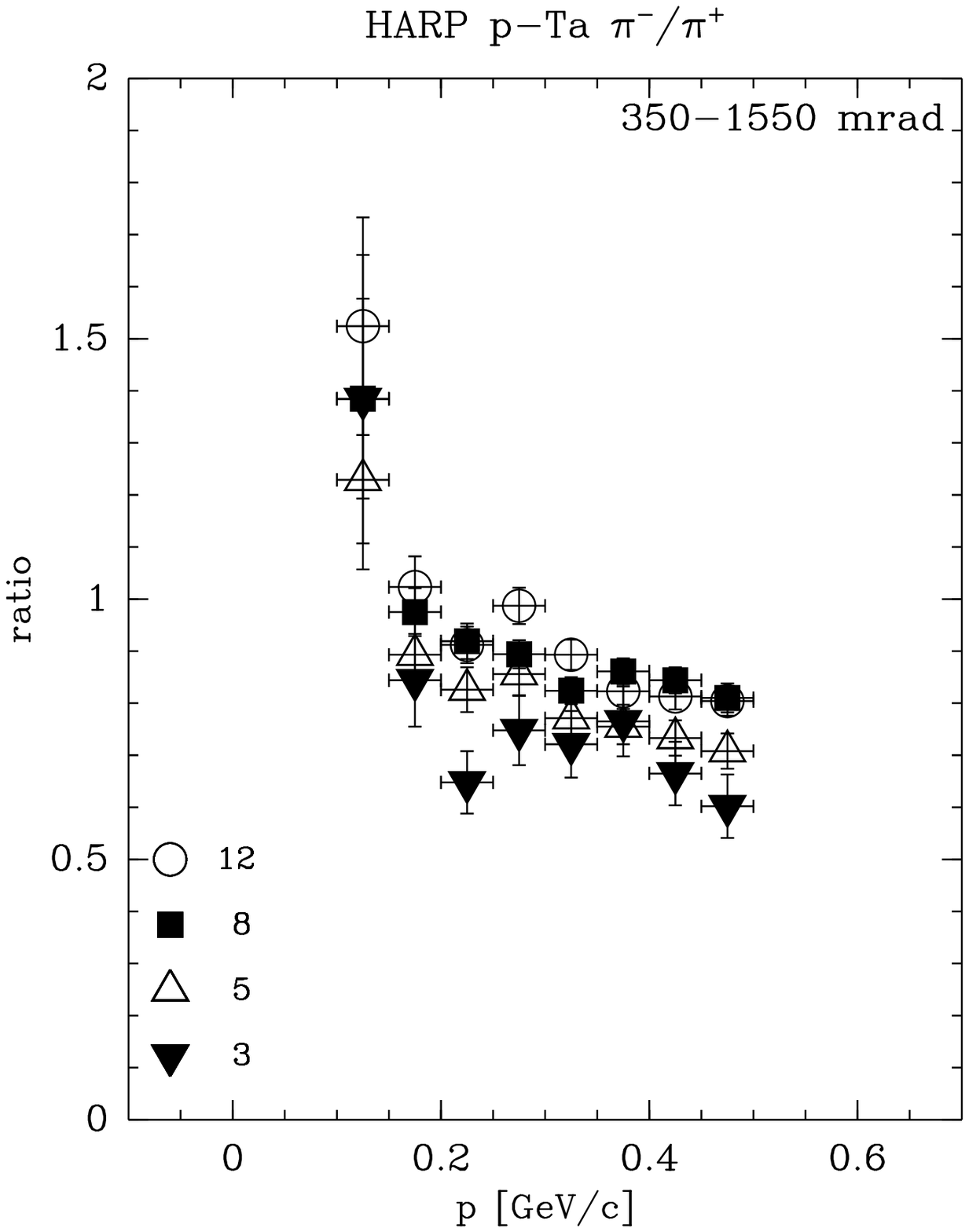,width=0.425\textwidth}
\end{center}
\caption{
The ratio of the differential cross-sections for \pim and \pip
 production in 
p--Ta interactions as a function of momentum integrated over the
forward angular region (shown in mrad).
The results are given for all incident beam momenta (filled triangles:
3~\GeVc; open triangles: 5~\GeVc; filled rectangles: 8~\GeVc; open
circles: 12~\GeVc).
}
\label{fig:xs-ratio}
\end{figure}

The increase of the pion yield per proton with increasing beam momentum
is visible in addition to a change of spectrum.
The spectra of the secondaries produced at small angles are harder
with increasing beam momentum. 
Also an asymmetry between \pip and \pim is observed at relatively
small angles with the beam in favour of a higher \pip rate.
At very large angles from the beam the spectra of \pip and \pim are
more symmetric.
The integrated \pim/\pip ratio in the forward direction is displayed in
Fig.~\ref{fig:xs-ratio} as a function of secondary momentum. 
In the largest part of the momentum range more \pip's are produced than
\pim's, with a smaller \pim/\pip ratio at lower incoming beam momenta.
One observes that the number of \pip's produced 
is smaller than the number of \pim's in the lowest momentum bin
(100~\MeVc--150~\MeVc).
This effect is only significant at the higher incoming beam momenta. 
We find a value of $1.52 \pm 0.21$ and $1.39 \pm 0.19$ for 12~\GeVc and
8~\GeVc, respectively, and 
$1.38 \pm 0.28$ and $1.23 \pm 0.17$ for 3~\GeVc and
5~\GeVc, respectively.
To exclude any detector-related effect one can use the
observation that the electrons and positrons in this momentum range
are predominantly originating from \piz decays and subsequent
$\gamma$ conversions.
Therefore their number and spectrum must be the same. 
It was verified that the ratio $e^+/e^-$ was equal to unity within a
statistical error of 2\%.  
To increase the sensitivity of this cross-check, data taken with other
targets, but within a few days from the tantalum runs reported here,
were also used.
It was also checked that the ratio of the efficiencies for positive
and negative pions predicted by the simulation did not show any
unexpected behaviour.  

The E910 collaboration makes a similar observation for their lowest
momentum bin (100 \MeVc -- 140 \MeVc) in p--Au collisions at 12.3~\GeVc
and 17.5~\GeVc incoming beam momentum and quotes a \pim to \pip yield
ratio 2--3~\cite{ref:E910}. 
They offer an interesting explanation in the form of $\Lambda^0$
production at rest which would enhance the \pim yield at low secondary
momenta.

An alternative analysis of the same data using different techniques is
described in Appendix B.

\subsection{Systematic errors}
\label{sec:syst}

The uncertainties are reported in some detail in
Table~\ref{tab:errors-3}.   
To obtain the entries in this table the double-differential
cross-sections were integrated in nine regions organized as a
three--by--three matrix in angle and momentum. 
(The ninth bin is not populated.) 
The angular ranges are 0.35~\rad~--~0.95~\rad, 0.95~\rad~--~1.55~\rad and
1.55~\rad~--~2.15~\rad, two bins in the forward direction and one 
backward bin.
The momentum ranges are 100~\MeVc~--~300~\MeVc, 300~\MeVc~--~500~\MeVc
and 500~\MeVc~--~700~\MeVc.

One observes that only for the 3~\GeVc beam is the statistical error 
similar in magnitude to the systematic error, while the statistical
error is negligible for the 8~\GeVc and 12~\GeVc beams.
The statistical error is calculated by error propagation as part of the
unfolding procedure. 
It takes into account that the unfolding matrix is obtained from the
data themselves and hence contributes also to the statistical error.
This procedure almost doubles the statistical error, but avoids an important 
systematic error which would otherwise be introduced by assuming a
cross-section model {\em a priori} to calculate the corrections. 

The largest systematic error corresponds to the uncertainty in the
absolute momentum scale, which was estimated to be 3\% using elastic
scattering (Section~\ref{sec:elastics}, Fig.~\ref{fig:el-p-mom-bias}).
It is difficult to better constrain this value, since it depends on
the knowledge of the beam momentum (known to 1\%) and the measurement
of the forward scattering angle in the elastic scattering interaction.
At low momentum in the relatively small angle forward
direction the uncertainty in the subtraction of the electron and
positron background due to \piz production is dominant.
This uncertainty is split between the variation in the shape of the
\piz spectrum and the normalization using the recognized electrons. 
The target region definition (cuts in \dzeroprime and \zzeroprime) and
the uncertainty in the PID efficiency and background from tertiaries
are of similar size and are not negligible.
Relatively small errors are introduced by the uncertainties in
the absorption correction, absolute knowledge of the angular and the
momentum resolution.
The correction for tertiaries (particles produced in secondary
interactions) is relatively large at low momenta and large angles. 
The fact that this region is most affected by this component is to be
expected. 

As already reported above, the overall normalization has an uncertainty
of 2\%, and is not reported in the table.

\begin{table}[tbp!] 
\small{
\begin{center}
\caption{Contributions to the experimental uncertainties. The numbers
  represent the uncertainty in percent of 
  the cross-section integrated over the angle and momentum region indicated.} 
\label{tab:errors-3}
\vspace{2mm}
\begin{tabular}{ l rrr | rrr | rr} \hline
\bf{Momentum range  (\MeVc)}&\multicolumn{3}{c|}{100 -- 300}
                            &\multicolumn{3}{c|}{300 -- 500}
                            &\multicolumn{2}{c}{500 -- 700} \\
\hline
\bf{Angle range (\rad)}&0.35--&0.95--&1.55--
                       &0.35--&0.95--&1.55--
                       &0.35--&0.95-- \\
\bf{Error source}      &0.95&1.55&2.15
                       &0.95&1.55&2.15
                       &0.95&1.55 \\
\hline
\bf{3 \GeVc beam}&&&&&&&&\\
\hline
Absorption               &  1.3 &  1.8 &  2.4 &  0.6 &  0.5 &  0.3 &  0.3 &  0.5 \\
Tertiaries               &  3.1 &  4.4 &  5.0 &  2.5 &  2.9 &  1.8 &  0.1 &  0.6 \\
Target region cut        &  3.2 &  1.0 &  1.1 &  2.1 &  0.6 &  2.8 &  0.8 &  2.0 \\
Efficiency               &  1.7 &  1.9 &  1.3 &  2.1 &  2.8 &  2.2 &  2.6 &  2.8 \\
Shape of $\pi^0$         &  8.6 &  1.9 &  0.0 &  0.1 &  0.0 &  0.1 &  0.0 &  0.0 \\
Normalization of $\pi^0$ &  5.5 &  1.9 &  0.9 &  0.2 &  0.1 &  0.0 &  0.0 &  0.0 \\
Particle ID              &  0.1 &  0.1 &  0.0 &  1.1 &  0.5 &  0.0 &  5.5 &  3.5 \\
Momentum resolution      &  2.7 &  1.5 &  1.6 &  0.5 &  0.1 &  0.6 &  0.4 &  0.3 \\
Momentum scale           &  7.0 &  4.4 &  3.6 &  1.2 &  4.0 &  4.4 &  7.0 & 11.3 \\
Angle bias               &  1.5 &  0.8 &  0.4 &  0.2 &  1.3 &  0.9 &  1.0 &  1.5 \\
\bf{Total systematics}   & 13.7 &  7.5 &  7.1 &  4.2 &  5.9 &  6.0 &  9.4 & 12.5 \\
\bf{Statistics}          &  5.0 &  3.9 &  4.9 &  3.9 &  5.3 & 10.6 &  5.4 & 10.2 \\
\hline
\bf{5 \GeVc beam}&&&&&&&&\\
\hline
Absorption               &  1.2 &  1.9 &  2.4 &  0.6 &  0.5 &  0.4 &  0.3 &  0.4 \\
Tertiaries               &  3.0 &  4.4 &  5.0 &  2.6 &  2.8 &  2.0 &  0.2 &  0.1 \\
Target region cut        &  2.8 &  0.9 &  1.2 &  1.0 &  0.6 &  0.6 &  0.1 &  0.2 \\
Efficiency               &  1.7 &  2.2 &  1.5 &  1.6 &  2.4 &  2.3 &  1.7 &  2.9 \\
Shape of $\pi^0$         &  6.9 &  0.6 &  0.6 &  0.2 &  0.1 &  0.1 &  0.0 &  0.0 \\
Normalization of $\pi^0$ &  6.0 &  2.1 &  1.0 &  0.2 &  0.1 &  0.0 &  0.0 &  0.0 \\
Particle ID              &  0.1 &  0.1 &  0.0 &  1.0 &  0.5 &  0.1 &  5.0 &  4.0 \\
Momentum resolution      &  2.3 &  1.9 &  1.8 &  0.1 &  0.6 &  0.9 &  0.3 &  0.9 \\
Momentum scale           &  6.1 &  4.8 &  4.1 &  1.3 &  2.6 &  5.5 &  4.2 &  9.7 \\
Angle bias               &  0.9 &  0.7 &  0.4 &  0.2 &  1.2 &  0.6 &  0.8 &  2.0 \\
\bf{Total systematics}   & 12.2 &  7.7 &  7.5 &  3.6 &  4.8 &  6.4 &  6.8 & 11.1 \\
\bf{Statistics        }  &  2.7 &  2.2 &  2.7 &  2.0 &  2.7 &  4.9 &  2.4 &  4.2 \\
\hline
\bf{8 \GeVc beam}&&&&&&&&\\
\hline
Absorption               &  1.3 &  1.9 &  2.2 &  0.6 &  0.6 &  0.4 &  0.3 &  0.5 \\
Tertiaries               &  2.5 &  4.4 &  4.2 &  2.6 &  3.3 &  2.1 &  0.2 &  0.3 \\
Target region cut        &  3.1 &  2.2 &  1.2 &  2.3 &  0.4 &  1.1 &  1.6 &  0.6 \\
Efficiency               &  1.4 &  1.9 &  1.4 &  1.3 &  2.1 &  1.9 &  1.5 &  2.5 \\
Shape of $\pi^0$         &  4.0 &  0.3 &  0.4 &  0.1 &  0.1 &  0.1 &  0.0 &  0.0 \\
Normalization of $\pi^0$ &  6.3 &  2.0 &  1.1 &  0.2 &  0.1 &  0.0 &  0.0 &  0.0 \\
Particle ID              &  0.1 &  0.1 &  0.0 &  1.3 &  0.6 &  0.2 &  5.5 &  3.6 \\
Momentum resolution      &  2.3 &  2.2 &  1.9 &  0.0 &  0.2 &  0.4 &  0.0 &  0.2 \\
Momentum scale           &  6.4 &  5.2 &  4.4 &  1.3 &  2.0 &  4.5 &  3.9 &  9.3 \\
Angle bias               &  0.7 &  0.6 &  0.3 &  0.4 &  1.2 &  0.7 &  1.0 &  1.1 \\
\bf{Total systematics}   & 11.1 &  8.3 &  7.1 &  4.2 &  4.6 &  5.5 &  7.2 & 10.4 \\
\bf{Statistics}          &  1.6 &  1.4 &  1.8 &  1.2 &  1.6 &  2.9 &  1.4 &  2.3 \\
\hline
\bf{12 \GeVc beam}&&&&&&&&\\
\hline
Absorption               &  1.1 &  1.8 &  2.1 &  0.5 &  0.5 &  0.5 &  0.3 &  0.3 \\
Tertiaries               &  0.8 &  3.5 &  4.3 &  0.8 &  2.5 &  1.8 &  1.3 &  0.0 \\
Target region cut        &  3.8 &  2.3 &  1.0 &  2.3 &  0.6 &  0.2 &  1.1 &  0.2 \\
Efficiency               &  1.6 &  2.3 &  2.5 &  1.1 &  2.4 &  2.3 &  1.2 &  2.4 \\
Shape of $\pi^0$         &  4.4 &  0.5 &  1.2 &  0.1 &  0.1 &  0.1 &  0.0 &  0.0 \\
Normalization of $\pi^0$ &  6.5 &  2.2 &  1.1 &  0.4 &  0.1 &  0.1 &  0.1 &  0.0 \\
Particle ID              &  0.0 &  0.0 &  0.0 &  1.1 &  0.5 &  0.0 &  5.3 &  3.8 \\
Momentum resolution      &  2.2 &  2.3 &  2.4 &  0.2 &  1.0 &  0.4 &  0.1 &  0.8 \\
Momentum scale           &  7.3 &  5.2 &  4.7 &  1.1 &  1.7 &  4.9 &  3.7 & 10.0 \\
Angle bias               &  0.5 &  0.6 &  0.1 &  0.5 &  1.1 &  0.7 &  0.9 &  1.8 \\
\bf{Total systematics}   & 11.8 &  8.0 &  7.8 &  3.2 &  4.2 &  5.8 &  6.8 & 11.1 \\
\bf{Statistics}          &  1.7 &  1.6 &  2.0 &  1.3 &  1.8 &  3.3 &  1.5 &  2.6 \\
\end{tabular}
\end{center}
}
\end{table}

\section{Comparisons with earlier data}
\label{sec:compare}

\begin{figure}[tbp]
\epsfig{figure=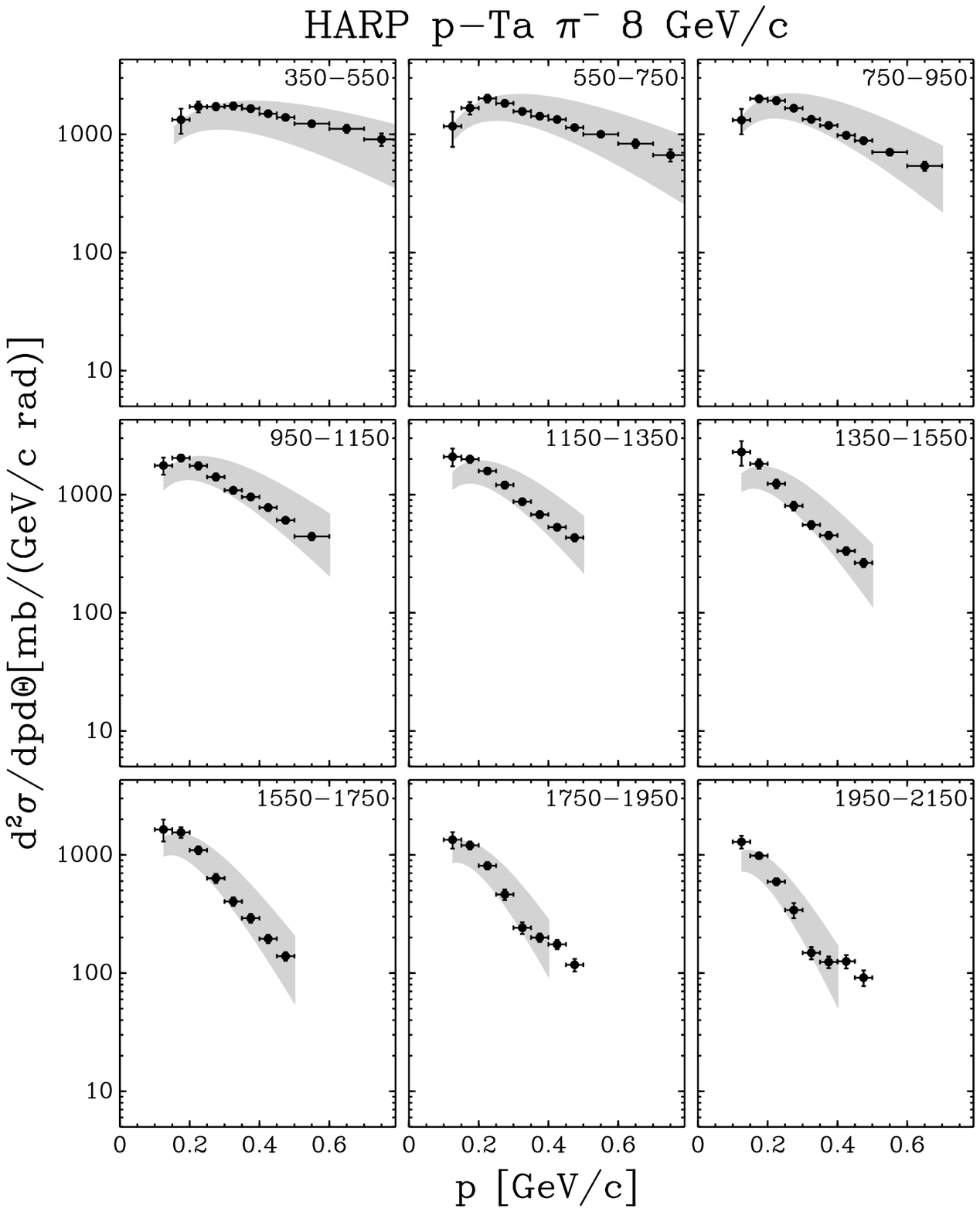,width=0.48\textwidth}
~
\epsfig{figure=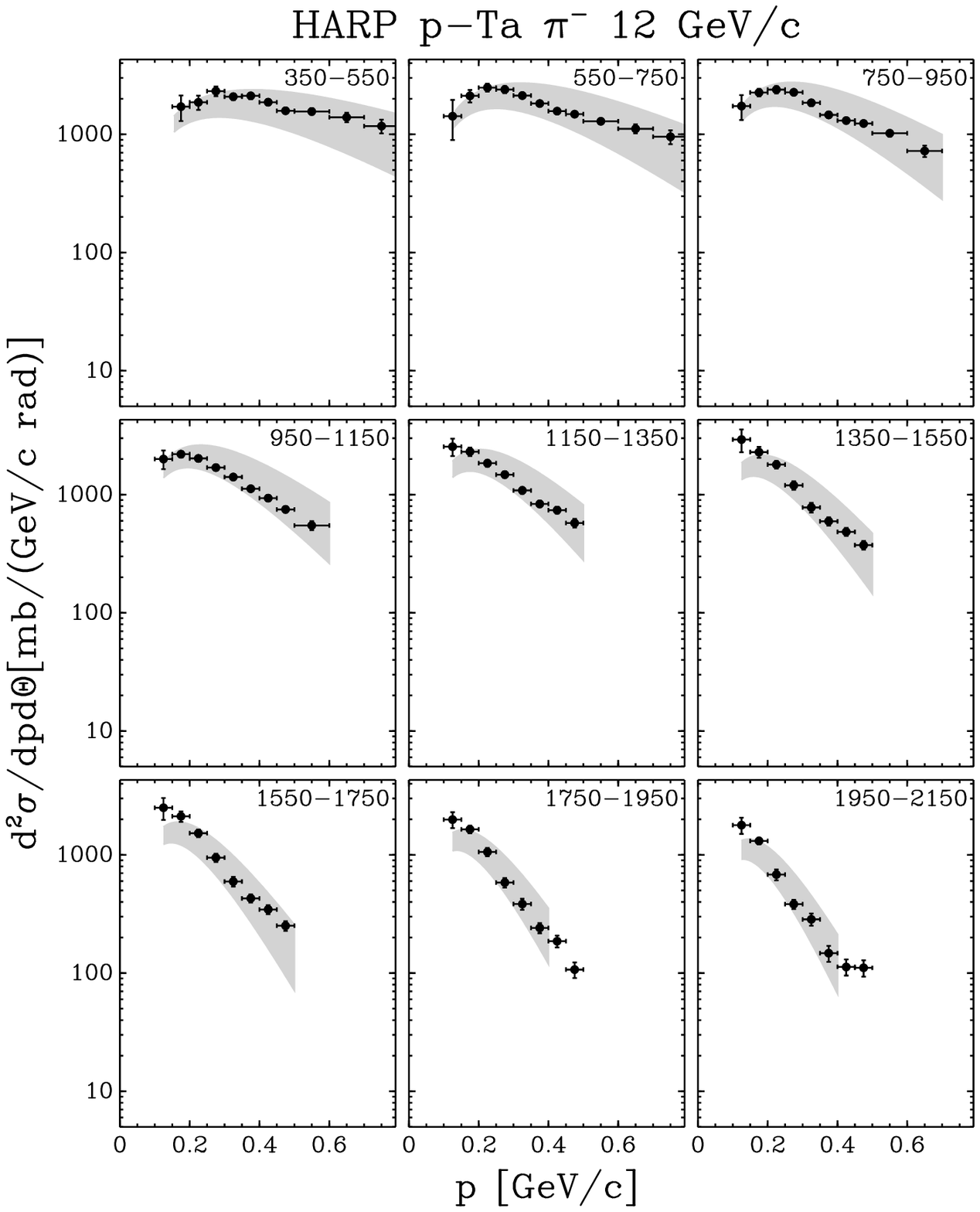,width=0.48\textwidth}
\caption{
Comparison of the HARP data with data from
Ref.~\cite{ref:armutliiski}.
The left panel shows the comparison of the parametrization of
the 10~\GeVc data of Ref.~\cite{ref:armutliiski} with the 8~\GeVc data
reported here; the right panel shows the comparison with the 12~\GeVc
data. 
The absolute normalization of the parametrization was fixed to the
data in both cases.
The band shows the range allowed by varying the slope parameters given
by \cite{ref:armutliiski} with two standard deviation and a 10\% variation
on the absolute scale.
}
\label{fig:compare}
\end{figure}

Very few p--Ta pion production data are available in the literature. 
Our data can only be compared with results from Ref.~\cite{ref:armutliiski}
where measurements of 
\pim production are reported in 10~\GeVc p--Ta interactions. 
The total number of \pim observed in the above reference is about
2600. 
No relevant \pip production data were found in the literature.
In the paper cited above no table of the double differential
cross-sections was provided, the measurements being given in
parametrized and graphical form only.
The authors of Ref.~\cite{ref:armutliiski} give the results as a
simple exponential in the invariant cross-section:
$\frac{E}{A} {\frac{{\mathrm{d}^3 \sigma}}{{\mathrm{d}p^3 }}}$,
where $E$ and $p$ are the energy and momentum of the produced
particle, respectively, and $A$ the atomic number of the target
nucleus. 
They parametrize their spectra in each angular bin with a function of
the form 
$f_{\pi^-}= c \ \exp{(-T/T_0)}$,
where $T$ is the kinetic energy of the produced particle and $T_0$ is
given by $T_0=T'/(1-\beta \ \cos{\theta})$.
The values of the parameters are 
$T'=(0.086\pm0.006) \ \GeVc$ and $\beta=0.78\pm0.03$.
Unfortunately, no absolute normalization is given numerically.
To provide a comparison with these data, the parametrization was
integrated over the angular bins used in our analysis and with an
arbitrary overall normalization overlaid over our 8~\GeVc and 12~\GeVc
results. 
The results of this comparison are shown in Fig.~\ref{fig:compare}.
The shaded band gives the excursion of the parametrization due to the
error in the slope parameters ($\pm 2\sigma$) with an additional assumed
10\% error on the absolute scale.
The latter additional error takes into account the fact that 
the errors on the slopes fitted to the individual angular
bins in the cited data are at least a factor of two larger than in the
exponential slope obtained from their global parametrization.
The agreement of our data with the simple parametrization is good.
Since the comparison is of similar quality for the two incoming beam
momenta, the lack of data with an exactly equal beam momentum does not
play a role.
To judge the comparison, one should keep in mind that the
statistics of Ref.~\cite{ref:armutliiski} is much smaller (2600 \pim)
than the statistics of the \pim samples in our 8~\GeVc and 12~\GeVc
data (38,000 and 29,000 \pim, respectively).  
The bands in the figure extend over the region where there is data
available from  Ref.~\cite{ref:armutliiski}.

\section{Implications for neutrino factory designs}
\label{sec:factory}

\begin{figure}[tbp]
\begin{center}
\epsfig{figure=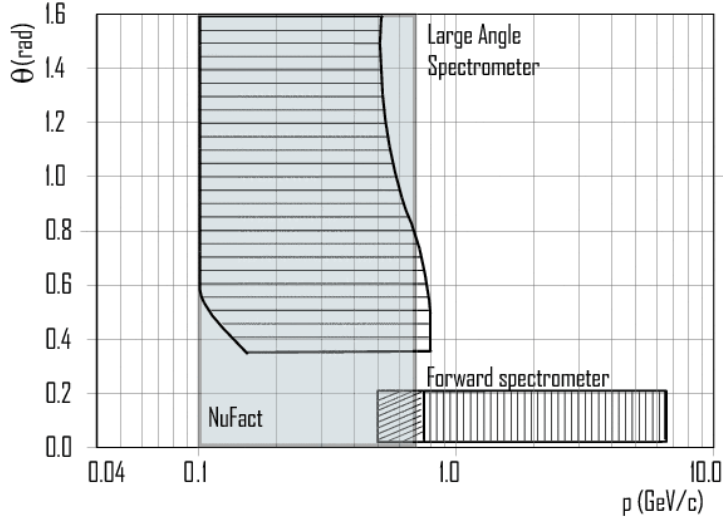,,bbllx=11,bblly=16,bburx=587,bbury=425,width=0.6\textwidth} 
\end{center}
\caption{
Kinematic region in the $p$--\tht plane covered by this analysis
compared to the maximum acceptance of an input stage of typical neutrino
factory designs.
The different neutrino factory designs have in addition to the limits
 shown different limits in \pt.
}
\label{fig:nufact-coverage}
\end{figure}

The data presented in this paper are particularly relevant for the
design of the input stage of future neutrino factories.
In addition, they will be valuable in validating and possibly improving
hadronic production models in a kinematic region where data are scarce. 
The kinematic coverage of the experiment is compared with the typical
range of the kinematical acceptance of neutrino factory designs in
Fig.~\ref{fig:nufact-coverage}.
It is shown that this experiment covers the full momentum range of
interest for production angles above 0.35~\rad. 
A small part of the small angle region can in principle be covered by
measurements with the  HARP forward spectrometer.
The analysis of the p--Ta data in the forward direction is in
progress. 
The analysis reported here covers the major part of pions produced in
the target and accepted by the focusing system of the input stage of a
neutrino factory. 
The importance of the knowledge of the smaller angles varies with the
different types of design being contemplated.
The effective coverage of the kinematic range can be defined as the
fraction of the number of muons transported by the input stage of a
neutrino factory design originating from decays for which the pion
production cross-section is within the kinematic range measured by the
present experiment. 
As an example, this effective coverage was evaluated for the ISS input
stage~\cite{ref:iss} to be 69\% for \pip and 72\% for \pim,
respectively~\cite{ref:fernow}, using a particular model for pion
production at an incoming beam momentum of 10.9~\GeVc~\cite{ref:brooks}. 

There are a number of options to obtain pion production rates for the
angular range below 0.35~\rad.
One option is to adjust hadron production models to the available data
and to use the extrapolation of these models in the unmeasured
region. 
Such tuning of models can also profit from the additional data
provided with the forward spectrometer.
In principle, the combination of the particle tracking in the large
angle and forward spectrometer can be developed and the region can be
extended towards angles near to the beam direction.
In that case the limits are given by the requirement $\pt > 50 \
\MeVc$ and by the minimum angle to remove through-going beam particles
($\approx 30~\mrad$).

\begin{figure}[tbp]
  \begin{center}
  \epsfig{figure=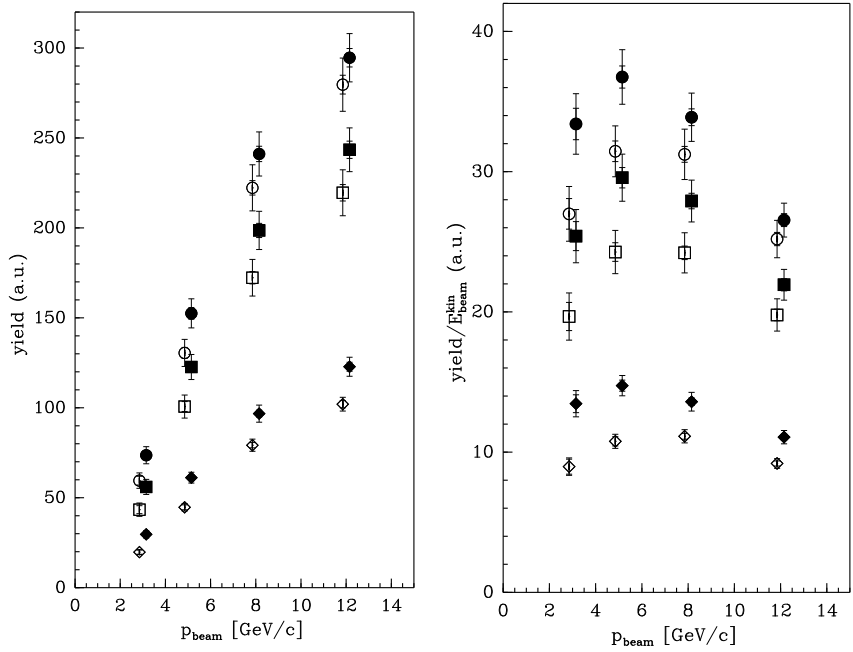,width=0.7\textwidth} 
  \end{center}
\caption{
Prediction of the \pip (closed symbols) and \pim (open symbols) yield
as a function of incident proton 
beam momentum for different designs of the neutrino factory focusing
stage. 
Shown are the integrated yields (left) and the integrated yields
normalized to the kinetic energy of the proton (right). 
The circles indicate the integral over the full HARP acceptance, the
squares are integrated over $0.35 \ \rad <\theta< 0.95 \ \rad$, while the diamonds
are calculated for the restricted angular range and
$250~\MeVc<p<500~\MeVc$.
The full error bar shows the overall (systematic and statistical)
error, while the inner error bar shows the error relevant for the
point--to-point comparison.  
For the latter error only the uncorrelated systematic uncertainties
were added to the statistical error.
}
\label{fig:nufact-yield}
\end{figure}

As an indication of the overall pion yield as a function of incoming
beam momentum, the \pip and \pim production cross-sections were
integrated over the full HARP kinematic range in the forward hemisphere
($100~\MeVc<p<700~\MeVc$ and $0.35 <\theta< 1.55$).
The results are shown in Fig.~\ref{fig:nufact-yield}\footnote{
Although the units are indicated as ``arbitrary'',
for the largest region, the yield is expressed as 
${{\mathrm{d}^2 \sigma}}/{{\mathrm{d}p\mathrm{d}\Omega }}$ in
mb/(\GeVc~sr). 
For the
other regions the same normalization is chosen, but now scaled with the
relative bin size to show visually the correct ratio of number of pions
produced in these kinematic regions. 
}.
The integrated yields are shown in the left panel and the integrated
yields normalized to the kinetic energy of the incoming beam particles
are shown in the right panel. 
The outer error bars indicate the total statistical and systematic errors.
If one compares the \pip and \pim rates for a given beam momentum or
if one compares the rates at a different beam momentum the relative
systematic error is reduced by about a factor two.
The relative uncertainties are shown as inner error bar.
It is shown that the pion yield increases with
momentum and that in our kinematic coverage the optimum yield is
between 5~\GeVc and 8~\GeVc.
However, these calculations should be completed with more realistic
kinematical cuts in the integration.
To show the trend the rates within restricted ranges are also given: a
restricted angular range ($0.35 <\theta< 0.95$)  and a range further
restricted in momentum ($250~\MeVc<p<500~\MeVc$).
The latter range may be most representative for the neutrino factory.

Of course this analysis only gives a simplified picture of the results.
One should note that the best result can be obtained by using the
full information of the double-differential cross-section and
by developing designs optimized specifically for each single beam
momentum. 
Then these optimized designs can be compared.

\section{Summary and Conclusions}
\label{sec:summary}

In this paper an analysis of the production of pions at
large angles with respect to the beam direction for protons of
3~\GeVc, 5~\GeVc, 8~\GeVc and 12~\GeVc impinging on a thin
(5\% $\lambda_{\mathrm{I}}$) tantalum target was described. 
The secondary pion yield was measured in a large angular and momentum
range and double-differential cross-sections were obtained.
A detailed error estimation has been discussed.
 
The use of a single detector for a range of beam momenta makes it
possible to measure the dependence of the pion yield on the beam
momentum with high precision.
These data can be used to make predictions for the fluxes of pions to
enable an optimized design of a future neutrino factory.

Very few pion production measurements in this energy range are reported
in the literature.
The only comparable data found in the literature agrees with the
results described in this paper.
Hadronic production models describing this energy range can now be
compared with the new results and, if needed, improved.
Data have been taken with different target materials (Be, C, Al, Cu, Sn
and Pb) for the beam momenta 3~\GeVc, 5~\GeVc, 8~\GeVc, 12~\GeVc and
15~\GeVc and will be presented in subsequent papers.
In particular, the data taken with a lead target will add valuable
information to the studies for the neutrino factory.
Also data with thick (one $\lambda_{\mathrm{I}}$) Ta and Pb targets have been
taken which would help modelling the neutrino factory yields.

\section{Acknowledgements}

We gratefully acknowledge the help and support of the PS beam staff
and of the numerous technical collaborators who contributed to the
detector design, construction, commissioning and operation.  
In particular, we would like to thank
G.~Barichello,
R.~Brocard,
K.~Burin,
V.~Carassiti,
F.~Chignoli,
D.~Conventi,
G.~Decreuse,
M.~Delattre,
C.~Detraz,  
A.~Domeniconi,
M.~Dwuznik,   
F.~Evangelisti,
B.~Friend,
A.~Iaciofano,
I.~Krasin, 
D.~Lacroix,
J.-C.~Legrand,
M.~Lobello, 
M.~Lollo,
J.~Loquet,
F.~Marinilli,
J.~Mulon,
L.~Musa,
R.~Nicholson,
A.~Pepato,
P.~Petev, 
X.~Pons,
I.~Rusinov,
M.~Scandurra,
E.~Usenko,
and
R.~van der Vlugt,
for their support in the construction of the detector.
The collaboration acknowledges the major contributions and advice of
M.~Baldo-Ceolin, 
L.~Linssen, 
M.T.~Muciaccia and A. Pullia
during the construction of the experiment.
The collaboration is indebted to 
V.~Ableev,
F.~Bergsma,
P.~Binko,
E.~Boter,
M.~Calvi, 
C.~Cavion, 
A.~Chukanov,  
M.~Doucet,
D.~D\"{u}llmann,
V.~Ermilova, 
W.~Flegel,
Y.~Hayato,
A.~Ichikawa,
A.~Ivanchenko,
O.~Klimov,
T.~Kobayashi,
D.~Kustov,
M.~Laveder,  
M.~Mass,
H.~Meinhard,
T.~Nakaya,
K.~Nishikawa,
M.~Pasquali,
M.~Placentino,
S.~Simone,
S.~Troquereau,
S.~Ueda and A.~Valassi
for their contributions to the experiment.
We would like to thank S.~Brooks, R.C.~Fernow and J.~Gallardo for their
help in evaluating the effective kinematic coverage of our data for the
neutrino factory input stage.

We acknowledge the contributions of 
V.~Ammosov,
G.~Chelkov,
D.~Dedovich,
F.~Dydak,
M.~Gostkin,
A.~Guskov, 
D.~Khartchenko, 
V.~Koreshev,
Z.~Kroumchtein,
I.~Nefedov,
A.~Semak, 
J.~Wotschack,
V.~Zaets and
A.~Zhemchugov
to the work described in this paper.

 The experiment was made possible by grants from
the Institut Interuniversitaire des Sciences Nucl\'eair\-es and the
Interuniversitair Instituut voor Kernwetenschappen (Belgium), 
Ministerio de Educacion y Ciencia, Grant FPA2003-06921-c02-02 and
Generalitat Valenciana, grant GV00-054-1,
CERN (Geneva, Switzerland), 
the German Bundesministerium f\"ur Bildung und Forschung (Germany), 
the Istituto Na\-zio\-na\-le di Fisica Nucleare (Italy), 
INR RAS (Moscow) and the Particle Physics and Astronomy Research Council (UK).
We gratefully acknowledge their support.
This work was supported in part by the Swiss National Science Foundation
and the Swiss Agency for Development and Cooperation in the framework of
the programme SCOPES - Scientific co-operation between Eastern Europe
and Switzerland.

\clearpage

\begin{appendix}

\section{Cross-section data}\label{app:data}
\begin{table}[hp!] 
\begin{center}
  \caption{\label{tab:xsec-p}
    HARP results for the double-differential $\pi^+$ production
    cross-section in the laboratory system,
    $d^2\sigma^{\pi^+}/(dpd\theta)$. Each row refers to a
    different $(p_{\hbox{\small min}} \le p<p_{\hbox{\small max}},
    \theta_{\hbox{\small min}} \le \theta<\theta_{\hbox{\small max}})$ bin,
    where $p$ and $\theta$ are the pion momentum and polar angle, respectively.
    The central value as well as the square-root of the diagonal elements
    of the covariance matrix are given.}
\vspace{2mm}
\begin{tabular}{rrrr|r@{$\pm$}lr@{$\pm$}lr@{$\pm$}lr@{$\pm$}l} 
\hline
$\theta_{\hbox{\small min}}$ &
$\theta_{\hbox{\small max}}$ &
$p_{\hbox{\small min}}$ &
$p_{\hbox{\small max}}$ &
\multicolumn{8}{c}{$d^2\sigma^{\pi^+}/(dpd\theta)$} 
\\
(rad) & (rad) & (\GeVc) & (\GeVc) &
\multicolumn{8}{c}{(barn/(\GeVc rad))}
\\
  &  &  & 
&\multicolumn{2}{c}{$ \bf{3 \ \GeVc}$} 
&\multicolumn{2}{c}{$ \bf{5 \ \GeVc}$} 
&\multicolumn{2}{c}{$ \bf{8 \ \GeVc}$} 
&\multicolumn{2}{c}{$ \bf{12 \ \GeVc}$} 
\\ 
\hline
 0.35 & 0.55 & 0.15 & 0.20& 0.10 &  0.08& 0.68 &  0.22& 1.19 &  0.28& 1.22 &  0.37\\ 
      &      & 0.20 & 0.25& 0.43 &  0.10& 0.86 &  0.15& 1.56 &  0.17& 1.87 &  0.24\\ 
      &      & 0.25 & 0.30& 0.49 &  0.08& 1.09 &  0.13& 1.78 &  0.15& 2.20 &  0.22\\ 
      &      & 0.30 & 0.35& 0.43 &  0.07& 1.34 &  0.11& 2.14 &  0.19& 2.37 &  0.14\\ 
      &      & 0.35 & 0.40& 0.57 &  0.07& 1.14 &  0.07& 2.02 &  0.12& 2.61 &  0.18\\ 
      &      & 0.40 & 0.45& 0.63 &  0.07& 1.18 &  0.09& 1.88 &  0.12& 2.39 &  0.12\\ 
      &      & 0.45 & 0.50& 0.62 &  0.07& 1.20 &  0.08& 1.93 &  0.15& 2.37 &  0.12\\ 
      &      & 0.50 & 0.60& 0.48 &  0.06& 1.14 &  0.07& 1.89 &  0.12& 2.25 &  0.14\\ 
      &      & 0.60 & 0.70& 0.24 &  0.05& 0.95 &  0.11& 1.60 &  0.17& 1.98 &  0.21\\ 
      &      & 0.70 & 0.80& 0.15 &  0.04& 0.54 &  0.11& 1.08 &  0.18& 1.46 &  0.22\\ 
\hline
 0.55 & 0.75 & 0.10 & 0.15& 0.35 &  0.16& 0.51 &  0.22& 0.97 &  0.34& 0.89 &  0.37\\ 
      &      & 0.15 & 0.20& 0.48 &  0.12& 1.02 &  0.18& 1.76 &  0.18& 1.83 &  0.29\\ 
      &      & 0.20 & 0.25& 0.67 &  0.09& 1.31 &  0.13& 2.05 &  0.17& 2.80 &  0.22\\ 
      &      & 0.25 & 0.30& 0.71 &  0.11& 1.14 &  0.09& 1.94 &  0.11& 2.46 &  0.14\\ 
      &      & 0.30 & 0.35& 0.65 &  0.10& 1.25 &  0.11& 1.95 &  0.12& 2.54 &  0.16\\ 
      &      & 0.35 & 0.40& 0.56 &  0.06& 1.20 &  0.08& 1.75 &  0.12& 2.58 &  0.15\\ 
      &      & 0.40 & 0.45& 0.49 &  0.05& 1.07 &  0.07& 1.70 &  0.10& 2.39 &  0.14\\ 
      &      & 0.45 & 0.50& 0.45 &  0.05& 0.99 &  0.07& 1.56 &  0.09& 2.17 &  0.13\\ 
      &      & 0.50 & 0.60& 0.30 &  0.04& 0.78 &  0.07& 1.34 &  0.11& 1.66 &  0.12\\ 
      &      & 0.60 & 0.70& 0.18 &  0.04& 0.47 &  0.07& 0.88 &  0.12& 1.19 &  0.13\\ 
      &      & 0.70 & 0.80& 0.086 &  0.023& 0.29 &  0.06& 0.54 &  0.10& 0.82 &  0.14\\ 
\hline
 0.75 & 0.95 & 0.10 & 0.15& 0.60 &  0.15& 0.80 &  0.19& 1.02 &  0.27& 1.03 &  0.30\\ 
      &      & 0.15 & 0.20& 0.76 &  0.11& 1.34 &  0.13& 1.99 &  0.14& 2.22 &  0.21\\ 
      &      & 0.20 & 0.25& 0.80 &  0.09& 1.58 &  0.13& 2.08 &  0.11& 2.30 &  0.14\\ 
      &      & 0.25 & 0.30& 0.61 &  0.07& 1.29 &  0.10& 1.92 &  0.11& 2.28 &  0.17\\ 
      &      & 0.30 & 0.35& 0.57 &  0.06& 1.09 &  0.09& 1.75 &  0.09& 2.05 &  0.12\\ 
      &      & 0.35 & 0.40& 0.48 &  0.06& 1.02 &  0.07& 1.47 &  0.08& 1.72 &  0.12\\ 
      &      & 0.40 & 0.45& 0.43 &  0.05& 0.82 &  0.05& 1.24 &  0.08& 1.61 &  0.09\\ 
      &      & 0.45 & 0.50& 0.32 &  0.04& 0.68 &  0.05& 1.09 &  0.06& 1.37 &  0.08\\ 
      &      & 0.50 & 0.60& 0.160 &  0.034& 0.45 &  0.05& 0.80 &  0.07& 1.03 &  0.08\\ 
      &      & 0.60 & 0.70& 0.061 &  0.018& 0.25 &  0.04& 0.48 &  0.07& 0.66 &  0.09\\ 
\hline
 0.95 & 1.15 & 0.10 & 0.15& 0.55 &  0.14& 0.98 &  0.19& 1.24 &  0.26& 1.46 &  0.31\\ 
      &      & 0.15 & 0.20& 0.85 &  0.08& 1.49 &  0.11& 2.08 &  0.14& 2.44 &  0.20\\ 
      &      & 0.20 & 0.25& 0.78 &  0.08& 1.37 &  0.09& 1.86 &  0.12& 2.47 &  0.12\\ 
      &      & 0.25 & 0.30& 0.61 &  0.06& 1.05 &  0.08& 1.66 &  0.10& 1.88 &  0.11\\ 
      &      & 0.30 & 0.35& 0.45 &  0.06& 0.77 &  0.05& 1.32 &  0.08& 1.59 &  0.10\\ 
      &      & 0.35 & 0.40& 0.34 &  0.04& 0.71 &  0.05& 1.02 &  0.07& 1.26 &  0.08\\ 
      &      & 0.40 & 0.45& 0.27 &  0.04& 0.55 &  0.04& 0.80 &  0.05& 0.99 &  0.06\\ 
      &      & 0.45 & 0.50& 0.16 &  0.04& 0.41 &  0.04& 0.62 &  0.04& 0.78 &  0.06\\ 
      &      & 0.50 & 0.60& 0.073 &  0.019& 0.22 &  0.04& 0.38 &  0.04& 0.48 &  0.05\\ 
\hline
\end{tabular}
\end{center}
\end{table}

\begin{table}[hp!] 
\begin{center}
\begin{tabular}{rrrr|r@{$\pm$}lr@{$\pm$}lr@{$\pm$}lr@{$\pm$}l} 
\hline
$\theta_{\hbox{\small min}}$ &
$\theta_{\hbox{\small max}}$ &
$p_{\hbox{\small min}}$ &
$p_{\hbox{\small max}}$ &
\multicolumn{8}{c}{$d^2\sigma^{\pi^+}/(dpd\theta)$} 
\\
(rad) & (rad) & (\GeVc) & (\GeVc) &
\multicolumn{8}{c}{(barn/(\GeVc rad))}
\\
  &  &  & 
&\multicolumn{2}{c}{$ \bf{3 \ \GeVc}$} 
&\multicolumn{2}{c}{$ \bf{5 \ \GeVc}$} 
&\multicolumn{2}{c}{$ \bf{8 \ \GeVc}$} 
&\multicolumn{2}{c}{$ \bf{12 \ \GeVc}$} 
\\ 
\hline
 1.15 & 1.35 & 0.10 & 0.15& 0.45 &  0.13& 1.05 &  0.21& 1.52 &  0.30& 1.83 &  0.36\\ 
      &      & 0.15 & 0.20& 0.85 &  0.09& 1.41 &  0.12& 2.18 &  0.17& 2.45 &  0.21\\ 
      &      & 0.20 & 0.25& 0.67 &  0.07& 1.19 &  0.08& 1.91 &  0.11& 2.11 &  0.15\\ 
      &      & 0.25 & 0.30& 0.46 &  0.05& 0.86 &  0.06& 1.28 &  0.10& 1.49 &  0.10\\ 
      &      & 0.30 & 0.35& 0.33 &  0.05& 0.66 &  0.05& 0.87 &  0.06& 1.06 &  0.08\\ 
      &      & 0.35 & 0.40& 0.233 &  0.031& 0.47 &  0.04& 0.66 &  0.04& 0.87 &  0.05\\ 
      &      & 0.40 & 0.45& 0.153 &  0.025& 0.30 &  0.04& 0.501 &  0.032& 0.67 &  0.04\\ 
      &      & 0.45 & 0.50& 0.079 &  0.019& 0.188 &  0.030& 0.374 &  0.027& 0.48 &  0.04\\ 
\hline
 1.35 & 1.55 & 0.10 & 0.15& 0.38 &  0.11& 1.16 &  0.25& 1.40 &  0.36& 1.90 &  0.37\\ 
      &      & 0.15 & 0.20& 0.79 &  0.11& 1.43 &  0.14& 1.94 &  0.22& 2.44 &  0.26\\ 
      &      & 0.20 & 0.25& 0.67 &  0.07& 1.01 &  0.10& 1.67 &  0.12& 2.06 &  0.14\\ 
      &      & 0.25 & 0.30& 0.44 &  0.05& 0.62 &  0.06& 1.08 &  0.08& 1.19 &  0.09\\ 
      &      & 0.30 & 0.35& 0.26 &  0.04& 0.42 &  0.04& 0.67 &  0.05& 0.85 &  0.06\\ 
      &      & 0.35 & 0.40& 0.152 &  0.025& 0.288 &  0.030& 0.47 &  0.04& 0.63 &  0.05\\ 
      &      & 0.40 & 0.45& 0.082 &  0.018& 0.170 &  0.021& 0.351 &  0.024& 0.43 &  0.04\\ 
      &      & 0.45 & 0.50& 0.042 &  0.011& 0.103 &  0.016& 0.252 &  0.026& 0.30 &  0.04\\ 
\hline
 1.55 & 1.75 & 0.10 & 0.15& 0.58 &  0.15& 1.11 &  0.24& 1.40 &  0.33& 1.67 &  0.41\\ 
      &      & 0.15 & 0.20& 0.78 &  0.09& 1.33 &  0.14& 1.65 &  0.17& 2.06 &  0.23\\ 
      &      & 0.20 & 0.25& 0.53 &  0.06& 0.95 &  0.08& 1.36 &  0.09& 1.69 &  0.13\\ 
      &      & 0.25 & 0.30& 0.32 &  0.05& 0.49 &  0.05& 0.75 &  0.06& 0.90 &  0.09\\ 
      &      & 0.30 & 0.35& 0.168 &  0.030& 0.341 &  0.032& 0.49 &  0.04& 0.57 &  0.05\\ 
      &      & 0.35 & 0.40& 0.081 &  0.018& 0.222 &  0.029& 0.327 &  0.030& 0.353 &  0.031\\ 
      &      & 0.40 & 0.45& 0.046 &  0.012& 0.119 &  0.020& 0.218 &  0.022& 0.228 &  0.023\\ 
      &      & 0.45 & 0.50& 0.030 &  0.009& 0.071 &  0.014& 0.149 &  0.018& 0.169 &  0.020\\ 
\hline
 1.75 & 1.95 & 0.10 & 0.15& 0.69 &  0.13& 0.93 &  0.19& 1.14 &  0.19& 1.39 &  0.23\\ 
      &      & 0.15 & 0.20& 0.68 &  0.07& 1.08 &  0.08& 1.27 &  0.09& 1.57 &  0.11\\ 
      &      & 0.20 & 0.25& 0.41 &  0.05& 0.74 &  0.06& 0.94 &  0.06& 1.14 &  0.08\\ 
      &      & 0.25 & 0.30& 0.24 &  0.04& 0.35 &  0.05& 0.49 &  0.05& 0.69 &  0.08\\ 
      &      & 0.30 & 0.35& 0.113 &  0.024& 0.182 &  0.023& 0.309 &  0.027& 0.31 &  0.05\\ 
      &      & 0.35 & 0.40& 0.057 &  0.018& 0.124 &  0.017& 0.187 &  0.019& 0.198 &  0.022\\ 
      &      & 0.40 & 0.45& 0.07 &  0.04& 0.087 &  0.015& 0.111 &  0.017& 0.152 &  0.019\\ 
      &      & 0.45 & 0.50& 0.019 &  0.009& 0.046 &  0.012& 0.065 &  0.012& 0.098 &  0.020\\ 
\hline
 1.95 & 2.15 & 0.10 & 0.15& 0.61 &  0.12& 0.82 &  0.14& 1.05 &  0.16& 1.33 &  0.20\\ 
      &      & 0.15 & 0.20& 0.53 &  0.06& 0.73 &  0.06& 1.02 &  0.05& 1.07 &  0.07\\ 
      &      & 0.20 & 0.25& 0.20 &  0.04& 0.45 &  0.04& 0.69 &  0.05& 0.75 &  0.06\\ 
      &      & 0.25 & 0.30& 0.118 &  0.027& 0.24 &  0.04& 0.33 &  0.04& 0.36 &  0.05\\ 
      &      & 0.30 & 0.35& 0.064 &  0.022& 0.098 &  0.021& 0.172 &  0.022& 0.162 &  0.021\\ 
      &      & 0.35 & 0.40& 0.016 &  0.009& 0.062 &  0.011& 0.098 &  0.015& 0.109 &  0.018\\ 
      &      & 0.40 & 0.45& 0.011 &  0.007& 0.050 &  0.012& 0.049 &  0.012& 0.064 &  0.014\\ 
      &      & 0.45 & 0.50& 0.011 &  0.009& 0.027 &  0.010& 0.025 &  0.007& 0.034 &  0.009\\ 
\hline

\end{tabular}
\end{center}
\end{table}

\begin{table}[hp!] 
\begin{center}
  \caption{\label{tab:xsec-n}
    HARP results for the double-differential $\pi^-$ production
    cross-section in the laboratory system,
    $d^2\sigma^{\pi^-}/(dpd\theta)$. Each row refers to a
    different $(p_{\hbox{\small min}} \le p<p_{\hbox{\small max}},
    \theta_{\hbox{\small min}} \le \theta<\theta_{\hbox{\small max}})$ bin,
    where $p$ and $\theta$ are the pion momentum and polar angle, respectively.
    The central value as well as the square-root of the diagonal elements
    of the covariance matrix are given.}
\vspace{2mm}
\begin{tabular}{rrrr|r@{$\pm$}lr@{$\pm$}lr@{$\pm$}lr@{$\pm$}l} 
\hline
$\theta_{\hbox{\small min}}$ &
$\theta_{\hbox{\small max}}$ &
$p_{\hbox{\small min}}$ &
$p_{\hbox{\small max}}$ &
\multicolumn{8}{c}{$d^2\sigma^{\pi^-}/(dpd\theta)$} 
\\
(rad) & (rad) & (\GeVc) & (\GeVc) &
\multicolumn{8}{c}{(barn/(\GeVc rad))}
\\
  &  &  & 
&\multicolumn{2}{c}{$ \bf{3 \ \GeVc}$} 
&\multicolumn{2}{c}{$ \bf{5 \ \GeVc}$} 
&\multicolumn{2}{c}{$ \bf{8 \ \GeVc}$} 
&\multicolumn{2}{c}{$ \bf{12 \ \GeVc}$} 
\\ 
\hline
 0.35 & 0.55 & 0.15 & 0.20& 0.31 &  0.13& 0.64 &  0.23& 1.33 &  0.32& 1.71 &  0.42\\ 
      &      & 0.20 & 0.25& 0.31 &  0.11& 0.99 &  0.14& 1.71 &  0.18& 1.87 &  0.26\\ 
      &      & 0.25 & 0.30& 0.30 &  0.09& 0.99 &  0.11& 1.72 &  0.12& 2.32 &  0.22\\ 
      &      & 0.30 & 0.35& 0.53 &  0.08& 0.85 &  0.08& 1.74 &  0.13& 2.08 &  0.14\\ 
      &      & 0.35 & 0.40& 0.48 &  0.07& 0.85 &  0.07& 1.65 &  0.10& 2.12 &  0.14\\ 
      &      & 0.40 & 0.45& 0.34 &  0.05& 0.76 &  0.06& 1.50 &  0.08& 1.87 &  0.11\\ 
      &      & 0.45 & 0.50& 0.28 &  0.04& 0.70 &  0.05& 1.39 &  0.07& 1.58 &  0.09\\ 
      &      & 0.50 & 0.60& 0.24 &  0.04& 0.77 &  0.06& 1.23 &  0.07& 1.56 &  0.10\\ 
      &      & 0.60 & 0.70& 0.23 &  0.04& 0.65 &  0.07& 1.12 &  0.09& 1.39 &  0.13\\ 
      &      & 0.70 & 0.80& 0.20 &  0.05& 0.48 &  0.07& 0.91 &  0.11& 1.18 &  0.16\\ 
\hline
 0.55 & 0.75 & 0.10 & 0.15& 0.45 &  0.18& 0.82 &  0.28& 1.17 &  0.39& 1.43 &  0.53\\ 
      &      & 0.15 & 0.20& 0.58 &  0.10& 1.28 &  0.18& 1.67 &  0.20& 2.12 &  0.25\\ 
      &      & 0.20 & 0.25& 0.40 &  0.07& 1.11 &  0.10& 2.02 &  0.16& 2.49 &  0.19\\ 
      &      & 0.25 & 0.30& 0.46 &  0.09& 1.17 &  0.10& 1.83 &  0.12& 2.39 &  0.17\\ 
      &      & 0.30 & 0.35& 0.37 &  0.06& 0.88 &  0.07& 1.56 &  0.10& 2.13 &  0.13\\ 
      &      & 0.35 & 0.40& 0.31 &  0.04& 0.85 &  0.07& 1.43 &  0.08& 1.82 &  0.09\\ 
      &      & 0.40 & 0.45& 0.31 &  0.05& 0.78 &  0.06& 1.34 &  0.07& 1.57 &  0.08\\ 
      &      & 0.45 & 0.50& 0.27 &  0.04& 0.61 &  0.05& 1.14 &  0.06& 1.48 &  0.07\\ 
      &      & 0.50 & 0.60& 0.26 &  0.04& 0.50 &  0.04& 1.00 &  0.05& 1.29 &  0.07\\ 
      &      & 0.60 & 0.70& 0.16 &  0.04& 0.43 &  0.04& 0.84 &  0.07& 1.12 &  0.10\\ 
      &      & 0.70 & 0.80& 0.086 &  0.027& 0.36 &  0.05& 0.67 &  0.08& 0.96 &  0.13\\ 
\hline
 0.75 & 0.95 & 0.10 & 0.15& 0.50 &  0.16& 1.08 &  0.22& 1.32 &  0.32& 1.74 &  0.41\\ 
      &      & 0.15 & 0.20& 0.60 &  0.08& 1.25 &  0.13& 2.00 &  0.13& 2.26 &  0.16\\ 
      &      & 0.20 & 0.25& 0.57 &  0.08& 1.14 &  0.09& 1.93 &  0.13& 2.39 &  0.17\\ 
      &      & 0.25 & 0.30& 0.61 &  0.07& 0.94 &  0.07& 1.67 &  0.09& 2.27 &  0.13\\ 
      &      & 0.30 & 0.35& 0.32 &  0.05& 0.84 &  0.07& 1.34 &  0.07& 1.85 &  0.11\\ 
      &      & 0.35 & 0.40& 0.33 &  0.04& 0.74 &  0.05& 1.19 &  0.06& 1.46 &  0.09\\ 
      &      & 0.40 & 0.45& 0.25 &  0.04& 0.57 &  0.04& 0.98 &  0.06& 1.31 &  0.07\\ 
      &      & 0.45 & 0.50& 0.182 &  0.028& 0.52 &  0.04& 0.88 &  0.06& 1.24 &  0.06\\ 
      &      & 0.50 & 0.60& 0.136 &  0.024& 0.46 &  0.04& 0.71 &  0.04& 1.02 &  0.06\\ 
      &      & 0.60 & 0.70& 0.080 &  0.018& 0.31 &  0.04& 0.54 &  0.05& 0.72 &  0.08\\ 
\hline
 0.95 & 1.15 & 0.10 & 0.15& 0.71 &  0.14& 1.27 &  0.21& 1.76 &  0.29& 2.00 &  0.36\\ 
      &      & 0.15 & 0.20& 0.64 &  0.07& 1.23 &  0.09& 2.04 &  0.14& 2.20 &  0.14\\ 
      &      & 0.20 & 0.25& 0.54 &  0.07& 1.07 &  0.08& 1.75 &  0.13& 2.02 &  0.13\\ 
      &      & 0.25 & 0.30& 0.53 &  0.07& 0.83 &  0.06& 1.41 &  0.10& 1.69 &  0.10\\ 
      &      & 0.30 & 0.35& 0.32 &  0.05& 0.74 &  0.06& 1.09 &  0.07& 1.41 &  0.08\\ 
      &      & 0.35 & 0.40& 0.37 &  0.05& 0.51 &  0.05& 0.96 &  0.05& 1.12 &  0.06\\ 
      &      & 0.40 & 0.45& 0.25 &  0.05& 0.376 &  0.029& 0.78 &  0.05& 0.93 &  0.06\\ 
      &      & 0.45 & 0.50& 0.143 &  0.026& 0.328 &  0.026& 0.61 &  0.04& 0.75 &  0.05\\ 
      &      & 0.50 & 0.60& 0.098 &  0.019& 0.264 &  0.025& 0.443 &  0.033& 0.55 &  0.05\\ 
\hline
\end{tabular}
\end{center}
\end{table}

\begin{table}[hp!] 
\begin{center}
\begin{tabular}{rrrr|r@{$\pm$}lr@{$\pm$}lr@{$\pm$}lr@{$\pm$}l} 
\hline
$\theta_{\hbox{\small min}}$ &
$\theta_{\hbox{\small max}}$ &
$p_{\hbox{\small min}}$ &
$p_{\hbox{\small max}}$ &
\multicolumn{8}{c}{$d^2\sigma^{\pi^-}/(dpd\theta)$} 
\\
(rad) & (rad) & (\GeVc) & (\GeVc) &
\multicolumn{8}{c}{(barn/(\GeVc rad))}
\\
  &  &  & 
&\multicolumn{2}{c}{$ \bf{3 \ \GeVc}$} 
&\multicolumn{2}{c}{$ \bf{5 \ \GeVc}$} 
&\multicolumn{2}{c}{$ \bf{8 \ \GeVc}$} 
&\multicolumn{2}{c}{$ \bf{12 \ \GeVc}$} 
\\ 
\hline
 1.15 & 1.35 & 0.10 & 0.15& 0.67 &  0.13& 1.15 &  0.21& 2.09 &  0.36& 2.55 &  0.43\\ 
      &      & 0.15 & 0.20& 0.58 &  0.07& 1.14 &  0.10& 2.00 &  0.14& 2.31 &  0.19\\ 
      &      & 0.20 & 0.25& 0.46 &  0.05& 0.96 &  0.07& 1.59 &  0.10& 1.84 &  0.12\\ 
      &      & 0.25 & 0.30& 0.26 &  0.04& 0.67 &  0.06& 1.21 &  0.08& 1.48 &  0.09\\ 
      &      & 0.30 & 0.35& 0.160 &  0.025& 0.56 &  0.05& 0.87 &  0.06& 1.09 &  0.07\\ 
      &      & 0.35 & 0.40& 0.168 &  0.027& 0.42 &  0.04& 0.68 &  0.04& 0.83 &  0.05\\ 
      &      & 0.40 & 0.45& 0.141 &  0.025& 0.293 &  0.026& 0.530 &  0.031& 0.74 &  0.05\\ 
      &      & 0.45 & 0.50& 0.085 &  0.018& 0.227 &  0.021& 0.432 &  0.030& 0.57 &  0.05\\ 
\hline
 1.35 & 1.55 & 0.10 & 0.15& 0.55 &  0.13& 1.18 &  0.21& 2.29 &  0.54& 2.92 &  0.64\\ 
      &      & 0.15 & 0.20& 0.53 &  0.07& 1.04 &  0.12& 1.82 &  0.17& 2.29 &  0.24\\ 
      &      & 0.20 & 0.25& 0.34 &  0.05& 0.76 &  0.08& 1.24 &  0.11& 1.79 &  0.13\\ 
      &      & 0.25 & 0.30& 0.31 &  0.04& 0.58 &  0.06& 0.80 &  0.07& 1.20 &  0.10\\ 
      &      & 0.30 & 0.35& 0.224 &  0.035& 0.40 &  0.04& 0.55 &  0.04& 0.78 &  0.07\\ 
      &      & 0.35 & 0.40& 0.144 &  0.026& 0.282 &  0.028& 0.452 &  0.033& 0.60 &  0.05\\ 
      &      & 0.40 & 0.45& 0.081 &  0.019& 0.206 &  0.022& 0.334 &  0.025& 0.48 &  0.04\\ 
      &      & 0.45 & 0.50& 0.052 &  0.012& 0.135 &  0.016& 0.265 &  0.021& 0.373 &  0.031\\ 
\hline
 1.55 & 1.75 & 0.10 & 0.15& 0.60 &  0.15& 1.23 &  0.27& 1.64 &  0.35& 2.50 &  0.53\\ 
      &      & 0.15 & 0.20& 0.51 &  0.08& 1.09 &  0.11& 1.55 &  0.16& 2.12 &  0.22\\ 
      &      & 0.20 & 0.25& 0.31 &  0.04& 0.70 &  0.07& 1.10 &  0.08& 1.52 &  0.11\\ 
      &      & 0.25 & 0.30& 0.185 &  0.032& 0.45 &  0.05& 0.63 &  0.06& 0.95 &  0.08\\ 
      &      & 0.30 & 0.35& 0.134 &  0.025& 0.295 &  0.032& 0.403 &  0.033& 0.59 &  0.06\\ 
      &      & 0.35 & 0.40& 0.103 &  0.022& 0.212 &  0.024& 0.292 &  0.025& 0.429 &  0.033\\ 
      &      & 0.40 & 0.45& 0.057 &  0.016& 0.134 &  0.016& 0.195 &  0.016& 0.345 &  0.031\\ 
      &      & 0.45 & 0.50& 0.040 &  0.012& 0.101 &  0.013& 0.139 &  0.012& 0.251 &  0.023\\ 
\hline
 1.75 & 1.95 & 0.10 & 0.15& 0.71 &  0.12& 1.27 &  0.22& 1.34 &  0.21& 1.99 &  0.31\\ 
      &      & 0.15 & 0.20& 0.48 &  0.06& 1.04 &  0.08& 1.20 &  0.09& 1.64 &  0.12\\ 
      &      & 0.20 & 0.25& 0.30 &  0.05& 0.55 &  0.05& 0.81 &  0.06& 1.06 &  0.09\\ 
      &      & 0.25 & 0.30& 0.180 &  0.033& 0.285 &  0.034& 0.46 &  0.05& 0.58 &  0.05\\ 
      &      & 0.30 & 0.35& 0.112 &  0.028& 0.199 &  0.024& 0.241 &  0.027& 0.38 &  0.04\\ 
      &      & 0.35 & 0.40& 0.054 &  0.017& 0.138 &  0.019& 0.200 &  0.016& 0.241 &  0.024\\ 
      &      & 0.40 & 0.45& 0.038 &  0.014& 0.096 &  0.014& 0.175 &  0.015& 0.186 &  0.021\\ 
      &      & 0.45 & 0.50& 0.026 &  0.012& 0.069 &  0.012& 0.118 &  0.014& 0.107 &  0.016\\ 
\hline
 1.95 & 2.15 & 0.10 & 0.15& 0.51 &  0.11& 1.23 &  0.18& 1.29 &  0.16& 1.78 &  0.28\\ 
      &      & 0.15 & 0.20& 0.44 &  0.06& 0.76 &  0.07& 0.98 &  0.06& 1.31 &  0.09\\ 
      &      & 0.20 & 0.25& 0.24 &  0.05& 0.38 &  0.04& 0.59 &  0.04& 0.68 &  0.07\\ 
      &      & 0.25 & 0.30& 0.080 &  0.030& 0.198 &  0.030& 0.34 &  0.05& 0.383 &  0.034\\ 
      &      & 0.30 & 0.35& 0.020 &  0.010& 0.128 &  0.019& 0.148 &  0.017& 0.285 &  0.034\\ 
      &      & 0.35 & 0.40& 0.019 &  0.010& 0.114 &  0.020& 0.124 &  0.014& 0.147 &  0.023\\ 
      &      & 0.40 & 0.45& 0.034 &  0.017& 0.056 &  0.015& 0.126 &  0.016& 0.113 &  0.017\\ 
      &      & 0.45 & 0.50& 0.037 &  0.018& 0.032 &  0.010& 0.091 &  0.014& 0.111 &  0.017\\ 
\hline
\end{tabular}
\end{center}
\end{table}
%


\section{Alternative analysis}
\label{sec:alternative}

The data taken in the 5~\GeVc beam have been analysed with an
alternative analysis which is described in detail in
Ref.~\cite{ref:silvia}. 
While the  unfolding procedure corrects for the efficiency, resolution
smearing and a number of backgrounds in an integrated manner, this
method makes sequential corrections for PID, energy-loss, efficiency 
and migration due to resolution smearing.

The alternative analysis proceeds with the following steps:
\begin{itemize}
\item
The beam particle selection is identical to the one in the analysis
described in this paper.
\item
The cut in the selection of the number of events accepted per
spill is applied at 50 events
reducing the sample\footnote{This cut was decided consistently with other
     cuts on the impact point to define as clean as possible data samples.}.
\item
The basic track selection is identical.  However, a stricter definition
of the target volume is used.
The cuts are applied at $| \dzeroprime | < 8.5 \ \mm$ and 
$-7.2 \ \mm < \zzeroprime \sin \theta < 12.8 \ \mm$, corresponding to
two standard deviations in the resolution.
This selection reduces the tertiary background, but the
efficiency correction is larger. 
\item
A PID selection is applied based on the \dedx of the particles and will be
described below in more detail.
The main difference is the method to determine the efficiency and
backgrounds and the choice of the cut (which is more efficient 
but has a lower purity).
This is one of the two main differences between the two methods.
\item
The correction for efficiency and absorption of secondary particles is
applied bin--by--bin as multiplicative correction.
This correction, although differently applied, is the same for both
methods. 
\item
The energy-loss correction is applied on a track--by--track basis, while
in the method described in this paper it was part of the unfolding
procedure. 
\item
The resolution smearing correction is simpler; it does not consider
the migration between angle bins (which is negligible), while it applies a
multiplicative correction to account for the momentum smearing.
It thus introduces a dependence on the assumed input spectra for this
correction which contributes to the systematic error. 
\item
The correction for \piz background follows the same assumptions, but
is quite different in implementation.
The relative size of the subtraction is smaller owing to the stricter
PID separation between pions and electrons (positrons).
\item
No subtraction for tertiary particles is applied.
Although this is an approximation, the stricter target volume
definition reduces this background to less than 2\%.
\end{itemize}

The various corrections have been applied using the same simulation
program as described in this paper.
The differences in the analyses, both of principle and technical
nature are large enough to provide a useful cross-check of the
methods. 
Since the main difference between the analyses is given by the
PID, this issue is described in somewhat more detail below.

The PID is based on a selection in \dedx as a function of the momentum.
The purity and the efficiency is evaluated
studying the \dedx spectra in the different momentum bins for each
angular bin separately 
by fitting two Landau distributions to each spectrum.
Particle separation between protons and pions can be achieved
with a purity of about 99\% up to 400~\MeVc (see
Fig.~\ref{fig:alternative:dedx300}). 
Above this value, efficiency and purity  are lower as shown in
Fig.~\ref{fig:alternative:dedx75}.
The two-component fits are used to determine these quantities
as a function of the momentum in angular bins. 

The electron contamination can be evaluated only for momenta less than
125 MeV/$c$. 
Above this value it is evaluated using a simulation. 
A similar assumption is made for the \piz spectrum as in the analysis
described in this paper.
Using momenta below 125 MeV/$c$ where the electrons can be identified,
the simulated data are normalized to obtain the same 
number of electrons and positrons as in the measured data.

\begin{figure}[tbp]
\begin{center}
\epsfig{figure=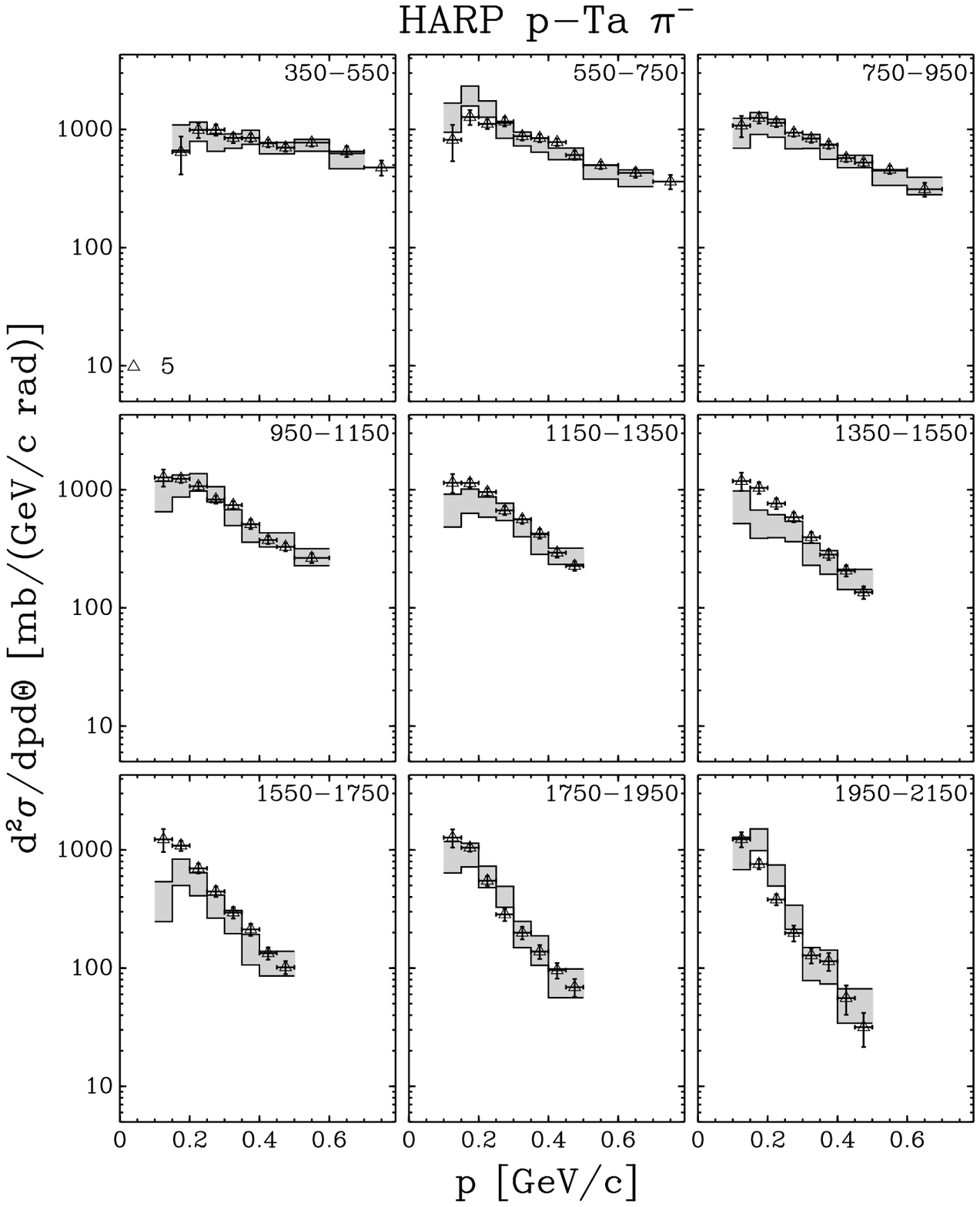,width=0.47\textwidth} 
~
\epsfig{figure=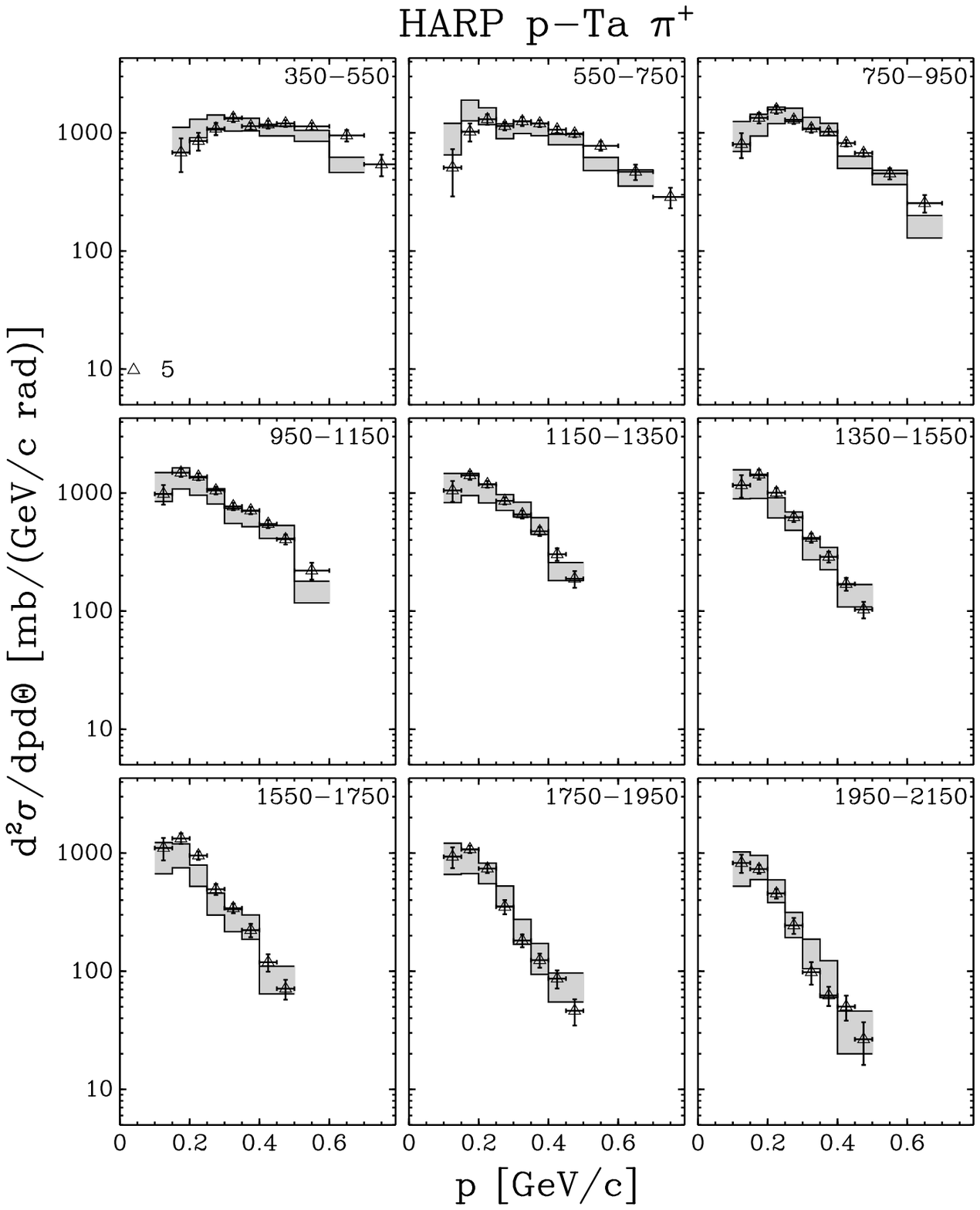,width=0.47\textwidth} 
\caption{
Comparison of the double-differential cross-sections measured for \pip
(right) and \pim (left) production in 
p--Ta interactions as a function of momentum displayed in different
angular bins using the two analyses.
The results of the alternative analysis are shown as shaded band and
are available only for 5~\GeVc incident beam momentum.
The width of the band represents an estimate of the uncorrelated error
 (one standard deviation) between the two methods.
The results of the standard analysis described in this paper are
 represented by data points.
}
\label{fig:xs-p-th-comp}
\end{center}
\end{figure}

The results of the alternative analysis are compatible with the
results reported in Fig.~\ref{fig:xs-p-th-pbeam-plus} and
\ref{fig:xs-p-th-pbeam-minus} within the quoted
systematic errors.
The comparison is shown in Fig.~\ref{fig:xs-p-th-comp}.
One observes good agreement between the two sets of spectra.
Taking into account the large number of differences between the two
approaches (event selection, track selection, energy-loss correction,
particle identification, background subtraction) this constitutes an
important cross-check of the correctness of the two analysis
approaches.

\end{appendix}


\end{document}